\documentstyle[12pt]{article}
\newlength{\pubnumber} 

\catcode`\@=11
\@addtoreset{equation}{section}

\def\section{\@startsection{section}{1}{\z@}{3.5ex plus 1ex minus .2ex}
 {2.3ex plus .2ex}{\large\bf}}
\def\subsection{\@startsection{subsection}{2}{\z@}{2.3ex plus .2ex}
 {2.3ex plus .2ex}{\bf}}

\begin{document}

\begin{titlepage}
\samepage{
\setcounter{page}{1}
\rightline{McGill/94-04}
\rightline{\tt hep-th/9402006}
\rightline{January 1994}
\rightline{Published:  {\it Nucl.\ Phys.}\/  {\bf B429} (1994) 533}
\vfill
\begin{center}
 {\Large \bf Modular Invariance, Finiteness,\\
     and Misaligned Supersymmetry:\\
  New Constraints on the Numbers of Physical\\
   String States\\}
\vfill
 {\large Keith R. Dienes\footnote{
   Present address:
      School of Natural Sciences, Institute for Advanced Study,
      Olden Lane, Princeton, NJ  08540  USA.  E-mail:
      dienes@guinness.ias.edu.}\\}
\vspace{.12in}
 {\it  Department of Physics, McGill University\\
  3600 University St., Montr\'eal, Qu\'ebec~H3A-2T8~~Canada\\}
\end{center}
\vfill
\begin{abstract}
  {\rm
    We investigate the generic distribution of bosonic and fermionic
    states at all mass levels in non-supersymmetric string theories,
    and find that a hidden ``misaligned supersymmetry'' must always
    appear in the string spectrum.
    We show that this misaligned supersymmetry is ultimately
    responsible for the finiteness of string amplitudes in the absence of
    full spacetime supersymmetry, and therefore the existence of
    misaligned supersymmetry provides a natural constraint
    on the degree to which spacetime supersymmetry can be broken in
    string theory without destroying the finiteness of string amplitudes.
    Misaligned supersymmetry also explains how the requirements
    of modular invariance and absence of physical tachyons generically
    affect the distribution of states throughout the string spectrum,
    and implicitly furnishes a two-variable generalization
    of some well-known results in the theory of modular functions. }
\end{abstract}
\vfill}
\end{titlepage}

\setcounter{footnote}{0}

\def\beq{\begin{equation}}
\def\eeq{\end{equation}}
\def\beqn{\begin{eqnarray}}
\def\eeqn{\end{eqnarray}}

\def\ie{{\it i.e.}}
\def\eg{{\it e.g.}}
\def\half{{\textstyle{1\over 2}}}
\def\third{{\textstyle {1\over3}}}
\def\quarter{{\textstyle {1\over4}}}

\def\Htw{{\tilde H}}
\def\chibar{{\overline{\chi}}}
\def\qbar{{\overline{q}}}
\def\ibar{{\overline{\imath}}}
\def\jbar{{\overline{\jmath}}}
\def\Hbar{{\overline{H}}}
\def\Qbar{{\overline{Q}}}
\def\abar{{\overline{a}}}
\def\alphabar{{\overline{\alpha}}}
\def\betabar{{\overline{\beta}}}
\def\tautwo{{ \tau_2 }}
\def\thetatwo{{ \vartheta_2 }}
\def\thetathree{{ \vartheta_3 }}
\def\thetafour{{ \vartheta_4 }}
\def\ttwo{{\vartheta_2}}
\def\tthree{{\vartheta_3}}
\def\tfour{{\vartheta_4}}
\def\ti{{\vartheta_i}}
\def\tj{{\vartheta_j}}
\def\tk{{\vartheta_k}}
\def\calF{{\cal F}}
\def\smallmatrix#1#2#3#4{{ {{#1}~{#2}\choose{#3}~{#4}} }}
\def\ab{{\alpha\beta}}
\def\Minv{{ (M^{-1}_\ab)_{ij} }}
\def\bone{{\bf 1}}
\def\ii{{(i)}}


\def\inbar{\,\vrule height1.5ex width.4pt depth0pt}

\def\IC{\relax\hbox{$\inbar\kern-.3em{\rm C}$}}
\def\IQ{\relax\hbox{$\inbar\kern-.3em{\rm Q}$}}
\def\IR{\relax{\rm I\kern-.18em R}}
 \font\cmss=cmss10 \font\cmsss=cmss10 at 7pt
\def\IZ{\relax\ifmmode\mathchoice
 {\hbox{\cmss Z\kern-.4em Z}}{\hbox{\cmss Z\kern-.4em Z}}
 {\lower.9pt\hbox{\cmsss Z\kern-.4em Z}}
 {\lower1.2pt\hbox{\cmsss Z\kern-.4em Z}}\else{\cmss Z\kern-.4em Z}\fi}

\hyphenation{su-per-sym-met-ric non-su-per-sym-met-ric}
\hyphenation{space-time-super-sym-met-ric}
\hyphenation{mod-u-lar mod-u-lar--in-var-i-ant}


\setcounter{footnote}{0}
\section{Introduction:  Motivation and Overview of Results}

The distribution of states in string theory is an important
but not particularly well-understood issue.
It is well-known, for example, that string theories generically
contain a variety of sectors, each contributing an infinite tower
of states from the massless level to the Planck scale, and it
is also a generic feature that the number of these states as a
function of the worldsheet energy $n$ grows asymptotically as $e^{C\sqrt{n}}$
where $C$ is the inverse Hagedorn temperature of the theory.
Beyond these gross features, however, not much is known.
For example, modular invariance presumably tightly constrains the numbers
of string states at all energy levels, but a precise formulation
of such a constraint is still lacking.  Similarly, the distribution
of bosonic and fermionic
string states at all energy levels is crucial in yielding the ultraviolet
finiteness for which string theory is famous, yet it is not clear
precisely how the actual distribution of such states level-by-level
conspires to achieve this remarkable result.
Of course, if the string theory in question exhibits
spacetime supersymmetry, both issues are rendered somewhat trivial:
there are necessarily equal numbers of bosonic and fermionic states
at every energy level, the one-loop partition function vanishes,
and the divergences from bosonic states are precisely cancelled by
those from fermionic states.
Yet how
does the string spectrum manage to maintain modular invariance and
finiteness in the {\it absence}\/ of spacetime supersymmetry?
Alternatively, to what extent can one break spacetime supersymmetry
in string theory without destroying these desirable features?

In this paper we shall provide some answers to these questions,
and in particular we shall demonstrate that even in the absence
of spacetime supersymmetry, string spectra generically
turn out to exhibit
a residual cancellation, a so-called ``misaligned supersymmetry''.
In fact, this property will turn out to be
completely general, and will describe the
distribution of bosonic and fermionic states in {\it any}\/ string
theory which is modular-invariant and free of physical tachyons.
Moreover, we will also see that this result can be interpreted
as the two-variable generalization of some well-known theorems
in the mathematical theory of modular functions.
Misaligned supersymmetry
is therefore a general result with many applications, and
in the rest of this introductory section
we shall outline the basic issues and results.
Details can then be found in subsequent sections.

\subsection{Motivation:  Some Background Issues in String Theory}

In order to gain insight into the relevant string
issues, we shall begin by discussing
some of the questions raised above.

\subsubsection{How does modular invariance
    constrain the numbers of states in string theory?}

Modular invariance is a powerful symmetry, arising at the one-loop
(toroidal) level in theories with two-dimensional
conformal symmetry as a consequence of the existence
on the torus of ``large'' diffeomorphisms which are not
connected to the identity.
It is usually expressed as
a constraint on the one-loop partition functions $Z(\tau)$
of such theories;  here $\tau$ is the complex torus modular
parameter, and modular invariance requires $Z(\tau)=Z(\tau+1)=
Z(-1/\tau)$.  This constraint is indeed quite restrictive,
limiting the combinations of conformal-field-theory characters
which can appear in the partition functions of modular-invariant theories.
However, the partition function
$Z(\tau)$ also contains within it information concerning the
net degeneracies of physical states at all energy levels in the theory.
How then does modular invariance constrain these degeneracies?

In order to formulate this question more mathematically,
let us first review some basic facts about one-loop partition functions.
Given a torus with modular parameter $\tau$ and a two-dimensional
field theory defined on that torus with left- and right-moving Hamiltonians
$H$ and $\Htw$ respectively, the one-loop partition function is defined
\beq
  Z(\tau) ~\equiv~ \sum_s \, {\rm Tr}_s\, (-1)^F \,q^H\, \qbar^\Htw ~.
\label{one}
\eeq
Here $q\equiv e^{2\pi i\tau}$, the sum is over the different sectors $s$
in the theory which are needed for modular invariance, the trace is over the
Fock space of excitations with different $H$ and $\Htw$ eigenvalues
in that sector, and $F$ is the spacetime fermion-number operator
in the theory ($=0$ for excitations which are spacetime bosons, $=1$
for spacetime fermions).  Since these traces are simply the various characters
$\chi_i$ and $\chibar_j$ of the relevant underlying worldsheet conformal
field theories, one can typically write $Z$ in the form
\beq
    Z ~=~ ({\rm Im}\,\tau)^{1-D/2} \, \sum_{i,j} \, N_{ij} \,
             \chi_i(q) \, \chibar_j(\qbar)
\label{two}
\eeq
where $\chi_i(q)$ and $\chibar_j(\qbar)$ are characters of the
chiral left- and right-moving
conformal field theories, $D$ is number of uncompactified spacetime dimensions,
the summation is over the various contributing
sector combinations $(i,j)$, and $N_{ij}$
is a coefficient matrix constrained by modular invariance.
Different choices for $N_{ij}$ correspond to different
theories.  Since the modular transformation properties of the characters
$\chi$ are typically well-known, the modular invariance of
(\ref{two}) is usually easy to verify.  However, it is often
useful to expand $Z$ as a double power series in $q$ and $\qbar$:
\beq
   Z(q,\qbar)~=~ ({\rm Im}\,\tau)^{1-D/2}\,\sum_{m,n}\,a_{mn}\,\qbar^m\,q^n~.
\label{three}
\eeq
Written in this form, each coefficient $a_{mn}$ is the net number of
states or degrees of freedom in the theory with $H$-eigenvalue $n$ and
$\Htw$-eigenvalue $m$;  typically $m$ and $n$ can have both integer
and non-integer values, and modular invariance requires
that $m=n$ (modulo 1) if $a_{mn}\not =0$.
By ``net'' we mean the number of
spacetime bosonic degrees of freedom {\it minus}\/ the number of those
which are spacetime fermionic.
While states with $m=n$ correspond to actual physical (``on-shell'')
particles in spacetime, those with $m\not= n$
(the so-called {\it unphysical}\/
or ``off-shell'' states) do not contribute to tree-level
amplitudes.
Thus, it is the quantities $\lbrace a_{nn}\rbrace$ for all $n\geq 0$
which give the net degeneracies of physical states at all mass
levels in the theory,
and which form our object of interest.

While it is usually quite
straightforward to derive the constraints on $N_{ij}$ arising
from modular invariance, it proves surprisingly difficult to see how these
translate into a constraint on the net degeneracies
$\lbrace a_{nn}\rbrace$.
Indeed, the generic behavior of $\lbrace a_{nn}\rbrace$ required
by modular invariance is almost completely unknown.
Our result will provide such a general constraint.

\subsubsection{How does the presence of unphysical tachyons
 affect the balance between bosonic and fermionic states in string theory?}

A slightly more physical way of addressing the same issue is to
focus instead on the tachyonic states which generically appear in
string theory.  Recall that
since the worldsheet Hamiltonians
$H$ and $\Htw$ correspond to the spacetime left- and right-moving
(mass)$^2$ of the states, we see that states with $m,n<0$ correspond
to spacetime tachyons.
A theory is thus free of physical tachyons if $a_{nn}=0$
for all $n<0$ in its partition function.
Note, in this regard, that the statement $a_{nn}=0$ for all $n<0$
actually implies the absence of {\it all}\/ tachyons whatsoever,
for there can never be fermionic physical tachyons
in a unitary string theory.  By contrast, a situation with no
net {\it un}\/physical tachyons merely implies that bosonic and
fermionic unphysical tachyons occur in equal numbers.

Now, it is well-known that any $D>2$ string theory in which
there are no net numbers of physical or unphysical tachyons
must necessarily have equal numbers of bosonic and fermionic states
at all mass levels:
\beqn
    ~&& {\rm no~net~physical~or~unphysical~tachyons}~~\nonumber\\
     ~&&~~~~~~~~~ \Longleftrightarrow ~~~
       {\rm equal~numbers~of~bosons~and~fermions~at~all~levels}~.~~~~
\label{relationone}
\eeqn
This is ultimately a consequence of modular invariance, which in this
simple case can be used to relate
the numbers of very low energy states such as tachyons to the numbers of
states at higher mass levels.
However, while the requirement that there be no {\it physical}\/ tachyons
is necessary for the consistency of the string in spacetime,
 {\it unphysical}\/ tachyons
cause no spacetime inconsistencies and are in fact unavoidable in
the vast majority of string theories (such as all non-supersymmetric heterotic
strings).  This is therefore the more general case.
The question then arises:
how do the bosonic and fermionic states
effectively redistribute themselves at all energy levels
in order to account for these unphysical tachyons?
To what extent is the delicate boson/fermion balance destroyed?

\subsubsection{To what extent can one break spacetime supersymmetry
without destroying the finiteness of string theory?}

A third way of asking essentially the same question is within the
framework of string finiteness and supersymmetry-breaking.
If we start with a string theory containing an unbroken spacetime
supersymmetry,
then there are an equal number of bosonic and fermionic states at
each mass level in the theory ({\it i.e.}, $a_{mn}=0$ for all $m$ and $n$),
and consequently we find $Z=0$.
This is of course trivially modular-invariant,
and the fact that such theories have $a_{nn}=0$ for all $n<0$
indicates that they also contain no physical tachyons.
These two conditions, however, are precisely those that
enable us to avoid certain ultraviolet and infrared divergences in
string loop amplitudes:  modular invariance
eliminates the ultraviolet divergence that would have appeared as $\tau\to 0$,
and the absence of physical tachyons ensures that there is no infrared
divergence as $\tau\to i\infty$.  For example,
the one-loop vacuum energy (cosmological constant) $\Lambda$ would
ordinarily diverge in field theory, but turns out to be {\it finite}\/
in any modular-invariant, tachyon-free string theory.
Indeed, these finiteness properties of string loop amplitudes are
some of the most remarkable and attractive features of string theory
relative to ordinary point-particle field theory.

If the spacetime supersymmetry is broken, however,
the partition function $Z$ will no longer vanish,
and bosonic states will no longer exactly cancel against fermionic
states level-by-level in the theory.
However, we would still like to retain the finiteness properties
of string amplitudes that arise in the supersymmetric theory.
What residual cancellation, therefore, must nevertheless survive
the supersymmetry-breaking process?  What weaker cancellation preserves
the modular invariance and tachyon-free properties which are necessary
for finiteness and string consistency?

\subsection{Overview of Misaligned Supersymmetry:  The Basic Ideas}

It turns out that misaligned supersymmetry provides an answer
to all of these questions:  it yields a constraint on the allowed
numbers of string states which arises from modular invariance; it describes
the perturbation of the boson/fermion balance due to the presence
of unphysical tachyons; and it serves as the residual cancellation which
is necessary for string finiteness.  Indeed, it furnishes us with a constraint
on those supersymmetry-breaking scenarios which maintain string finiteness,
essentially restricting us to only those scenarios in
which a {\it misaligned}\/ supersymmetry survives.
In the remainder of this section we shall therefore briefly describe the
basic features of this misaligned supersymmetry.
Further details may be found in subsequent sections.

The basic idea behind misaligned supersymmetry is quite simple.
As we have said, ordinary supersymmetry may characterized
by a complete cancellation of the physical state degeneracies
$a_{nn}$ for all $n$, and this in turn implies
that there are equal numbers of bosons and fermions at all mass levels
in the theory.  In the more general case of {\it misaligned}\/
supersymmetry, each
of these features is changed somewhat.  First, the object which experiences
a cancellation is no longer the actual net state degeneracies
$a_{nn}$, but rather a new object called
the ``sector-averaged'' state degeneracies
and denoted $\langle a_{nn}\rangle$.  This is will be defined below.
Second, just as the cancellation of the actual net degeneracies $a_{nn}$
implied equal numbers of bosonic and fermionic states at every
mass level in the theory, the cancellation of the sector-averaged number
of states  $\langle a_{nn}\rangle$ will instead turn out to
imply a subtle boson/fermion {\it oscillation} in which,
for example, any surplus of bosonic states
at any energy level of the theory implies the existence of a larger surplus
of fermionic states at the next higher level, which in turn implies
an even larger surplus of bosonic states at an even higher level, and
so forth.
Such an oscillation is quite dramatic and highly constrained, and its
precise form will be discussed below.

\subsubsection{The ``Sector-Averaged'' Number of States
     $\langle a_{nn}\rangle$}

We begin by describing the ``sector-averaged'' number
of states $\langle a_{nn}\rangle$ and its corresponding
cancellation.  In order to do this, let us first
recall how states are typically arranged in string theory.

As we have mentioned,
the generic string spectrum consists of a collection of infinite
towers of states:  each tower corresponds to a different $(i,j)$ sector
of the underlying string worldsheet theory, and consists of
a ground state with a certain vacuum energy $H_{ij}$
and infinitely many higher excited states with energies $n=H_{ij}+\ell$
where $\ell\in \IZ$.
The crucial observation, however, is that the different sectors in
the theory will in general be {\it misaligned}\/
relative to each other, and start out with
different vacuum energies $H_{ij}$ (modulo 1).
For example, while one sector may contain states with integer energies $n$,
another sector may contain states with $n\in\IZ+1/2$, and another
contain states with $n\in \IZ+1/4$.
Each sector thus essentially contributes a separate set of states to the
total string spectrum, and we can denote the net degeneracies from
each individual sector $(i,j)$ as $\lbrace a_{nn}^{(ij)}\rbrace$,
where $n\in \IZ + H_{ij}$.  Of course, for each sector $(i,j)$ in the theory,
these state degeneracies are simply the $m=n$ coefficients
within a power-series expansion of
the corresponding partition-function characters:
\beq
  \chi_i(q)\, \chibar_j(\qbar) ~=~ \sum_{m,n}  \,a_{mn}^{(ij)} ~\qbar^m\,q^n~.
\label{fiveb}
\eeq
Thus, the numbers of physical states $a_{nn}$ in the entire theory
can be decomposed
into the separate contributions $\lbrace a_{nn}^{(ij)}\rbrace$
from each relevant
sector $(i,j)$ in the partition function (\ref{two}).

Now, the number of physical states at each mass level of a theory
uniquely determines many properties of that theory,
and in particular one  such property which may easily be determined
is its {\it central charge} (or equivalently its Hagedorn temperature).
Specifically,
for each sector $(i,j)$, it is well-known that
$\lbrace a_{nn}^{(ij)}\rbrace$ must grow exponentially with $n$,
\beq
         a_{nn}^{(ij)}~\sim~ A \,n^{-B}\,e^{C_{\rm tot}\sqrt{n}}~
     ~~~~~{\rm as}~~~ n\to\infty~;
\label{Hagform}
\eeq
here $A$ and $B$ are constants, and $C_{\rm tot}=1/T_H$ is the inverse Hagedorn
temperature of the theory.  Note that this inverse Hagedorn temperature
$C_{\rm tot}$ receives separate contributions from the left- and right-moving
underlying theories, $C_{\rm tot}= C_{\rm left}+C_{\rm right}$, with each
contribution directly related to the
central charge of the corresponding theory via
\beq
       C_{\rm left,\,right}= 4\pi\sqrt{{c_{\rm left,\,right}\over 24}}~.
\label{Cc}
\eeq
Thus, given (\ref{Hagform}), the total inverse Hagedorn temperature
$C_{\rm tot}$ of the theory can be easily determined by analyzing the
growth of the degeneracies $a_{nn}^{(ij)}$:
\beq
   C_{\rm tot}~=~ C_{\rm left}+C_{\rm right} ~=~
      \lim_{n\to\infty} {\log a_{nn}^{(ij)}\over \sqrt{n}}~.
\label{cdef}
\eeq
It does not matter which of the contributing
sectors $(i,j)$ is selected for this purpose,
since each yields the same value $C_{\rm tot}$.

For each sector $(i,j)$ in the theory,
let us now take the next step and imagine analytically
continuing the set of numbers $\lbrace a_{nn}^{(ij)}\rbrace$
to form a smooth function $\Phi^{(ij)}(n)$  which not only reproduces $\lbrace
a_{nn}^{(ij)}\rbrace$ for the appropriate values $n\in \IZ+H_{ij}$,
but which is continuous as a function of $n$.
Clearly these functions $\Phi^{(ij)}(n)$ must not only contain
the above leading Hagedorn-type exponential dependence,
\beq
    \Phi^{(ij)}(n)~=~ A\,n^{-B}\,e^{C_{\rm tot}\sqrt{n}}~+~...~,
\label{philead}
\eeq
but must also contain all
of the subleading behavior as well so that exact results can be obtained
for the relevant values of $n$.
Indeed, these functions $\Phi^{(ij)}(n)$ may be regarded as the
complete (and exact) asymptotic expansions for the
$\lbrace a_{nn}^{(ij)}\rbrace$,
and there exist straightforward and well-defined procedures
for generating these functions \cite{HR,KV}.
Note that in general these functions are quite complicated, and contain
infinitely many subleading terms:  some of these are polynomially suppressed
relative to the leading term (\ref{philead}), and others are
exponentially suppressed.
These functions $\Phi^{(ij)}(n)$ will be described in detail in
Sect.~2.

Given that such functions exist, however,
the ``sector-averaged'' number of states is then defined
quite simply as a sum of these functions over all sectors in the theory,
\beq
     \langle a_{nn} \rangle ~\equiv~ \sum_{ij} \, N_{ij} \,\Phi^{(ij)}(n)~,
\label{sectoraverage}
\eeq
where $N_{ij}$ are the coefficients in (\ref{two}).
This sector-averaged quantity $\langle a_{nn}\rangle$ therefore differs quite
strongly from any of the actual physical-state degeneracies
$a^{(ij)}_{nn}$ which arise within a given sector,
and differs as well from the total physical-state degeneracies $a_{nn}$ which
appear in (\ref{three}).
Instead, $\langle a_{nn}\rangle$ is a continuous function which
represents their ``average'' as defined in (\ref{sectoraverage}).

Our main result, then, is that
although each sector has a number of states
$a_{nn}^{(ij)}$ which grows in accordance with (\ref{Hagform}),
this sector-averaged number of states $\langle a_{nn}\rangle$
must grow {\it exponentially}\/ more slowly.  Specifically, if we define
$C_{\rm eff}$ in analogy to $C_{\rm tot}$,
\beq
     C_{\rm eff}~\equiv~
     \lim_{n\to\infty}
      {\log \,\langle a_{nn}\rangle\over \sqrt{n}}~,
\label{ceff}
\eeq
then we must have
\beq
        C_{\rm eff} ~<~ C_{\rm tot}~.
\label{csmaller}
\eeq
We shall prove this result in Sect.~3.
Moreover, we shall in fact conjecture
that $C_{\rm eff}$ vanishes identically,
\beq
          C_{\rm eff}~{\buildrel {?}\over =}~ 0~,
\label{cvanishing}
\eeq
with $\langle a_{nn}\rangle$ experiencing at most polynomial growth.
This conjecture will be discussed in Sects.~3 and 5.

These, then, are the cancellations implicit in ``misaligned supersymmetry'',
required by modular invariance and necessary for string finiteness.
Indeed, from (\ref{csmaller}), we see that the cancellation
governing the string spectrum is sufficiently strong that
not only must the leading Hagedorn terms (\ref{philead}) cancel,
but so must all of the subleading terms within the
$\Phi^{(ij)}(n)$ which are polynomially suppressed.
The severity of this
cancellation (\ref{csmaller}) thus implies that all traces
of the original central charge of the theory are effectively removed in
this sector-averaging process, with the sector-averaged number of
states $\langle a_{nn} \rangle$ growing with $n$ as though derived from
an underlying theory with a different central charge.
Furthermore, if the conjecture (\ref{cvanishing}) is correct,
then in fact {\it all}\/ exponential growth of $\langle a_{nn}\rangle$
must be cancelled, whether leading or subleading.
This implies that $\langle a_{nn}\rangle$ should experience at most
polynomial growth, and for spacetime dimensions $D\geq 2$ we shall
in fact argue that even this polynomial growth should
be cancelled.  Thus in these cases we expect
$\langle a_{nn}\rangle$ to actually {\it vanish}\/ as $n\to\infty$.

\subsubsection{Boson/Fermion Oscillations}

The cancellation (\ref{csmaller}) has far-reaching implications,
and in particular implies a corresponding ``misaligned supersymmetry'' with
boson/fermion oscillations.
We can perhaps most easily see how this emerges by considering
a particular example, a toy string theory containing only two sectors.
Let us therefore focus on the following model partition function
\beq
      Z_{\rm toy}~=~ ({\rm Im}\,\tau)^{1-D/2}\,
      \biggl\lbrace N_1 \,[A(q)]^\ast B(q) \,+\,
      N_2 \,[C(q)]^\ast D(q) \biggr\rbrace~
\label{toyZ}
\eeq
where $A,B,C,$ and $D$ are any four characters corresponding to
any chiral worldsheet conformal field theory,
with chiral vacuum energies $H_{A,B,C,D}$ respectively.
For concreteness, let us assume
that $H_A,H_B \in {\IZ}$, and  $H_C,H_D\in {\IZ}+\half$.
We thus have two separate towers of states in this theory,
with one sector $(AB)$ contributing states
with degeneracies $\lbrace a_{nn}^{(AB)}\rbrace$
situated at integer energy levels $n$,
and another sector $(CD)$ contributing states with
deneracies $\lbrace a_{nn}^{(CD)}\rbrace$ at energy levels $n\in \IZ+1/2$.
Let us furthermore suppose that these are the only two sectors in the
theory (so that $Z_{\rm toy}$ is modular invariant), and that
this theory has no physical tachyons (which requires that
either $H_A$ or $H_B$ is non-negative, and that either
$H_C$ or $H_D$ is non-negative).
Then if $\Phi^{(AB)}(n)$ and $\Phi^{(CD)}(n)$ are
respectively the complete asymptotic expansions which
correspond to the degeneracies $a_{nn}^{(AB)}$ and $a_{nn}^{(CD)}$
for these two sectors,
then our result (\ref{csmaller}) asserts that
a sufficient number of leading terms in $\Phi^{(AB)}$ and
$\Phi^{(CD)}$ must
cancel exactly so that the rate of exponential growth of their
sum is reduced.  Explicitly, denoting all of these leading terms as
$\tilde \Phi^{(AB)}(n)$ and $\tilde \Phi^{(CD)}(n)$ respectively,
we must have
\beq
      N_1 \,\tilde \Phi^{(AB)}(n) ~+~N_2 \,\tilde \Phi^{(CD)}(n)~=~0~.
\label{toyresult}
\eeq

It is important to realize that this result does not imply any direct
cancellation between bosonic and fermionic states in this theory,
for (\ref{toyresult}) represents merely a cancellation of
the {\it functional forms}\/ $\tilde \Phi^{(AB)}(n)$
and $\tilde \Phi^{(CD)}(n)$.
Indeed, despite the result (\ref{toyresult}),
the total physical-state degeneracies $\lbrace a_{nn}\rbrace$
for this theory do not vanish for any particular $n$.
Rather, due to the misalignment between the two sectors in this
hypothetical example, the actual values taken by the total
partition-function coefficients $a_{nn}$ as $n\to\infty$ are
\beq
          a_{nn}~\sim~ \cases{
       N_1 \,\tilde \Phi^{(AB)}(n)  & for $n\in \IZ$\cr
       N_2 \, \tilde\Phi^{(CD)}(n) & for $n\in \IZ+1/2$~. \cr}
\label{toyactualvalues}
\eeq
Thus we see that there exists no
single value of $n$ for which the actual physical degeneracy
$a_{nn}$ is described by the vanishing sum $N_1\tilde\Phi^{(AB)}
+N_2\tilde\Phi^{(CD)}$.

Perhaps even more interestingly, this result implies that we cannot
even {\it pair}\/ the states situated at {\it corresponding}\/ levels
in the $(AB)$ and $(CD)$ sectors,
for while the net number of states at the $\ell^{\rm th}$
level of the $(AB)$ sector is given by $N_1\Phi^{(AB)}(\ell)$,
the net number of states at the $\ell^{\rm th}$ level
of the $(CD)$ sector is given by
$N_2\Phi^{(CD)}(\ell+\half) = -N_1\Phi^{(AB)}(\ell+\half)$.
The two sectors thus ``sample'' these cancelling functions at
different energies $n=H_{ij}+\ell$, and it is only by considering these
state degeneracies as general functions of $n$ ---
or equivalently by considering
the ``sector averaged'' quantity $\langle a_{nn}\rangle$ ---
that the cancellation indicated in (\ref{toyresult}) becomes apparent.

In Fig.~1, we have sketched a likely scenario for this toy model,
plotting (as functions of energy $n$)
both the physical-state degeneracies $a_{nn}$,
and their ``sector-average'' $\langle a_{nn}\rangle$.
Note that for the actual degeneracies $a_{nn}$,
we are plotting $\pm\log_{10}(|a_{nn}|)$ where
the minus sign is chosen if $a_{nn}<0$ ({\it i.e.}, if there is a surplus
of fermionic states over bosonic states at energy $n$).
Although $a_{nn}$ takes values only at the discrete energies $n\in{\IZ}/2$,
we have connected these points in order of increasing $n$ to stress the
fluctuating oscillatory behavior that $a_{nn}$ experiences as the energy
$n$ is increased.
Note that for $n\in {\IZ}$, the values of $a_{nn}$ are
all positive;  these are the state degeneracies $a_{nn}^{(AB)}$
from the $(AB)$ sector alone.  Similarly, the negative values of
$a_{nn}$ appear for $n\in{\IZ}+1/2$, and represent the individual
contributions $a_{nn}^{(CD)}$ from the $(CD)$ sector.
We have also superimposed the typical behavior of $\langle a_{nn}\rangle$,
which, given its definition as a sum of asymptotic forms, is a continuous
function of $n$.  While the values of the individual $a_{nn}$ experience
exponential growth with $C=C_{\rm tot}$, the cancellation (\ref{csmaller})
guarantees that $\langle a_{nn} \rangle$
grows exponentially more slowly than $a_{nn}$,
with a rate $C_{\rm eff}<C_{\rm tot}$.
The sketch in Fig.~1 assumes that $C_{\rm eff}=C_{\rm tot}/4$,
but if (\ref{cvanishing}) is true, then of course
$\langle a_{nn}\rangle$ experiences no exponential growth at all
and remains flat.

As can be seen from Fig.~1,
the immediate consequence of the cancellation (\ref{csmaller})
is that the net number of actual physical states $a_{nn}$
can at most {\it oscillate around zero}\/ as the energy $n$ increases
and as the contributions from different sectors with differently aligned
values of $n$ contribute either positively or negatively to the
partition function $Z$.
The wavelength of this oscillation is
clearly $\Delta n=1$, corresponding to the
energy difference between adjacent states in the {\it same}\/ sector,
while the amplitude of this oscillation of course grows exponentially
with $n$ (since each individual $\Phi^{(ij)}(n)$ retains its leading
Hagedorn-like behavior).
Thus, although the net number of states $a_{nn}$ diverges as $n\to \infty$,
it can do so only in a very tightly constrained
manner, so that the {\it functional forms}\/ which describe this
leading asymptotic behavior have a sum which vanishes.

We will see that this oscillatory behavior is in fact a generic consequence
of our result, and appears in any modular-invariant theory which is
free of physical tachyons.
Thus, in analogy to (\ref{relationone}), we now have
\beq
    {\rm no~physical~tachyons}~~~\Longleftrightarrow~~~
        {\rm boson/fermion~oscillations}~.
\label{relationtwo}
\eeq
Conversely, the existence of such oscillations in the spectrum
of a given unknown theory may well be taken as a signature of an underlying
modular invariance \cite{cudelldien}.
Moreover, as we shall demonstrate in Sect.~5, these boson/fermion
oscillations also serve as the general mechanism by which the
finiteness of modular-invariant, tachyon-free theories is
reflected level-by-level in the degeneracies of physical states.
Of course, a non-trivial corollary of the result (\ref{relationtwo})
is that the absence of physical tachyons {\it requires}\/ the presence
of spacetime fermions.
This explains why the bosonic string was doomed to have physical
tachyons in its spectrum, and why the cure to this problem (a GSO projection)
could only be implemented in the context of an enlarged theory
({\it e.g.}, the superstring or heterotic string)
which also contained spacetime fermions.

Finally, of course, this result can also be interpreted as providing
a tight constraint on the general pattern of supersymmetry-breaking
in string theory.  As we discussed above, in a space\-time-super\-sym\-metric
(and therefore tach\-yon-free) theory, bosonic and fermionic sectors
contribute precisely equal numbers of states at each individual energy $n$;
consequently we have $a_{nn}=0$ level-by-level, and the amplitude
of our ``oscillation'' in this special case is zero.
However, if the supersymmetry is to be broken in such a way that
physical tachyons are not introduced and modular invariance is
to be maintained (as we would demand in any physically sensible theory),
then our result implies that one can at most ``misalign'' this bosonic and
fermionic cancellation, introducing a mismatch between the bosonic and
fermionic state degeneracies at each level in such a way that any surplus of
bosonic states at any energy level is compensated for by an even greater
surplus of fermionic states at a neighboring higher energy level, leading to
an even greater bosonic surplus at an even higher energy, and so forth.
The magnitudes of these surpluses are of course tightly constrained,
since the cancellation of the corresponding leading functional forms must be
preserved.  Thus, we see that this residual cancellation, this hidden
``misaligned'' supersymmetry, is another unavoidable consequence of modular
invariance, and it would be interesting to see which classes of
physical super\-sym\-metry-break\-ing scenarios are thereby precluded.
For example, we can already rule out any supersymmetry-breaking scenario
in which the energies of, say, the fermionic states are merely shifted
relative to those of their bosonic counterparts by an amount $\Delta n$;
rather, we require a mechanism which somehow also simultaneously
creates (or eliminates) a certain number $\Phi(n+\Delta n)-\Phi(n)$
of states at each level $n$ so that the state degeneracies at the shifted
energies are still described the same (cancelling) functional forms.
Such a mechanism would clearly be highly non-trivial.
In particular, ``misaligned supersymmetry'' is as yet
only a result concerning the {\it numbers}\/ of bosonic and
fermionic states in string theory, and no dynamical symmetry operators
or currents relating these misaligned states have been constructed.

We have now completed our overview of our main results;  the rest of
this paper will provide details and examples, and is organized as follows.
In Sect.~2 we shall first review the asymptotic expansions upon which
our results rest, and in Sect.~3 we will prove our main theorem
(\ref{csmaller}) and demonstrate that (\ref{csmaller}) in fact serves
as the two-variable generalization of a well-known theorem in modular
function theory.  In Sect.~4 we shall then provide some explicit examples
of these cancellations and the corresponding ``misaligned supersymmetry'',
and in Sect.~5 we shall discuss how this phenomenon may ultimately
be responsible for the finiteness of string loop amplitudes
by considering the case of the one-loop cosmological constant.
We will also discuss our conjecture that in fact $C_{\rm eff}=0$.
Our final comments concerning various extensions and applications
will be presented in Sect.~6.

Because our proofs and calculations will be presented in great
detail, this paper is somewhat lengthy.
We have therefore organized the rest of this paper in such a way that the
reader unconcerned with the details of the asymptotic expansions
$\Phi^{(ij)}(n)$
and willing to accept the result (\ref{csmaller})
can omit Sects.~2--4 without loss of continuity, and proceed
directly to Sect.~5.


\vfill\eject
\setcounter{footnote}{0}
\section{Asymptotic Expansions for the Numbers of States}


In this section we shall briefly review the methods
(originally due to Hardy and Ramanujan \cite{HR}
and recently generalized by Kani and Vafa \cite{KV})
for deriving the asymptotic functions describing the physical-state
degeneracies.  We will concentrate on only those broad features of
the derivation which will be relevant for our later work, leaving many
of the technical details to be found in \cite{KV}.

The problem may be stated mathematically as follows.
We are given a set of functions $\chi_i(\tau)$, $i=1,...,N$, forming
an $N$-dimensional representation of the modular group with modular
weight $k\in \IZ/2$.  This means that for any transformation
$M=\smallmatrix{a}{b}{c}{d}\in SL(2,\IZ)$
and any $\tau$ in the fundamental domain $\calF\equiv
\lbrace \tau:\, |{\rm Re}\,\tau|\leq 1/2,\, {\rm Im}\,\tau>0,\,
|\tau|\geq 1\rbrace $ of the modular group, we have
\beq
   \chi_i(M\tau) ~=~  (c\tau+d)^k \,
      \sum_{j=1}^N\, M_{ij}\,\chi_j(\tau)~
\label{Mtrans}
\eeq
where $M\tau\equiv (a\tau+b)/(c\tau+d)$ and where $M_{ij}$ is
an $N\times N$ matrix in the representation space.
We assume each function $\chi_i$ to have a $q$-expansion of the form
\beq
   \chi_i(q) ~=~ q^{H_i}\,\sum_{n=0}^\infty\,a_n^\ii \, q^n~
\label{chiform}
\eeq
where $q\equiv e^{2\pi i\tau}$ and all $a_n^\ii\geq 0$;
thus each $\chi_i$ is an
eigenfunction of the transformation $T:\,\tau\to \tau+1$ with
eigenvalue $\exp(2\pi i H_i)$.
We will also assume that our functions $\chi_i$ are normalized
so that each $a_0^\ii=1$.
Such functions $\chi_i$ arise, for example, as
the characters ${\rm Tr}\, q^H$ of the various highest-weight sectors of
conformal field theories with Hamiltonians $H$;
the quantities $H_i$ in (\ref{chiform}) are then interpreted
as the sector vacuum energies,
which are related to the central charge $c$ of the
conformal field theory and its various highest weights $h_i\geq 0$ via
\beq
    H_i ~=~ h_i ~-~c/24 ~.
\label{Hch}
\eeq
The $a_n^\ii$, on the other hand, are interpreted as the state degeneracies
of the $i^{\rm th}$ sector at excitation number $n$, and
our goal is to derive asymptotic expansions for these degeneracies
$a_n^\ii$ as functions of $n$.
It is well-known that the {\it leading}\/ asymptotic behavior of these
degeneracies is of the Hagedorn form
\beq
  a_n^\ii  ~\sim~ A_i\,n^{-B_i}\,e^{C_i\sqrt{n}} ~~~{\rm as}~n\to\infty
\label{Hagedornform}
\eeq
where $A_i$, $B_i$, and $C_i$ are constants, and in fact we will see that
\beq
      C_i \,=\, \sqrt{{{2c}\over 3}} \, \pi ~,~~~
      B_i \,=\, {\textstyle{3\over 4}} - \half\,k~,~~~
      A_i \,=\, {1\over{\sqrt{2}}}\,(e^{\pi i k/2}  S_{i1})\,
      \left( {c\over {24}}\right)^{B_i-1/2}~.
\label{Hagedornvalues}
\eeq
Here $S_{i1}$ is the $(i,1)$ element of
the representation matrix [as defined in (\ref{Mtrans})]
corresponding to the modular transformation
$S\equiv\smallmatrix{0}{-1}{1}{~0}$,
where $j=1$ denotes the vacuum-sector or identity-sector character
for which $H_j= -c/24$ (or $h_j=0$).
Note that the product $e^{\pi i k/2}  S_{i1}$ is always real and non-zero.
There are, however, an infinite number of additional
subdominant and subleading terms in the complete
(and often {\it exact}) asymptotic expansions of
the physical-state degeneracies,
and we will be seeking these complete expansions.

The derivation proceeds in two basic steps.
The first is to invert (\ref{chiform}), extracting
the degeneracies $a_n^\ii$ through a contour
integral:
\beq
  a_n^\ii~=~ {1\over{2\pi i}} \, \oint_{C} \, dq \, {{\chi_i(q)}\over
     {q^{n+H_i+1}}} ~
\label{contour}
\eeq
where the contour $C$ is any closed counter-clockwise loop
encircling the origin once and remaining entirely within the unit
disk $|q|\leq 1$ (so that $\chi_i(q)$ remains convergent).
It is convenient to take $C$ to be a circle of radius $1$.
While this integral can in principle be evaluated for any radius,
it proves advantageous to take the radius near $1$,
for in this limit the
contour integral will be dominated by contributions from the regions near
certain points on the unit circle $|q|=1$, and
these contributions will be relatively simple to evaluate.
In particular, we see from the general forms (\ref{chiform}) that
the characters $\chi_i(q)$ develop essential singularities
at all points on the unit circle for which
$\tau\in \IQ$, for at these points there exist an infinite
number of values of $n$ for which $q^n=1$, causing $\chi(q)$ to diverge.
This can often also be seen by writing the infinite sums (\ref{chiform})
as infinite products, for such product representations (when they
exist) typically include the factor
\beq
     \prod_{n=1}^\infty \,{1\over{1-q^n}}
\label{infprodfactor}
\eeq
which diverges whenever there exists an $n\geq 1$ such that $q^n=1$.
Thus, we can evaluate the contour integral
(\ref{contour}) in the radius $=1$ limit by
carefully dissecting our contour into arcs near
each rational point on the unit circle, and
summing the separate arc integrals.

The second step involves performing this dissection most
efficiently.  While the divergences at
each rational point $q_{\alpha\beta}\equiv\exp(2\pi i \beta/\alpha)$
with $\alpha>\beta\geq 0$
are necessarily of infinite degree, we see that if $\alpha$ and $\beta$
are chosen relatively prime (so that $\beta/\alpha\equiv \tau_\ab$
is expressed in lowest form),
the divergences will be ``stronger'' at points for which $\alpha$ is smaller.
For example, the divergence of $\chi(q)$ at $q_{\alpha\beta}\equiv
\exp(2\pi i\beta/\alpha)$ is
``twice'' as strong as that at $q_{2\alpha,\beta}$ in the sense that there
are twice as many values of $n$ in (\ref{chiform}) or (\ref{infprodfactor})
for which
$q^n=1$.  Indeed, the dominant contribution to the integral (\ref{contour})
is that near the point $q_{1,0}=1$, and
we will see that this contribution alone is sufficient to
yield (\ref{Hagedornform}).

We can thus dissect our contour most efficiently by summing the
contributions near points $q_{\alpha\beta}$ in order of increasing
$\alpha\geq 1$, choosing
a terminating maximum value $\alpha_{\rm max}$ depending
on the overall accuracy desired.
Then, for each value $\alpha$, $1\leq\alpha\leq \alpha_{\rm max}$,
we consider those $\beta$ ($1\leq \beta < \alpha$)
which are relatively prime to $\alpha$, allowing $\beta=0$ only for
$\alpha=1$.  For example, we
have $\beta=1,3$ for $\alpha=4$, and $\beta=1,2,...,6$ for $\alpha=7$.
This procedure ensures that we have organized the rational
points according to their relative contributions to the integral
(\ref{contour}).

Finally, in order to evaluate these contributions from each such
rational point $q_\ab$, we make use of the fact that these points
can be transformed to $q'_\ab=0$ by modular transformations.
Thus while it may at first seem difficult to evaluate the numerator
$\chi_i(q\approx q_{\alpha\beta})$,
modular transformations allow us to relate this to the much simpler
quantities $\chi_j(q'\approx 0)$.
In particular, let $M_{\alpha\beta}\in SL(2,\IZ)$ be a modular
transformation transforming $q=q_\ab$ to $q'_\ab=0$.
Although such a transformation is not unique (because the {\it phase}\/ of the
resulting $q'_\ab=0$ is unspecified), any such chosen transformation
can be written in the form
\beq
   M_\ab ~\equiv~ \pmatrix{ -\beta' & r \cr -\alpha & \beta \cr} ~,~~~~~
    r ~=~  (1+\beta\beta')/\alpha
\label{Mabdef}
\eeq
where $\beta'\in \IZ$ is the free parameter indicating our specific
choice.\footnote{
    This freedom to adjust the phase of $q'_\ab$, or equivalently
    to adjust ${\rm Re}\,\tau'_\ab$, corresponds to
    the freedom to multiply $M_\ab$ from the left
    by arbitrary additional factors of $T \equiv \smallmatrix{1}{1}{0}{1}$.
    Thus we see that all of the allowed values of
    $\beta'$ are equal modulo $\alpha$.  All final results will
    nevertheless be invariant under such shifts in $\beta'$.}
One can then use (\ref{Mtrans}) to relate $\chi_i(q\approx q_\ab)$ to
the various $\chi_j(q'\approx 0)$, each of which
behaves as $(q')^{H_j}$.  Explicitly,
\beq
       \chi_i(q\approx q_{\alpha\beta})~\approx~
       (-\alpha \tau+\beta)^{-k}\sum_{j=1}^N ~(M_{\alpha\beta}^{-1})_{ij}
         \,(q'\approx 0)^{H_j}~.
\label{chirelation}
\eeq
This result can then be substituted into
each separate arc integral near each rational point,
and in this form these arc integrals can be evaluated
(ultimately yielding Bessel functions).
These separate contributions from each rational point can then
be summed to yield an approximation for the whole contour integral
(\ref{contour}).

The details of this analysis are somewhat intricate, and are
given in Refs.~\cite{HR} and \cite{KV}.
The result one finds, however, is relatively simple:
\beq
     a_n^\ii ~=~ \sum_{\alpha=1}^{\alpha_{\rm max}}
     \,\left( {{2\pi}\over\alpha} \right)
    \,\sum_{j=1}^N  \, Q_{ij}^{(\alpha)} \, f_j^{(\alpha)}(n)
\label{finalone}
\eeq
where
\beq
  Q_{ij}^{(\alpha)} ~\equiv~ e^{\pi i k/2}\, \sum_\beta\,\Minv~
    \exp \left[ 2\pi i \left(
      {{\beta'}\over\alpha} \,H_j \,-\, {\beta\over\alpha} \,H_i
      \right)\right]~
   \exp \left( -2\pi i \,{{\beta\over\alpha}} \,n\right)~
\label{Qdef}
\eeq
and where
\beq
   f_j^{(\alpha)}(n)~\equiv~
   \left( \sqrt{{{H_j}\over{n+H_i}}} \,\right)^{1-k}\,
    J_{k-1}\left({{ 4\pi}\over\alpha}
       \sqrt{ H_j(n+H_i)} \right) ~.
\label{fdef}
\eeq
Here the sum in (\ref{Qdef}) is over those values
of $\beta$, $1\leq \beta <\alpha$ (with $\beta=0$ only for $\alpha=1$),
such that $\alpha$ and $\beta$ are relatively prime.  The modular
transformations $M_\ab$ are defined in (\ref{Mabdef}), and their representation
matrices are defined generally in (\ref{Mtrans}).
It is in fact straightforward to show \cite{KV}
that the quantity $Q_{ij}^{(\alpha)}$ (as well as $a_n^\ii$ itself)
is not only real for all $i$, $j$, and $\alpha$, but also independent of the
particular choices for $\beta'$ in each term of (\ref{Qdef}),
as asserted earlier.
Indeed, different values of $\beta'$ correspond
to extra factors of $T$ in $M_\ab$, and these are offset by the
explicit $\beta'$-dependent phase in (\ref{Qdef}).
The functions $J_\nu(x)$ appearing in (\ref{fdef}) are the
Bessel functions of the first kind.

While (\ref{finalone}) is written in a form suitable for those values of $j$
for which $H_j\geq 0$, there are often other values of $j$ for which $H_j<0$;
indeed, if the $\chi_j$ are the characters of the different highest-weight
sectors of a conformal field theory with positive central charge, then the
existence of an identity-sector character with $H_j<0$ is guaranteed.
In these cases, the argument of the Bessel function becomes imaginary.
However, introducing the Bessel
function of first kind with imaginary argument,
\beq
   I_\nu(x) ~\equiv~ e^{-i\pi\nu/2} \,J_\nu(e^{i\pi/2}x)~,
\label{besselimag}
\eeq
we find that for these cases we can rewrite (\ref{fdef}) as
\beq
   H_j>0:~~~~~~~
   f_j^{(\alpha)}(n)~=~\left( \sqrt{{{|H_j|}\over{n+H_i}}} \,\right)^{1-k}\,
         I_{k-1}\left({{ 4\pi}\over\alpha}
       \sqrt{ |H_j|(n+H_i)} \right) ~.
\label{finaloneimag}
\eeq
Thus the reality of the $a_n^\ii$ is preserved.
Similarly, there may also exist cases for which $H_j=0$.  In these
cases $f_j^{(\alpha)}(n)$ takes the simple form
\beq
   H_j=0:~~~~~~~
   f_j^{(\alpha)}(n)~=~
    \left\lbrack {\,2\pi \over \alpha}\,(n+H_i) \right\rbrack^{k-1}~,
\label{finalonezero}
\eeq
and once again the contribution to $a_n^\ii$ is real.

The foregoing discussion has only touched upon the basic method of
derivation, and in particular there are many places in the complete
derivation where approximations are introduced in order to achieve
the above result.  This implies, of course, that the above
asymptotic expansions are at best only approximate.
A detailed analysis of the error terms has been performed in Ref.~\cite{KV},
however, and we shall present here only the final results.
First, it is shown in \cite{KV} that the minimum error is achieved
if, for a given value of $n$, the $\alpha$-summation in (\ref{finalone})
is terminated at values
\beq
      \alpha_{\rm max} ~\sim~ {\cal O} (\sqrt{n})~.
\label{alphamaxsize}
\eeq
As is typical with such asymptotic expansions, subsequent ``higher-order''
terms only increase the error.
Given this termination, then, it is shown in \cite{KV}
that asymptotic expansions are correct to within
\beq
     {\rm error} ~\sim~ {\cal O} (n^p) ~~~~~~{\rm where}~~~~
        p=\cases{k/2 &for $k\leq 0$\cr
                  k  &for $k>0$.\cr}
\label{error}
\eeq
Thus, since we know that $a_n^\ii$ is always an integer, we see that
the truncated asymptotic expansion
(\ref{finalone}) becomes {\it exact}\/ for $k\leq 0$ and sufficiently large
$n$:
we simply take the integer closest to the sum in (\ref{finalone}).  For $k=0$,
on the other hand, the sum (\ref{finalone}) can differ from the true $a_n^\ii$
by only an overall additive constant.  Such agreement is indeed remarkable,
indicating that in most physically relevant cases (as we shall see), the
asymptotic series (\ref{finalone}) can be indeed
taken to truly represent the physical-state degeneracies $a_n^\ii$.

Let us now get some feeling for how the result (\ref{finalone}) works
in practice by demonstrating that (\ref{finalone})
implies the leading Hagedorn exponential behavior (\ref{Hagedornform}).
The Bessel function with imaginary argument $I_\nu(x)$ has
the asymptotic ($|x|\to \infty$) exponentially-growing
behavior \cite{Abramowitz}:
\beq
      I_\nu(x) ~=~ {{e^x}\over{\sqrt{2 \pi x}}}\,
        \left[ 1 - {{(\mu-1)}\over{8x}} + {{(\mu-1)(\mu-9)}\over{2!\,(8x)^2}}
     - {{(\mu-1)(\mu-9)(\mu-25)}\over{3!\, (8x)^3}} + ...\right]
\label{Besselasymptotic}
\eeq
with $\mu\equiv 4\nu^2$, whereas the ordinary Bessel function
$J_\nu(x)$ has the damped asymptotic behavior
\beqn
    J_\nu(x) ~=~ \sqrt{{2\over{\pi x}}} ~\Biggl\lbrace &&
 ~\left[ 1 - {{(\mu-1)(\mu-9)}\over{2!\,(8x)^2}} + ...\right] \, \cos\theta
      \nonumber\\
     ~-~ && ~\left[
   {{\mu-1}\over{8x}} ~-~
    {{(\mu-1)(\mu-9)(\mu-25)}\over{3!\,(8x)^3}} + ...\right] \, \sin\theta
     \biggr\rbrace
\label{otherbesselasymp}
\eeqn
with $\theta\equiv x-\half \pi \nu - \quarter\pi$.
Thus, we see that the dominant, exponentially-growing contributions
to $a_n^\ii$ come
from those terms in (\ref{finalone}) for which $H_j<0$.  In particular, the
minimum value of $H_j$ is $H_j= -c/24$ (which occurs for $h_j=0$,
 \ie, for the identity sector), and thus the strongest exponential growth
generally arises for this value of $j$ and for $\alpha=1$.
We will henceforth denote this sector as $j=1$.
The simplest choice for $M_{10}$ is $M_{10}=\smallmatrix{~0}{1}{-1}{0}= -S$
where $S\equiv\smallmatrix{0}{-1}{1}{~0}$, whereupon it follows that
$\beta'=0$ and $M^{-1}_{10}= S$.  Hence
$Q_{ij}^{(1)}= e^{\pi i k/2}  S_{ij}$ (a general result), and for
$n\gg H_i$ we find
\beq
    a_n^\ii ~=~ 2\pi\, \left(\sqrt{{{24\,n}\over{c}}}\,\right)^{k-1} \,
   (e^{\pi i k/2} \,S_{i1}) ~ I_{k-1}\left( 4\pi\sqrt{
     {c\over{24}}\,n}\,\right) ~+~... ~.
\label{stepfour}
\eeq
Use of (\ref{Besselasymptotic})
then yields (\ref{Hagedornform}) and (\ref{Hagedornvalues}).
Note that since $j=1$ in (\ref{stepfour}) corresponds
to the identity sector with $h_j=0$, we are guaranteed that $S_{i1}\not= 0$
for all $i$.  Perhaps the easiest way to see this is via the general
relationship between the $S$-matrix and the fusion rules of any
conformal field theory \cite{Verlinde};  since each highest-weight sector
must fuse with the identity to give back itself,
we must have $S_{i1}\not= 0$.
Indeed, the fusion rules of a given conformal field theory
\beq
           \lbrack \phi_i \rbrack~\times~
           \lbrack \phi_j \rbrack~=~
           \sum_k \,N_{ijk}\,
           \lbrack \phi_k \rbrack~
\label{fusionrules}
\eeq
can be deduced from this $S_{ij}$-matrix directly via
the Verlinde formula
\beq
                 N_{ijk}~=~ \sum_\ell \,
     {  S_{i\ell}  S_{j\ell} S_{\ell k} \over S_{\ell 1}}~,
\label{Verlinderesult}
\eeq
and we see that we must have $S_{i1}\not=0$ for all $i$
in order for sensible fusion rules to be obtained.

It is also a straightforward matter to classify the sorts of {\it subleading}\/
terms which appear in addition to the dominant Hagedorn
term (\ref{Hagedornform}).
First there are the subleading terms {\it within}\/ the dominant
exponential term
\beq
      Q\,e^{C_{\rm max}\sqrt{n+H}} \biggl[A_1 \,(n+H)^{-B_1} \,+\,
     A_2 \,(n+H)^{-B_2} \,+\, A_3\,(n+H)^{-B_3} \,+\,...\biggr]~;
\label{completedominant}
\eeq
here $Q$ signifies the coefficients defined in (\ref{Qdef}),
the values of $B_n$ increase by half-integer steps, and this series
terminates only if $2k$ is an odd integer.
This dominant exponential term arises from the identity sector, as noted
above, and will appear for the character of any sector connected to the
identity
sector via the $S$ modular transformation.
Next, there are the {\it subdominant}\/ exponentially growing terms;
these are again of the same form as (\ref{completedominant}),
but have exponential growth parameters $C'_{\rm max}$, $C''_{\rm max}$,
 {\it etc.}
which are smaller than
$C_{\rm max}$ and which depend on the spectrum of vacuum energies $H_i$.
Such terms arise from sectors with $-c/24<H_i<0$.
Then there are the sets of terms, again of the form
(\ref{completedominant}),
with exponential growth parameters $C=\half C_{\rm max},\third C_{\rm max},...,
\half C'_{\rm max},\third C'_{\rm max},...$, {\it etc.};  these terms arise
from the $H_i<0$ sectors as well.  Finally there are
the {\it non-growing}\/ terms
which arise from those sectors with $H_i \geq 0$:  these vanish in the
$n\to \infty$ limit, and are otherwise essentially negligible
as contributors to $a_n^\ii$.  The remarkable fact, however, is
that for $k<0$ and large $n$, the sum of {\it all}\/ of these terms reproduces
the values of $a_n^\ii$ {\it exactly}.

This concludes our review of the derivation of the asymptotic expansions
describing the physical-state degeneracies of conformal field theories.
In the remainder of this section, we shall, for completeness,
evaluate the first few coefficients $Q^{(\alpha)}_{ij}$ and demonstrate
that they are indeed real.
We will also discuss what modifications
are necessary to the above formulas if our
characters $\chi_i$ are {\it not}\/ normalized
with $a_{0}^{(i)}=1$.

The fundamental observation in evaluating the
coefficients $Q_{ij}^{(\alpha)}$ is to recognize that
although the general formula for $Q_{ij}^{(\alpha)}$ is given in (\ref{Qdef}),
certain choices for $\beta'$ within each term of (\ref{Qdef})
will enable the $Q_{ij}^{(\alpha)}$'s to assume particularly simple
forms.  If we define the modular transformations
\beq
    S~\equiv~ \pmatrix{ 0 & -1 \cr 1 & 0 \cr}~=~-S^{-1} ~~~~~{\rm and}~~~~~
     T ~\equiv~\pmatrix{ 1 & 1 \cr 0 & 1 \cr}~,
\label{STdefs}
\eeq
then any modular-transformation $M_\ab$
can be expressed as a product of $S$ and $T$ transformations.
Note that the representation matrices as defined in (\ref{Mtrans})
satisfy $(AB)_{ij}=\sum_k A_{ik}B_{kj}$;  furthermore, $T_{ij}$
is diagonal in the space of characters:  $T_{ij}=\exp(2\pi iH_j)\delta_{ij}$.
Also note that $(-{\bf 1})_{ij}=e^{\pi i k} \delta_{ij}$, so that
while the two $SL(2,\IZ)$ matrices $M$ and $-M$
both describe the same modular transformation, they in general
have different representation matrices in the space of
characters:  $(-M)_{ij}=e^{\pi i k} M_{ij}$.
This ambiguity is avoided by ensuring that $M_\ab$ is of the
form (\ref{Mabdef}).
Now, for $\alpha=1$, we have $\beta=0$, and we have already seen
that the simplest choice $\beta'=0$ can be accommodated by
choosing $M_{10}= -S$.  This then leads, as before, to the general result
\beq
    Q_{ij}^{(1)}~=~e^{\pi i k/2}\, S_{ij}~.
\label{Qone}
\eeq
Note that this quantity is manifestly real, since under the $S$-transformation
one finds that modular functions transform as
\beq
    \chi_i\,(-1/\tau) ~=~ e^{-\pi i k/2}\,
     \tau^k\, \sum_{ij} {\tilde{S}}_{ij}\,\chi_j(\tau)
\label{stepfive}
\eeq
with real coefficients ${\tilde{S}}_{ij}$, and according to the definitions
(\ref{Mtrans}) and (\ref{STdefs})
we find that $S_{ij}= e^{-\pi i k/2} {\tilde{S}}_{ij}$.
Proceeding to the $\alpha=2$ case, we have only $\beta=1$,
and by choosing $M_{21}=T^{-1} ST^{-2}ST^{-1}$ we have $\beta'= -1$.
Thus $M^{-1}_{21}=TST^2ST$, and
since in general we have $(TXT)_{ij}=\exp[2\pi i(H_i+H_j)]\,X_{ij}$,
we obtain the result
\beqn
   Q_{ij}^{(2)} ~&=&~ e^{\pi i k/2} \, (ST^2S)_{ij} \,
       \exp\left[ \,\pi i\, (n+H_i+H_j)\right] ~ \nonumber\\
          ~&=&~ e^{\pi i k/2} \, (ST^2S)_{ij} \,
       \exp\left[ \,\pi i\, (H_i+H_j)\right] ~ (-1)^n~.
\label{Qtwo}
\eeqn
In order to see that (\ref{Qtwo}) is manifestly real, we note that this
same choice for $M_{21}$ can also be written as $-ST^2S$.
Thus (\ref{Qtwo}) can equivalently be written as
\beq
   Q_{ij}^{(2)} ~=~ e^{\pi i k/2} \, (-ST^{-2} S)_{ij} \,
       \exp\left[ -\pi i\, (H_i+H_j)\right] ~ (-1)^n~,
\label{stepsix}
\eeq
and since we find
\beqn
    e^{\pi i k/2} \,(-ST^{-2}S)_{ij} ~&=&~
     e^{\pi i k/2}  \, e^{\pi i k}\,
    \sum_k\, e^{-4\pi iH_k}\, S_{ik} S_{kj} \nonumber\\
  &=&~ e^{-\pi i k/2} \,\left[ \sum_k\,
     e^{4\pi iH_k}\, S_{ik} S_{kj}\right]^\ast \nonumber\\
  &=&~ \left[ e^{\pi i k/2}  \, (ST^2S)_{ij} \right]^\ast
\label{stepseven}
\eeqn
[where we have used the fact that
$S_{ij}^\ast= e^{\pi i k} S_{ij}$],
we see that (\ref{stepsix}) is merely the complex conjugate
of (\ref{Qtwo}).
Now, for $\alpha=3$, there are two values of $\beta$ to consider:  $\beta=1$
and $\beta=2$.   Choosing the transformations $M_{31}=-ST^3 S$
and $M_{32}= T^{-1} ST^{-3} ST^{-1}$ respectively yields
$\beta'=-1$ and $\beta'= -2$, giving
\beqn
   Q_{ij}^{(3)} ~&=&~~\phantom{+~} e^{\pi i k/2} \, (-ST^{-3}S)_{ij}
        \,\exp\left[
        -{{2\pi i}\over 3}\, (n+ H_i+H_j)\right] \nonumber\\
   ~&\phantom{=}&~~+~ e^{\pi i k/2} \, (TST^3ST)_{ij} \,\exp\left[
        -{{4\pi i}\over 3}\, (n+ H_i+H_j)\right] ~.
\label{stepeight}
\eeqn
Similar manipulations show, however, that these two terms are
complex conjugates of each other.  We thus obtain the simpler result
\beq
     Q_{ij}^{(3)}~=~ 2\,{\rm Re}\, \left\lbrace e^{\pi i k/2} \,(ST^3S)_{ij}\,
    \exp\left[ {{2\pi i}\over 3}\, (n+ H_i+H_j)\right] \right\rbrace~.
\label{Qthree}
\eeq
The final results for the $\alpha=4,5,{\rm ~and~}6$ cases
can be determined in the same manner:
\beqn
    Q_{ij}^{(4)}~&=&~ 2\,{\rm Re}\, \left\lbrace e^{\pi i k/2} \,(ST^4S)_{ij}\,
    \exp\left[ {{\pi i}\over 2}\, (n+ H_i+H_j)\right] \right\rbrace~,
  \nonumber\\
    Q_{ij}^{(5)}~&=&~ 2\,{\rm Re}\, \left\lbrace e^{\pi i k/2} \,(ST^5S)_{ij}\,
    \exp\left[ {{2\pi i}\over 5}\, (n+ H_i+H_j)\right] \right\rbrace~
   \nonumber\\
    ~&&~~~+~ 2\,{\rm Re}\, \left\lbrace e^{-\pi i k/2}\,(ST^2ST^{-2}S)_{ij}\,
    \exp\left[ {{4\pi i}\over 5}\, (n+ H_i-H_j)\right] \right\rbrace~,
    \nonumber\\
    Q_{ij}^{(6)}~&=&~ 2\,{\rm Re}\, \left\lbrace e^{\pi i k/2} \,(ST^6S)_{ij}\,
    \exp\left[ {{\pi i}\over 3}\, (n+ H_i+H_j)\right] \right\rbrace~.
\label{Qfourfivesix}
\eeqn

Finally, let us consider what modifications must be made to these
asymptotic expansion formulas in the case that our characters $\chi_i$
are {\it not}\/ normalized to $a_0^{(i)}=1$ for all $i$.
Of course, given a particular expansion derived for a normalized
character $\chi_i$, we need simply multiply this result by $a_0^{(i)}$
in order to obtain the expansion for the corresponding {\it un}\/normalized
character.  However, this assumes that the above values of
$Q_{ij}^{(\alpha)}$ are
taken to be those calculated for the {\it normalized}\/ characters;
in particular, the representation matrices $M_{ij}$ for modular
transformations $M$ which appear in the above results
for $Q_{ij}^{(\alpha)}$
are assumed to be those calculated for the normalized characters.
The new matrices $\widehat {M_{ij}}$ for the {\it un}\/normalized characters
have, however, elements which are trivially rescaled relative to those for
the normalized characters:
\beq
    \widehat {M_{ij}}~=~  a_0^{(i)}\, M_{ij}\, (a_0^{(j)})^{-1}~.
\label{rescaledM}
\eeq
This implies
\beq
  Q_{ij}~=~  (a_0^{(i)})^{-1}\, \widehat{Q_{ij}}\, a_0^{(j)}~,
\label{rescaledQ}
\eeq
and thus, rewriting the new result in terms of
appropriately {\it un}\/normalized quantities,
we see that the overall factor $a_0^{(i)}$ cancels and
leaves only an extra factor of $a_0^{(j)}$ inside the $j$-summation.
We therefore have two equivalent options when working with unnormalized
characters $\chi_i$:  we can either calculate an asymptotic expansion by
using {\it normalized}\/ matrices $M_{ij}$ in the definition
for $Q_{ij}^{(\alpha)}$ and multiplying the entire result by $a_0^{(i)}$,
or we can use {\it unnormalized}\/ matrices in the
definition for $Q_{ij}^{(\alpha)}$ and insert an
extra factor of $a_0^{(j)}$ inside the summation over $j$.
We shall use both approaches in Sect.~4.

\vfill\eject
\setcounter{footnote}{0}
\section{The Main Theorem}

In this section we shall prove our main result, given
in (\ref{csmaller}).
It turns out that this result can be viewed as
the two-variable generalization of a well-known
theorem in modular function theory which applies
to functions $f(q)$ of a single modular parameter $q$.
We shall therefore first provide a proof of this one-variable
theorem which makes use of the asymptotic expansions discussed
in Sect.~2.
The proof of the two-variable case will then be relatively straightforward.

\subsection{The One-Variable Case}

The one-variable theorem that we wish to prove states that any function
\beq
        f(\tau)~=~ q^H \sum_{n=0}^\infty a_n q^n~,~~~~~~
         q\equiv e^{2\pi i\tau}
\label{function}
\eeq
which transforms as an eigenfunction with
modular weight $k$ under the $S$ and $T$ modular transformations
\beqn
       f(\tau+1) ~&=&~ \exp(2\pi iH) \,f(\tau)~,\nonumber\\
       f(-1/\tau) ~&=&~ \sigma \,\tau^k\,   f(\tau)~,~~~~~~~|\sigma|=1
\label{eigenfunction}
\eeqn
must vanish identically ({\it i.e.}, $a_n=0$ for all $n$) if
\beq
      k ~<~ 12\,H~.
\label{condition}
\eeq
In general, $k\in {\IZ}/2$ and $\sigma^8=1$.
Thus, since
the Dedekind eta-function $\eta= q^{1/24}\prod_{n=1}^\infty(1-q^n)$
satisfies (\ref{eigenfunction}) with $H=1/24$, $k=1/2$,
and $\sigma =\exp(-\pi i/4)$, one can always multiply or divide
$f$ by a sufficient number of factors of $\eta$ to cast this theorem into
its more familiar form:   any function $f$ which
is modular-{\it invariant}\/ (\ie, $H\geq 0$, $H\in {\IZ}$, $\sigma=1$)
must vanish identically if it has {\it negative}\/ modular weight $k$.
An example of such a function meeting all of these conditions
is
\beq
       J~\equiv~ {\vartheta_3}^4 - {\vartheta_2}^4 - {\vartheta_4}^4~
\label{Jdef}
\eeq
where the $\vartheta_i$ are the classical Jacobi $\vartheta$-functions;
since $J$ has $H=1/2$, $k=2$, and $\sigma = -1$ (or equivalently,
since $J/\eta^{12}$ is modular-invariant with negative modular
weight $k=-4$), $J$ vanishes identically.
While the proof of this general theorem is
standard (see, for example, \cite{Koblitz}), our goal here is to
see how this ``cancellation'' can be understood from the point of view of
the asymptotic expansions presented in Sect.~2.

Note that since $f$ is assumed to satisfy (\ref{eigenfunction}), $f$
forms a one-dimensional representation of the modular group,
with $T_{ij}=\exp(2\pi iH)$ and $S_{ij}=\sigma$.
Using the defining relations of the modular group
$ S^2= (ST)^3= -{\bf 1}$ and the fact
that $(-{\bf 1})_{ij}= e^{\pi ik}\delta_{ij}$,
it is easy to show that any such
one-dimensional representation ``matrices'' $S_{ij}$ and $T_{ij}$
can satisfy the defining relations only if $k-12H\in 4{\IZ}$.
This is therefore a general constraint on any such modular-eigenfunctions
$f$ (such as $\eta$), and the theorem applies only to those special
cases (such as $J$) for which $k-12H<0$.

In order to use the machinery of the asymptotic expansions presented in
Sect.~2, we shall first explicitly organize $f$ as a linear combination of
non-vanishing, linearly-independent characters $\chi_i(q)$:
\beq
        f~=~ \sum_i \,c_i\,\chi_i~=~
        \sum_i \,c_i\, q^{H_i}\,\sum_{n=0}^\infty  \,a_n^{(i)} q^n~.
\label{flinearcomb}
\eeq
Here the $\chi_i$ are members of a single
system of characters closed under modular transformations,
and therefore satisfy (\ref{Mtrans})
and (\ref{chiform}) with normalizations $a_0^{(i)}=1$.
It is necessary to proceed in this manner because $f$ by itself will be shown
to vanish, and therefore $f$ itself cannot be properly normalized (there exists
no $a_0\not= 0$).  We can also assume that $f$ contains a sufficient
number of $\eta$-function factors so that it is modular-invariant:
this implies that the vacuum energies are all integers, $H_i \in {\IZ}$,
and that the coefficients satisfy
\beq
              \sum_{i=1}^N \, c_i \, M_{ij}~=~ c_j~
\label{coeffcondition}
\eeq
for all matrices $M_{ij}$ which represent modular transformations
in the $\chi_i$ system of characters.  Furthermore, as discussed above,
the condition (\ref{condition}) also allows us to choose the number of
$\eta$-function factors so that, without loss of generality,
\beq
          k~<~0~~~~~~{\rm and}~~~~~~~
          H_i ~\geq~ 0~~ {\rm if}~~c_i\not= 0~.
\label{notachyons}
\eeq
Given this decomposition of $f$,
then, the procedure for our proof will be quite straightforward:
we shall simply calculate the asymptotic expansions
of each $\chi_i$ individually, and add them together.\footnote{
    It is possible, of course, that several of the $\chi_i$
    appearing in (\ref{flinearcomb}) correspond to sectors
    with vacuum energies $H_i$ which are equal modulo 1 and
    {\it negative}\/;  the resulting function $f$ would still have
    a positive overall vacuum energy $H\geq 0$ provided that all
    tachyonic contributions from these sectors cancel in
    the sum (\ref{flinearcomb}).  However, in such cases we can always
    form linear combinations of these characters $\chi_i$ so that
    the new ``characters'' are still eigenfunctions of $T$ and have
    {\it positive}\/ net vacuum energies $H_i$.
    Thus, even in such cases, a decomposition into characters
    (or linear combinations of characters) exists for which
    (\ref{notachyons}) is satisfied.  The effects of working with
    such linear combinations of characters will be discussed in Sect.~4,
    and we shall assume for the remainder of this section that
    no such linear combinations are necessary.}

The asymptotic expansions for each $a_n^{(i)}$ in (\ref{flinearcomb})
are given in (\ref{finalone}), but before these expansions
can be added together to form an asymptotic expansion
for the combined coefficients $a_n$ in (\ref{function}), we need
to make one important adjustment:
we first need to take into account the fact that each character
$\chi_i$ in principle has a different vacuum energy $H_i$,
and that therefore the degeneracy of the
$n^{\rm th}$ level in a sector with vacuum energy $H_i$
should really be added to the degeneracy of the $(n-1)^{\rm th}$ level
in a sector which vacuum energy $H_i+1$, {\it etc}.
Indeed, it is only the combination $n'\equiv n+H_i$ which
has physical meaning as the ``energy'' of the state, invariant across
all sectors of the theory.  We can therefore, as a first step,
rewrite the asymptotic
expansions from Sect.~2 in terms of this properly shifted
variable $n'$ (and henceforth drop the prime on $n$):
\beq
     a_n^\ii ~=~ \sum_{\alpha=1}^{\alpha_{\rm max}}
     \,\left( {{2\pi}\over\alpha} \right)
    \,\sum_{j=1}^N  \, Q_{ij}^{(\alpha)} \, f_j^{(\alpha)}(n)
\label{resulta}
\eeq
where
\beq
  Q_{ij}^{(\alpha)} ~\equiv~ e^{\pi i k/2}\, \sum_\beta\,\Minv~
    \exp \left[ {2\pi i\over \alpha} \left(
      \beta'\,H_j \,-\, \beta\,n\right) \right]~
\label{resultQ}
\eeq
and where
\beq
   f_j^{(\alpha)}(n)~\equiv~\cases{
   \left( \sqrt{ H_j/n} \,\right)^{1-k}\,
    J_{k-1}\left(  4\pi \sqrt{ H_j n}/\alpha \right) & for $H_j>0$ \cr
   \left( \sqrt{ |H_j|/n} \,\right)^{1-k}\,
    I_{k-1}\left(  4\pi \sqrt{ |H_j| n}/\alpha \right) & for $H_j<0$ \cr
   \left( 2\pi n/\alpha \right)^{k-1} & for $H_j=0$~.\cr}
\label{resultf}
\eeq
While this variable shift $n\to n'\equiv n+H_i$ is somewhat trivial for
the present one-variable case (in particular,
both the original and final $n$'s are integers since
each $H_i\in{\IZ}$), we will see that this
step is far more subtle for the two-variable case.

In terms of these shifted variables, then, the $q$-expansion
for $f$ becomes $f= \sum_n \,a_n\,q^n$ where $a_n$ is now the
number of states at energy $n$, and the asymptotic expansion
for $a_n$ is simply given by
\beq
        a_n~=~ \sum_{i,\alpha,j} \left({2\pi\over\alpha}\right)
             c_i\, Q_{ij}^{(\alpha)}\,f_j^{(\alpha)}(n)~.
\label{StepOne}
\eeq
However, having performed the shift of variables,
we now see from (\ref{resultf})
that $f_j^{(\alpha)}(n)$ is independent of $i$.
We therefore find that the sum over $i$ factors,
\beq
        a_n~=~ \sum_{\alpha,j} \left({2\pi\over\alpha}\right)
             \left(\sum_i c_i\, Q_{ij}^{(\alpha)}\right)
       \,f_j^{(\alpha)}(n)~,
\label{StepTwo}
\eeq
and that within this factored sum there is yet another factorization:
\beq
       \sum_i c_i Q_{ij}^{(\alpha)} ~=~
         e^{\pi i k/2} \sum_\beta
        \left( \sum_i c_i (M_{\alpha\beta}^{-1})_{ij}\right)
         \exp\left[ {2\pi i\over\alpha} \left(\beta' H_j -\beta n\right)
            \right] ~.
\label{StepThree}
\eeq
However, we recall from Sect.~2 that $M_{\alpha\beta}^{-1}$ is a modular
transformation, and that $(M_{\alpha\beta}^{-1})_{ij}$ is the representation
matrix corresponding to this transformation in the space
of characters $\chi_i$.  From (\ref{coeffcondition}), therefore,
we see that
\beq
              \sum_i \, c_i \, (M_{\alpha\beta}^{-1})_{ij}~=~ c_j~
\label{StepFour}
\eeq
for all $\alpha$ and $\beta$, yielding simply
\beq
       \sum_i c_i Q_{ij}^{(\alpha)} ~=~
         e^{\pi i k/2}  \, c_j\,
         \sum_\beta \,
     \exp\left[ {2\pi i\over\alpha} \left(\beta' H_j -\beta n\right)
            \right] ~.
\label{StepFive}
\eeq

It may seem strange at this point that $\beta'$ still appears in
(\ref{StepFive}), since {\it a priori}\/ all knowledge of the original
choice of $\beta'$ [which was implicit in our choice of modular
transformation $M_{\alpha\beta}$ in (\ref{Mabdef})] has been removed.
The important point to realize is that
for any given values of $\alpha$ and $\beta$,
there are only certain allowed values of $\beta'$ which may be chosen,
and (as discussed in the footnote in Sect.~2) these values of $\beta'$
are all equal modulo $\alpha$.
Thus, since $H_j\in{\IZ}$ for all $j$,
the result in (\ref{StepFive}) is independent of the particular
choice of $\beta'$, and depends instead on only the value of $\beta'$
modulo $\alpha$.  This value is uniquely determined from knowledge
of $\alpha$ and $\beta$ alone, and therefore does not require
knowledge of the chosen modular transformation $M_{\alpha\beta}$.

Thus, combining the above results, we have
\beq
        a_n~=~ \sum_{\alpha,j} \left({2\pi\over\alpha}\right)
         e^{\pi i k/2}  \,
         \left\lbrace \sum_\beta \,
     \exp\left[ {2\pi i\over\alpha} \left(\beta' H_j -\beta n\right)
            \right] \right\rbrace
        c_j\, f_j^{(\alpha)}(n)~.
\label{StepSix}
\eeq
However, since the $c_j$ are the original coefficients of $f$,
we recognize that $c_j\not= 0$ only if $H_j\geq 0$.
Thus,
the only functions $f_j^{(\alpha)}(n)$ which contribute
to the asymptotic expansion of $a_n$ are those for which $H_j\geq 0$.
However, for $k<0$, all of these functions vanish in
in the $n\to\infty$ limit: if $H_j>0$, then
$f_j^{(\alpha)}(n)\sim n^{k/2-3/4}$ [recall (\ref{resultf}) and
(\ref{otherbesselasymp})],
while if $H_j=0$, then $f_j^{(\alpha)}(n)\sim n^{k-1}$.
We therefore find
\beq
          a_n ~\to~ 0~~~~~~{\rm as}~ n\to \infty~,
\label{StepSeven}
\eeq
and since the {\it error} in the asymptotic expansion
also vanishes for $k<0$ [recall (\ref{error})],
for sufficiently large $n$ this can be taken to be an exact result:
\beq
         a_n~=~0~~~~~~{\rm for~sufficiently~large}~~n~.
\label{StepEight}
\eeq
It is then a simple matter, using the modular transformation
$\tau\to -1/\tau$ and a Poisson resummation,
to demonstrate that {\it all}\/ coefficients $a_n$ must vanish exactly.

As an example, let us consider the function $J$ given in (\ref{Jdef}).
As indicated in (\ref{flinearcomb}), this function can be written
as a sum of characters which meet all of the necessary conditions:
\beq
     {J\over \eta^{12}}~=~
   {1\over \eta^{8}} \,\left\lbrace
    ({\chi_0})^7 \chi_{1/2} + 7\,({\chi_0})^5 ({\chi_{1/2}})^3 +
              7\, ({\chi_0})^3 ({\chi_{1/2}})^5
     + {\chi_0} ({\chi_{1/2}})^7 - (\chi_{1/16})^{8}\right\rbrace~.
\label{Jdecomp}
\eeq
Here $\chi_h$ with $h\in\lbrace 0,1/2,1/16\rbrace$ are
the characters of the $c=1/2$ Ising model,
and recalling that $1/\eta$ is the character of an uncompactified boson
with $c=1$, we see that
each of the five terms in the sum (\ref{Jdecomp}) is
therefore a distinct character in the large $c=12$ conformal
field theory which consists of a tensor product
of eight uncompactified bosons and eight Ising models.  Indeed,
each of these characters has $k=-4$, with
individual vacuum energies $H_i=\sum{h}-c/24=0,1,2,3,0$ respectively.
The conditions (\ref{notachyons}) are therefore satisfied.
Indeed, since each of the terms in (\ref{Jdecomp}) is the character $\chi_i$
of a $c=12$ conformal field theory, their separate degeneracies
$a_n^{(i)}$  each grow asymptotically $\sim \exp(C\sqrt{n})$ with
$C=2\sqrt{2}\pi$, in accordance with (\ref{Hagedornvalues}).
However, the {\it sum}\/ of these asymptotic expansions, and indeed each
total degeneracy $a_n$, vanishes identically.  The physical interpretation
of this fact is that the final term in (\ref{Jdecomp}) is the contribution of
a {\it fermionic}\/ Ramond sector (as indicated by the fact that each
factor $\chi_{1/16}$ corresponds to an $h=1/16$ Ising-model spin field),
and thus cancels the contributions of the four bosonic (Neveu-Schwarz) sectors
to yield a supersymmetric theory.
The result $J=0$ is often called the Jacobi identity, and the partition
functions of supersymmetric theories are usually proportional to $J$.

An interesting corollary of this proof
concerns the case of functions $f$ which are modular-invariant
with $H\geq 0$ but which have {\it positive}\/ modular weight $k$.
Such functions of course do not vanish, and we would {\it a priori}\/
expect exponential growth in their coefficients.
However, as we can see from the above derivation,
this is not the case:  the relevant functions
$f_j^{(\alpha)}(n)$ can grow at most {\it polynomially}\/ with $n$,
and moreover the {\it error}\/ in the asymptotic expansion for a given
$a_n$ can also grow at most polynomially ($\sim n^k$).
Thus, surprisingly, the maximum possible growth for such cases
is polynomial rather than exponential.  For example, the $E_8$
character $ch(E_8)={\vartheta_2}^8+ {\vartheta_3}^8+ {\vartheta_4}^8$
has $H=0$ and $k=4$, and indeed the coefficients
in its $q$-expansion grow only as fast as $n^{4}$.

A similar but more delicate situation exists for
modular-invariant functions which have $k=0$.
In this case the relevant functions $f_j^{(\alpha)}(n)$
all vanish in the asymptotic limit $n\to\infty$,
and indeed one again obtains (\ref{StepSeven}).
However, the {\it error}\/ in the asymptotic expansions
can be as great as an overall additive {\it constant}.
Thus we have an interesting situation in which the values of $a_n$ experience
no growth at all, but are not necessarily zero.
However, one can show that if the $a_n$'s experience no growth and
$f(-1/\tau)=f(\tau)$, then the only possible solution is
 $a_n\propto \delta_{n,0}$ if $H=0$, or all $a_n=0$ if $H>0$.
Thus, if $k=0$ and $H\geq 0$, we have $f={\rm constant}$, and
this constant is non-zero only if $H=0$.
An example of the latter situation is the modular-invariant
function
\beq
K~\equiv~ {\vartheta_2\vartheta_3\vartheta_4\over 2\,\eta^3}~=~
     \left[({\chi_0})^2 - ({\chi_{1/2}})^2\right]^2\,(\chi_{1/16})^2
\label{Kdef}
\eeq
with $H=k=0$;
by the above result this must equal a constant, and indeed one
finds $K=1$.  These results are also in accord with
known theorems in modular function theory.

\subsection{The Two-Variable Case}

We shall now proceed to examine the more physically relevant
case of modular functions of two variables, $Z(q,\overline{q})$,
to see if an analogous result can be obtained.

At the outset, there are a number of differences between this and
the simpler one-variable case.
The fact that we have both a holomorphic variable $q$ and an anti-holomorphic
variable $\qbar$ means that
modular-invariant functions $Z(q,\qbar)$ can in principle be formed
with {\it two}\/ separate systems of characters, $\chi$ and $\chibar$:
\beq
      Z~=~ ({\rm Im}\,\tau)^k\,\sum_{i,\ibar}\,
          \chibar^{\,\ast}_\ibar \,N_{\ibar i}\,\chi_i~=~
       ({\rm Im}\,\tau)^k\,\sum_{i,\ibar}\,
            N_{\ibar i}\,q^{H_i}\,\qbar^{\Hbar_\ibar} \,\sum_{m,n}
           \abar_m^{(\ibar)} \,a_n^{(i)}\,\qbar^m q^n~.
\label{Zexplicit}
\eeq
In fact, partition functions of this general type appear for any
string theory (such as all heterotic string theories) with unequal
underlying left- and right-moving worldsheet conformal field theories,
and the $(\ibar,i)$ summation within (\ref{Zexplicit}) will indeed be finite
if these worldsheet theories are rational.
It is therefore necessary, in general, to keep track of {\it two} separate
systems of characters, $\chi$ and $\chibar$, related to each other
in (\ref{Zexplicit}) only through
the requirement of modular invariance.
Hence, throughout this section, we shall use the following
notation:  a bar above any variable will refer to the right-moving
system of characters $\chibar$, and complex conjugation will instead be
indicated explicitly with an asterisk.  Of course, $\qbar=q^\ast$.

As in the one-variable case, we will demand that our function $Z$
is modular-invariant with modular weight $k$.  Explicitly,
this means that $Z(\tau+1)=Z(-1/\tau)=Z(\tau)$ where
the individual characters $\chi$ and $\chibar$ satisfy (\ref{Mtrans})
with modular weight $k$.  [Note that the factor $({\rm Im}\,\tau)^k$
in (\ref{Zexplicit}) is necessary so that $Z(-1/\tau)=Z(\tau)$.]
It is then straightforward to show that
the coefficients $N_{\ibar i}$ in (\ref{Zexplicit}) must satisfy
\beq
          \sum_{i,\ibar}~
   M_{\ibar\jbar}^{\,\ast}\,N_{\ibar i}\,M_{ij}~=~
      N_{\jbar j}
\label{coeffconditiontwo}
\eeq
for all pairs of matrices $M_{ij}$ and $M_{\ibar\jbar}$ which
respectively represent the same modular transformation $M$ in
the $\chi_i$ and $\chibar_\ibar$ systems.
This condition is of course the two-variable analogue of
(\ref{coeffcondition}).
Note that for the special case $M=T$, (\ref{coeffconditiontwo}) implies
the so-called {\it level-matching}\/ condition
\beq
       H_i ~-~ \Hbar_\ibar ~\in~{\IZ}~~~~~~~{\rm if}~~N_{\ibar i}\not= 0~.
\label{levelmatching}
\eeq

This condition is the source of the second fundamental difference from the
one-variable case.  For the one-variable case, the analogous
modular-invariance condition required simply that all $H_i\in{\IZ}$;
essentially the ``anti-holomorphic characters'' were all $\chibar=1$,
with $\Hbar_\ibar=0$. Now, however, we see that we need no longer have
integer vacuum energies $H_i$ and $\Hbar_\ibar$, and in fact for most
physical situations non-integer values of $H_i$ and $\Hbar_\ibar$ do appear.
Indeed, all that is required is that any pairs $H_i$ and $\Hbar_\ibar$ coupled
together via the $N_{\ibar i}$ matrix be equal modulo $1$.
It is this fact which permits a ``misalignment'' of the various
sectors in such a left/right theory.  Recall that each character $\chi_i$ and
$\chibar_\ibar$ corresponds to an entire tower of chiral states, with integer
spacing between adjacent levels in each tower.  Thus, while the level-matching
condition (\ref{levelmatching}) guarantees that a given pair of left- and
right-moving characters are properly aligned relative to each other (yielding
a set of properly constructed left/right states), a given {\it pair}\/ of
characters corresponding to a given sector of the theory can nevertheless be
``misaligned'' relative to another pair.  Thus, the different sectors will in
general exhibit a whole spectrum of alignments, each corresponding to a
different value of $H_i=\Hbar_\ibar$ (modulo 1), and any analogous
cancellation which we expect to observe between the different sectors of the
theory must now somehow take into account the fact that these sectors
are misaligned.

Given the modular-invariance condition (\ref{coeffconditiontwo}),
we now must impose the conditions analogous to those of the one-variable
case in (\ref{notachyons}).
The first of these is straightforward:  we shall once again consider
those partition functions $Z$ for which the modular weight $k$ is
negative.
Indeed, in most physical situations the weight $k$ is related to the
number of uncompactified spacetime dimensions $D$ via
\beq
                 k~= ~ 1~-~D/2~,
\label{kDrelation}
\eeq
so this condition corresponds to the physically interesting
cases with $D>2$.
(We will nevertheless show at the end of this section that
our results apply for {\it all}\/ spacetime dimensions $D$, and thereby
include the cases with non-negative values of $k$ as well.)
Generalizing the second condition in (\ref{notachyons}) is
more subtle, however, for two distinct possibilities present themselves:
\beq
       H_i~\geq~0 ~~~~{\rm and}~~~~ \Hbar_\ibar~\geq~0  ~~~~~~{\rm if}~~
             N_{\ibar i}\not= 0~,
\label{choiceone}
\eeq
or
\beq
       H_i~\geq~0 ~~~~{\rm or}~~~~ \Hbar_\ibar~\geq~0  ~~~~~~{\rm if}~~
             N_{\ibar i}\not= 0~.
\label{choicetwo}
\eeq
The first of these two options is the strictest generalization of the
one-variable condition (\ref{notachyons}), and for $k<0$ leads to the
analogous result $Z=0$ ({\it i.e.}, equal
numbers of bosonic and fermionic states at all mass levels).\footnote{
  Similarly, for $k=0$ we would actually have $Z=$ constant,
  with a net number of physical states surviving at the massless level only.}
This is therefore the appropriate condition for theories with
spacetime supersymmetry,
for such theories, in agreement with (\ref{choiceone}),
necessarily lack both physical and unphysical tachyons.
By contrast, the second option (\ref{choicetwo})
is far weaker, and merely requires that there exist
no {\it physical}\/ tachyons;  unphysical tachyons are still permitted.
As we will see, this case corresponds to theories {\it without}\/ spacetime
supersymmetry, theories in which the numbers of bosons and fermions
at all mass levels are {\it a priori}\/ unequal.
The condition (\ref{choicetwo}) nevertheless still prohibits the appearance
of {\it physical}\/ tachyons:  these are the tachyons which cause
fundamental physical inconsistencies in spacetime,
and which lead to divergent amplitudes.
Thus (\ref{choicetwo}) is the weakest condition that can
be imposed for physically sensible theories,
and while the existence of unphysical tachyons
will of course prevent $Z$ from vanishing altogether,
our goal is to determine if the absence of physical
tachyons nevertheless leads to any (weaker) cancellation.\footnote{
    As in the one-variable case, we note that a decomposition of the
    partition function $Z$ into characters satisfying (\ref{choicetwo})
    exists even in those cases in which certain sectors are individually
    tachyonic, provided their tachyonic contributions are cancelled (or
    ``GSO-projected'') in the sum (\ref{Zexplicit}).
    Thus (\ref{choicetwo}) indeed represents a general condition for
    functions $Z$ without physical tachyons, and in these particular cases
    linear combinations of characters may be necessary when
    constructing a suitable character-decomposition of $Z$.
    In Sect.~4 we shall discuss in detail the effects of working with
    such linear combinations, and in the remainder of this section we
    shall assume for simplicity that no such linear combinations are needed.}

Our derivation now proceeds exactly as for the one-variable case:
we shall derive an asymptotic expansion for each term in the
partition function (\ref{Zexplicit}), corresponding to a different
sector of the theory, and then add these asymptotic expansions together.

Our first step is to adjust our summation variables $m$
and $n$ in (\ref{Zexplicit}) so that they correspond to the invariant
energy of the corresponding state, and not just the number of
integer-excitations above the varying vacuum energies $H_i$ and $\Hbar_\ibar$.
This will enable us to expand the partition function (\ref{Zexplicit})
in the simple form
\beq
         Z ~=~ ({\rm Im}\,\tau)^k \,\sum_{m,n}\, a_{mn}\, \qbar^m q^n~
\label{Zcombined}
\eeq
and so read off the total degeneracy $a_{nn}$ of
physical states with a given energy $n$.
As in the one-variable case, this is trivially accomplished
by {\it shifting}\/ the summation variables $m$ and $n$
in each $(\ibar,i)$ sector so that these different vacuum energies are
properly incorporated:  $n\to n+H_i$, $m\to m+\Hbar_\ibar$.
Unlike the one-variable case, however,
this operation has drastic consequences.
Of course, since the values of $H_i$ and $\Hbar_\ibar$ are non-integral,
the new summation variables $m$ and $n$ in (\ref{Zcombined})
are also non-integral, and each sector $(\ibar,i)$
will contribute states to only those degeneracy counts $a_{mn}$ with
$m=n=H_i=\Hbar_\ibar$ (modulo 1).
This much is non-problematic.
However, it is then no longer appropriate to simply {\it add}\/
the asymptotic expansions as functions of $n$ in order to
obtain a value for a given $a_{nn}$;
indeed, for any $n$,
one should properly add together only those asymptotic expansions
which correspond to sectors with vacuum energies $H_i$ and $\Hbar_\ibar$
satisfying $H_i=\Hbar_\ibar=n$ (modulo 1).

Despite this fact,
we shall nevertheless proceed to add together the asymptotic expansions
from {\it all}\/ the sectors, just as was done in the one-variable case.
While the physical significance of this addition may not yet be
apparent, we will find that a powerful result with important
physical consequences can nevertheless be obtained.
We will refer to this addition operation as ``sector-averaging'',
and denote the resulting ``sector-averaged'' degeneracy
at energy $n$ as $\langle a_{nn} \rangle$.
Our results from this point forward shall therefore apply
to $\langle a_{nn}\rangle$ rather than to any particular
value of $a_{nn}$ in (\ref{Zcombined}),
and after obtaining our ultimate result for $\langle a_{nn}\rangle$
we shall discuss the consequences for the true state degeneracies
$a_{nn}$.

Thus, explicitly, our calculation will be formulated as follows.
First, for each sector $(i,\ibar)$ in (\ref{Zexplicit}),
we shift the variables $m$ and $n$ so that the vacuum energies
$H_i$ and $\Hbar_\ibar$ are properly incorporated and $m=n=H_i=\Hbar_\ibar$
(modulo 1):
\beq
          \chibar_\ibar \, \chi_i ~=~ \sum_{m,n}\, a_{mn}^{(\ibar i)}~
       \qbar^m \,q^n~.
\label{sectordegs}
\eeq
The quantities $a_{mn}^{(\ibar i)}$ then represent the
net degeneracies of physical and unphysical states in the $(\ibar,i)$
sector, related to the separate chiral degeneracies $a_n^{(i)}$ and
$\abar_m^{(\ibar)}$ via
\beq
        a_{mn}^{(\ibar i)}~=~ \abar_m^{(\ibar)}\,a_n^{(i)}~.
\label{product}
\eeq
Let us now draw a notational distinction
between a particular degeneracy $a_n^{(i)}$ and its corresponding
 {\it asymptotic expansion},
denoting this latter function [given
on the right side of (\ref{resulta})]
as $\phi^{(i)}(n)$.
Thus, further denoting by
$\Phi^{(\ibar i)}(n)$ the function describing
the physical degeneracies $a_{nn}^{(\ibar i)}$, we have
\beq
        \Phi^{(\ibar i)}(n)~=~ [\overline{\phi}^{(\ibar)}(n)]^\ast\,
        \phi^{(i)}(n)~.
\label{productphi}
\eeq
The operation of ``sector averaging'' is then rigorously defined via the
modular-invariant summation of these functional forms:
\beq
      \langle a_{nn}\rangle~\equiv~
      \sum_{i,\ibar}\,N_{\ibar i}\,
        \Phi^{(\ibar i)}(n)~=~
      \sum_{i,\ibar}\,N_{\ibar i}\,
     [\overline{\phi}^{(\ibar)}(n)]^\ast\,
        \phi^{(i)}(n)~.
\label{sectoraveraging}
\eeq
Although the asymptotic expansions $\phi^{(i)}(n)$ are {\it a priori}\/
real, we have explicitly indicated their complex conjugations (where
appropriate) in (\ref{productphi}) and (\ref{sectoraveraging}).

Substituting the asymptotic expansions given in (\ref{resulta}),
we then find
\beq
      \langle a_{nn} \rangle ~=~
      \sum_{i,\ibar}\,N_{\ibar i}\,
      \sum_{\alpha,\alphabar} \left( {4\pi^2\over \alpha \alphabar}\right)
      \sum_{j,\jbar} \,Q_{ij}^{(\alpha)}\,
      \left(\Qbar_{\ibar\jbar}^{(\alphabar)}\right)^\ast\,
       f_j^{(\alpha)}(n)
       \,\overline{f}_\jbar^{(\alphabar)}(n)~,
\label{stepNine}
\eeq
and once again the summations over $i$ and $\ibar$ factor explicitly:
\beq
      \langle a_{nn} \rangle ~=~
      \sum_{\alpha,\alphabar,j,\jbar}
     \left( {4\pi^2\over \alpha \alphabar}\right)
      \left\lbrace
    \sum_{i,\ibar}
      \,\left(\Qbar_{\ibar\jbar}^{(\alphabar)}\right)^\ast\, N_{\ibar i}\,
      Q_{ij}^{(\alpha)}\right\rbrace
       \, f_j^{(\alpha)}(n)
       \,\overline{f}_\jbar^{(\alphabar)}(n)~.
\label{stepTen}
\eeq
Indeed, just as in the one-variable case, there is now an additional
factorization within the $(i,\ibar)$ summation:
\beqn
    \sum_{i,\ibar}
      \,\left(\Qbar_{\ibar\jbar}^{(\alphabar)}\right)^\ast\, N_{\ibar i}\,
      Q_{ij}^{(\alpha)}~&=&~
   \sum_{\beta,\betabar}\,\exp\left\lbrack
     { 2\pi i \over\alpha} \left(
     \beta'H_j - \beta n\right)
    -{2\pi  i\over \alphabar} \left(
     \betabar' \,\Hbar_\jbar - \betabar \,n\right)
      \right\rbrack
             ~\times \nonumber\\
    &&~~~~~~~~~~
      \times ~\left\lbrace \sum_{i,\ibar}~
       \left(M_{\alphabar\betabar}^{-1}\right)_{\ibar\jbar}^\ast\,
           N_{\ibar i}\,
          (M_{\alpha \beta}^{-1})_{ij}\right\rbrace~.
\label{stepEleven}
\eeqn
Note that these two last steps implicitly assume
that the variable $n$ is independent of $i$, which is the essence of
the ``sector averaging'' discussed above.  Even though different
sectors $(\ibar,i)$ correspond to different values of $n$ modulo $1$,
we are here adding the sum of {\it functional forms} of $n$.

The above factorization now enables us to make one important
simplification.  Using (\ref{coeffconditiontwo}), we can rewrite
the second line of (\ref{stepEleven}):
\beq
      \sum_{i,\ibar}~
       \left(M_{\alphabar\betabar}^{-1}\right)_{\ibar\jbar}^\ast\,
           N_{\ibar i}\,
          (M_{\alpha \beta}^{-1})_{ij}
   ~=~ \left( N\,M_{\alphabar\betabar}\,M_{\alpha\beta}^{-1}\right)_{\jbar j}~;
\label{MNMtoNMM}
\eeq
such rewritings are useful in that they generally allow us to focus
on the modular-transformation representation matrices in only one
character system ({\it e.g.}, the holomorphic system, as formulated
above) rather than on both simultaneously.
Thus, collecting our results in this final form, we have
the following total expansion for $\langle a_{nn} \rangle$:
\beq
      \langle a_{nn} \rangle ~=~
      \sum_{\alpha,\alphabar,j,\jbar}
     \left( {4\pi^2\over \alpha \alphabar}\right)
   \, {\cal P}_{\jbar j}^{(\alphabar,\alpha)}\,
       f_j^{(\alpha)}(n) \,\overline{f}_\jbar^{(\alphabar)}(n)~
\label{finalexpansion}
\eeq
where
\beq
  {\cal P}_{\jbar j}^{(\alphabar,\alpha)}~\equiv~
   \sum_{\beta,\betabar}\,\exp\left\lbrack
     { 2\pi i \over\alpha} \left(
     \beta'H_j - \beta n\right)
    -{2\pi  i\over \alphabar} \left(
     \betabar' \,\Hbar_\jbar - \betabar \,n\right)
      \right\rbrack
   \left( N\,M_{\alphabar\betabar}\,M_{\alpha\beta}^{-1}\right)_{\jbar j}~.
\label{Pdef}
\eeq

Let us now examine the {\it leading}\/ term in this
expansion for $\langle a_{nn} \rangle$.
Since the leading term for each chiral character is obtained
for $\alpha=1$, the leading term in (\ref{finalexpansion})
is that for which $\alpha=1$ and $\alphabar=1$.
However, in this case we find that
${\cal P}^{(\alphabar,\alpha)}_{\jbar j}$
assumes a particularly simple form:  since the only allowed values
of $\beta$ and $\betabar$ are $\beta=\betabar=0$,
we have simply
\beq
       {\cal P}_{\jbar j}^{(1,1)}~=~ (N M_{10} M_{10}^{-1})_{\jbar j}~=~
           N_{\jbar j}~.
\label{Poneone}
\eeq
Thus, we find that the leading term
in the expansion of $\langle a_{nn}\rangle$ is simply
\beq
      \langle a_{nn} \rangle ~=~
      4\pi^2\, \sum_{j,\jbar} \,
          N_{\jbar j} \, f_j^{(1)}(n)
      \, \overline{f}_\jbar^{(1)}(n)~+~...
\label{leadingterm}
\eeq

This result is sufficient to prove our main theorem.
Since the partition function $Z$ in (\ref{Zexplicit})
is assumed to satisfy (\ref{choicetwo}),
we see that the only values of $j$ and $\jbar$ which contribute
to the sum in (\ref{leadingterm}) are those for which
either $H_j\geq 0$ or $\Hbar_\jbar \geq 0$.
This means that there exists no term in (\ref{leadingterm}) which
contains a product of two functions
$f_j^{(1)}(n)$ and $\overline{f}_\jbar^{(1)}(n)$ with both
experiencing maximum growth.
In particular, the term in the asymptotic expansion with the greatest
possible growth would {\it a priori}\/ have been given by
$f_{j=1}^{(1)}(n) \overline{f}_{\jbar=1}^{(1)}(n)$
where $j=1$ and $\jbar=1$ respectively indicate
the vacuum sectors of the separate chiral theories,
yet we see that such a maximally-growing term is
necessarily absent in (\ref{leadingterm}).
The absence of such a term in the expansion for the
sector-averaged $\langle a_{nn} \rangle$ is of course
the direct consequence of the
absence of a corresponding physical tachyon in the partition function $Z$.

By contrast, let us consider the degeneracies
$a_{nn}^{(\ibar i)}$ of each individual sector $(\ibar, i)$
 {\it before}\/ sector-averaging, as given in (\ref{sectordegs})
and (\ref{product}).  In analogy to (\ref{stepTen}),
we find that these $a_{nn}^{(\ibar i)}$ have expansions
\beq
      a_{nn}^{(\ibar i)}  ~=~
      \sum_{\alpha,\alphabar,j,\jbar}
     \left( {4\pi^2\over \alpha \alphabar}\right)
      \,\left(\Qbar_{\ibar\jbar}^{(\alphabar)}\right)^\ast\,
      Q_{ij}^{(\alpha)}
       \, f_j^{(\alpha)}(n)
       \overline{f}_\jbar^{(\alphabar)}(n)~,
\label{stepTwelve}
\eeq
and using (\ref{Qone}), we see that the
leading terms in (\ref{stepTwelve}) are
\beq
      a_{nn}^{(\ibar i)}  ~=~
      \sum_{j,\jbar}
     4\pi^2 \, S_{\ibar \jbar} \,S_{ij}
       \, f_j^{(\alpha)}(n)
       \overline{f}_\jbar^{(\alphabar)}(n)~+~...
\label{stepThirteen}
\eeq
The terms in (\ref{stepTwelve}) with maximal growth are again
 {\it a priori}\/ given by $f_{j=1}^{(1)}(n) \overline{f}_{\jbar=1}^{(1)}(n)$
where $j=\jbar=1$ indicate the vacuum sectors of the separate chiral theories.
However, recall that for valid conformal field theory characters,
$S_{i1}$ and $S_{\ibar 1}$ are necessarily {\it non}\/-zero \cite{Verlinde}.
Thus, for each of the individual sectors $(\ibar,i)$,
this term with maximal growth must exist in the expansion for
the degeneracies, and it is only in the
process of ``sector averaging'' that this term is cancelled.

This is precisely the content of our main result, as expressed
in (\ref{csmaller}).
Since the leading term in (\ref{stepThirteen}) contains the factor
$f_{j=1}^{(1)}(n) \overline{f}_{\jbar=1}^{(1)}(n)$,
the growth of each $a_{nn}^{(\ibar i)}$ as $n\to\infty$ is
\beq
        a_{nn}^{(\ibar i)}~\sim~
      \exp\left\lbrace 4\pi \left(\sqrt{|H_1|n}
                      +\sqrt{|\overline{H}_1|n}\right)\right\rbrace
   ~\equiv~ \exp\left(C_{\rm tot} \sqrt{n}\right)
\label{annCtot}
\eeq
where
\beq
          C_{\rm tot} ~=~ C_{\rm left}+C_{\rm right}~\equiv~ 4\pi\left(
       \sqrt{{c_{\rm left}\over 24}} +
       \sqrt{{c_{\rm right}\over 24}}  \right)~.
\label{Ctotc}
\eeq
However, the growth in $\langle a_{nn}\rangle$ is necessarily
less rapid, since the rate of exponential growth in (\ref{annCtot})
can come only from the dominant term $f_{j=1}^{(1)}(n)
\overline{f}_{\jbar=1}^{(1)}(n)$.
Thus, defining the total effective ``sector-averaged'' value of $C$
through
\beq
        \langle a_{nn}\rangle ~\sim~
           \exp\left(C_{\rm eff} \sqrt{n}\right)~~~~~~~~{\rm as}~n\to\infty~,
\label{Ceffdef}
\eeq
we have
\beq
           C_{\rm eff} ~<~ C_{\rm tot}~.
\label{lessthan}
\eeq

Since the exact value of $C_{\rm eff}$ in (\ref{Ceffdef}) is
in principle determined by the largest remaining {\it subdominant}\/ term in
the complete $\langle a_{nn}\rangle$ expansion,
let us now focus briefly on those subdominant terms which might be relevant.
Such terms can have a variety of origins, and we shall outline them below.
\begin{itemize}
\item First, there are of course subdominant terms
   within (\ref{leadingterm}), for the presence of {\it unphysical}\/
   tachyons within the partition function $Z$ requires
   that there exist non-vanishing terms in (\ref{leadingterm}) for which
   $H_j<0$ and $\overline{H}_\jbar\geq 0$ (or vice versa).
   It is in fact a generic property that there exist such unphysical tachyons
   in non-supersymmetric theories, for modular invariance alone can be used
   to demonstrate that any $D>2$ theory which lacks both physical {\it and}\/
   unphysical tachyons will have a vanishing partition function.
   For example, in the case of generic non-supersymmetric string theories
   containing gravitons, it is not difficult to show\footnote{
     The proof runs as follows.
     Such unphysical tachyonic states are bosonic, and are essentially the
     graviton state {\it without}\/ the excitation of the left- (or right-)
     moving mode of the coordinate boson.  Since the constraint equations
     for these unphysical tachyons are therefore those of the graviton itself,
     such a state is guaranteed to survive all Fock-space projections.
     The only way to cancel the contribution of this state to the partition
     function $Z$ is to have a corresponding {\it fermionic}\/ state with the
     same (unphysical tachyonic) left/right energy distribution.  However,
     such fermionic states are likewise related to the physical
     {\it gravitino}\/ states, and therefore any constraint equations
     which project out the gravitino state (thereby rendering the theory
     non-supersymmetric) must project out this fermionic tachyon as well.
     Thus, for non-supersymmetric theories containing a graviton,
     there always exist sectors in the partition function with
     $H_i=-c_{\rm left}/24$ and $\Hbar_\ibar=0$ (or vice versa) whose
     contributions to $Z$ are not cancelled.}
  that such unphysical tachyons will appear in sectors with
  $H_j= -c_{\rm left}/24$ and $\overline{H}_\jbar=0$ (or vice versa).
  These unphysical tachyons then imply exponential growth (\ref{Ceffdef})
  with $C_{\rm eff}=C_{\rm left}\equiv 4\pi\sqrt{c_{\rm left}/24}$
  (or $C_{\rm eff}=C_{\rm right}\equiv 4\pi\sqrt{c_{\rm right}/24}$) only.
  Note that this is, in general, the greatest exponential growth
  that can arise from the leading terms (\ref{leadingterm}).
\item  Similarly, there may be other strong subdominant exponentials
  which arise from terms in the general expansion (\ref{finalexpansion})
  with either $\alpha>1$ or $\alphabar>1$.  The greatest of these
  is {\it a priori}\/ the case when $(\alphabar,\alpha)=(1,2)$ or
  $(2,1)$.  Since $M_{10}^{-1}$ and $M_{21}^{-1}$ are different
  modular transformations, a simplification analogous to that in
  (\ref{Poneone}) is not possible, and therefore in principle
  {\it any} combination of functions
  $f_j^{(1)}(n) \overline{f}_\jbar^{(2)}(n)$ or
  $f_j^{(2)}(n) \overline{f}_\jbar^{(1)}(n)$ is possible.
  Whether such terms actually appear in general depends on the details of the
  types of unphysical tachyons that a given partition function contains,
  and therefore,
  unlike the first class of subleading exponentials discussed above,
  such terms are not necessarily universal.
  Note that these terms may nevertheless yield stronger exponential
  growth than those in the first class discussed above.  In particular,
  general terms such as
  $f_j^{(\alpha)}(n) \overline{f}_\jbar^{(\alphabar)}(n)$
  with $H_j<0$ and $\Hbar_\jbar<0$ yield values
\beq
    C_{\rm eff}~=~4\pi\left(
     {\sqrt{|H_j|}\over \alpha} +
       {\sqrt{|\Hbar_\jbar|}\over \alphabar}\right)~,
\label{possibleterm}
\eeq
  and, depending on the particular values of
  $H_j$ and $\Hbar_\jbar$ involved, such values of $C_{\rm eff}$ may exceed
  the values $C_{\rm left}$ or $C_{\rm right}$ which arise from
  from the first class of exponentials.
  Of course, terms leading to (\ref{possibleterm}) are possible only if
  ${\cal P}_{\jbar j}^{(\alphabar,\alpha)}$
  in (\ref{Pdef}) is non-zero for appropriate values of $(\jbar,j)$.
  We shall discuss these subleading terms in more detail in Sect.~4.3.
\item  Finally, there is another potential source of exponential growth,
  this one arising from the asymptotic expansion {\it error}\/ terms.
  Recall that in the one-variable case, no exponential growth could
  ever arise from the error terms, for these terms are at most polynomial.
  In the present two-variable case, however, there is a separate error term
  from each of the left and right-moving asymptotic expansions,
  and thus the error term from the left-moving expansion
  can multiply an exponential term from the right-moving expansion,
  and vice versa.  The maximum exponential growth this can produce,
  however, is $C=C_{\rm left}$ or $C_{\rm right}$ respectively,
  and this is the same growth which arises from the subdominant terms
  in the first class discussed above.
\end{itemize}

Thus, we see that the largest value of $C_{\rm eff}$ depends crucially
on the details of the underlying non-supersymmetric theory,
and is determined by the structure and energy distribution of its
unphysical tachyons.
In fact, we shall discuss the precise value of $C_{\rm eff}$
at several points in this and later sections.
The important point, however, is that none of these subdominant terms
can ever reproduce the leading exponential growth that is experienced
by each $a_{nn}^{(\ibar i)}$ individually.
Thus the existence of these subdominant terms can never alter
our main result (\ref{lessthan}).

As discussed in the Introduction, the result (\ref{lessthan}) for
$\langle a_{nn}\rangle$ has profound implications for the values
of the actual degeneracies $a_{nn}^{(\ibar i)}$.
We have already seen that each $a_{nn}^{(\ibar i)}$ necessarily experiences
stronger exponential growth than does the sector-averaged quantity
$\langle a_{nn}\rangle$, and indeed we have
\beq
   \lim_{n\to\infty}~{\langle a_{nn}\rangle \over a_{nn}^{(\ibar i)}}~=~0~.
\label{ratio}
\eeq
This of course implies that the dominant exponentials in the different
$a_{nn}^{(\ibar i)}$ expansions must exactly cancel in the sector-averaging
summation, with some sectors having positive dominant terms and others
negative (respectively, with some bosonic and others fermionic).
However, this does not necessarily imply any cancellation in the
net number of actual {\it states}\/ in the theory.  Recall that each sector
$(\ibar,i)$ in principle has a different alignment, and contributes states
to the spectrum at only those energies $n$ satisfying $n=H_i=\Hbar_\ibar$
(modulo 1).  Thus two sectors --- for example, one
with states at $n=0$ (modulo 1) and the other with
$n=1/2$ (modulo 1) --- can have equal and opposite dominant terms in
their asymptotic expansions without having any cancellation
between their actual states at any given mass level.  Indeed, even the
 {\it number}\/ of states at the {\it different}\/ mass levels
will not be equal:  if $\Phi(n)$ represents this leading asymptotic term,
then while the degeneracies of the first sector will be given
asymptotically by the values $\lbrace \Phi(\ell),~\ell\in{\IZ}\rbrace$,
those of the second sector will be given asymptotically by the different
values $\lbrace -\Phi(\ell+\half),~\ell\in{\IZ}\rbrace$.
It is only by considering the {\it functional forms}\/ $\Phi$ that
cancellations are apparent.

Our result thus implies that as the invariant energy $n$
is increased, the net number of states $a_{nn}$
at energy $n$ exhibits an oscillation:
first this number will be positive (implying more bosonic states than
fermionic states), then negative (implying more fermionic states
than bosonic states at the next mass level), and then positive again.
Indeed, the ``wavelength'' of this repeating oscillation is $\Delta n=1$,
corresponding to energy difference between adjacent states in the same
sector.  While the amplitude of this oscillation grows exponentially,
the oscillation asymptotically becomes {\it symmetric}\/ between
positive and negative values.
In general, of course, there are more than two sectors in the theory, and
while some groups of sectors will be aligned relative to each other, others
will
be misaligned.  Thus there may exist a potentially large set of values
of $n$ (modulo 1) at which states will be found, and
the pattern of oscillation which we have described
may be quite complicated within each ``wavelength'' $\Delta n$.
We will see explicit examples of this in Sect.~4.
What is guaranteed by our result, however, is that the leading
terms must cancel as we sum over sectors,
and that therefore the behavior of the net number of states at energy $n$
must execute this increasingly symmetric
oscillatory behavior as $n$ is increased.

In physical terms, this result amounts to a strong constraint on
the degree to which supersymmetry may be broken in a modular-invariant
theory without introducing damaging physical tachyons.  Quite simply,
the supersymmetry can be at most ``misaligned'':   introducing
any surplus of {\it bosonic}\/ states at any mass level in the theory
necessarily implies the simultaneous creation a larger surplus
of {\it fermionic}\/ states at a higher mass level,
which in turn implies an even larger surplus of {\it bosonic}\/ states at
an even higher mass level, {\it etc}.
Since modular invariance and the absence of physical tachyons
are precisely the conditions for {\it finite}\/ string amplitudes,
this scenario now provides a glimpse of how this finiteness arises
level-by-level in the actual spectrum of states that contribute
to single- and multi-loop processes.
Indeed, we shall discuss the relationship between
this ``misaligned supersymmetry'' and the value of one
such one-loop amplitude, namely the one-loop
cosmological constant, in Sect.~5.

Finally, let us discuss the role played in the two-variable theorem by the
modular weight $k$.  As we recall from Sect.~2, the most important
part of the asymptotic expansions which depend on the modular weight $k$ are
the {\it error}\/ terms:  for $k<0$ they asymptotically vanish like a
(negative) power of $n$, and for $k>0$ they grow at most polynomially
with $n$.  In the one-variable case,
this was sufficient to imply that any tachyon-free modular-invariant function
$f(q)$ would have to vanish if $k<0$, and contain at most polynomially growing
coefficients if $k>0$.
In the two-variable case, however, no such implication follows,
primarily because the error terms from each chiral half
of the theory multiply the exponentially growing terms
from the other half of the theory.  This has the benefit of
effectively rendering our result independent of the value of $k$.
Thus, our theorem concerning the reduction of the
effective growth rate from $C_{\rm tot}$ to $C_{\rm eff}$
is unaffected by the actual value of the modular weight $k$,
and our two-variable theorem (as well as its primary consequence,
the misaligned supersymmetry) remain valid {\it regardless}\/ of the
spacetime dimension $D$.

This theorem is therefore quite powerful from a purely physical
standpoint, implying a misaligned supersymmetry for {\it all}\/
spacetime dimensions $D$.  From a mathematical standpoint, however,
we see that
its robustness essentially results from the weakness of its assertion.
Indeed, our theorem asserts merely that $C_{\rm eff}<C_{\rm tot}$,
yet the true analogue of the one-variable result would instead be
the more powerful claim that
\beq
          C_{\rm eff}~{\buildrel {?}\over =}~ 0~.
\label{conjecturefirst}
\eeq
Note that this would imply that {\it all}\/ exponential
growth must vanish in the sum (\ref{finalexpansion}), with
the contributions from subleading terms cancelling the contributions
from unphysical tachyons.
Moreover, in analogy with the one-variable result,
we would assert that in the limit $n\to\infty$,
the behavior of the asymptotic expansions $\langle a_{nn}\rangle$
must be
\beqn
       \langle a_{nn}\rangle ~{\buildrel {?}\over \sim}~
         \cases{ 0 & for $k <1$ \cr
         n^{k-1}   & for $k\geq 1$~,}
\label{conjecturesecond}
\eeqn
since the strongest growth that can arise from any of the functions
$f_j^{(\alpha)}(n)$ with $H_j \geq 0$ is $\sim n^{k-1}$ for $k>1/2$.
For $k\leq 1/2$, by contrast, all such functions $f_j^{(\alpha)}$ vanish
as $n\to\infty$.

Surprisingly, it turns out
that there is independent evidence for {\it both}\/ of these conjectures
(\ref{conjecturefirst}) and (\ref{conjecturesecond}),
evidence which comes from an analysis of the finiteness properties of
certain one-loop amplitudes in string theories which lack physical tachyons.
This evidence will be discussed in Sect.~5.
Therefore, while we have proven that $C_{\rm eff}<C_{\rm tot}$,
we shall in fact conjecture that (\ref{conjecturefirst})
and (\ref{conjecturesecond}) hold as well.

It is nevertheless evident from the above discussion we can never prove
these conjectures by following the procedure presented this section,
a procedure which consists of multiplying together the separate chiral
asymptotic expansions for $a_n^{(i)}$ and $\abar_n^{(\ibar)}$ in order to
obtain an expansion for their product $a_{nn}^{(\ibar i)}$.
Indeed, we shall see in Sect.~4.3 that even these separate
chiral asymptotic expansions have significant shortcomings when
applied to situations (like those encountered for $\langle a_{nn}\rangle$)
in which the energy $n$ is treated as a continuous variable.
Thus, what is necessary is a fundamentally new type of
asymptotic expansion, one which is calculated directly for
$\langle a_{nn}\rangle$.
We shall briefly outline how this might be derived in Sect.~6.


\vfill\eject
\setcounter{footnote}{0}
\section{Two Examples}


In this section, we provide two examples to illustrate the
general results proven in Sect.~3.
We will also illustrate certain techniques which often enable
great simplifications when analyzing specific partition functions
and their misaligned supersymmetry properties.
While the first example is chosen for its simplicity, the
second is more typical of the sorts of partition functions
which arise for heterotic string theories compactified to four dimensions.
This second example will also be relevant to Sect.~5, where
we relate ``misaligned supersymmetry'' to the one-loop cosmological constant.

\subsection{First Example}

As our first simple example, let us consider the modular-invariant,
tachyon-free, purely real partition function:
\beq
    Z ~\equiv ~{\textstyle{1\over 128}}\,
        ({\rm Im}\,\tau)^{-1/2}\,
     |\eta|^{-12} \sum_{{i,j,k=2}\atop{i\not= j\not=k}}^4
       \,(-1)^{i+1} ~|\ti|^2 ~\left\lbrack
  \,\overline{\tj}^2\tk^2 \,+\,
    (-1)^{i+1}\,\overline{\tk}^2\tj^2\right\rbrack^2~.
\label{exampleZ}
\eeq
Like the typical partition functions of string theory,
this function has negative modular weight ($k=-1/2$),
and contains both unphysical tachyons and massless states.
Here the $\vartheta$-functions are the classical Jacobi
theta functions, which are related to the Ising-model characters
$\chi_0$, $\chi_{1/2}$, and $\chi_{1/16}$ via
\beqn
        \chi_0~&=&~ \half\,\left( \sqrt{\vartheta_3\over\eta} +
        \sqrt{\vartheta_4\over\eta}\,  \right)\nonumber\\
        \chi_{1/2}~&=&~ \half\,\left( \sqrt{\vartheta_3\over\eta} -
        \sqrt{\vartheta_4\over\eta}\, \right)\nonumber\\
        \chi_{1/16}~&=&~  \sqrt{\vartheta_2\over 2\eta}~.
\label{thetaIsing}
\eeqn
Recall that the $c=1/2$ Ising model is the conformal field theory
corresponding to a single real (Majorana) worldsheet fermion.

Our first step is to cast this partition function $Z$
in the form (\ref{Zexplicit}), so that we can determine the
relevant set of characters and their mixings under modular
transformations.
Note that our partition function (\ref{exampleZ})
consists of a sum of terms, each of which contains five
$\vartheta$-factors and five $\overline{\vartheta}$-factors.
Using the relations (\ref{thetaIsing}) and recalling that $\eta^{-1}$
is the character of an uncompactified $c=1$ boson, this would suggest
that the relevant characters $\chi_i$ and $\chibar_i$ are
those of a $c=6$ theory formed as a tensor product of
one uncompactified boson and ten (chiral) Ising models
(or ten Majorana fermions).
Since each Ising model factor in this tensor-product
has three separate sectors, there are in principle
$3^{10}$ separate sectors in this $c=6$ theory, and therefore
$3^{10}$ individual characters $\chi_i$ and $\chibar_i$.
Even if we do not distinguish between the {\it ordering}\/ of the factors,
we still face the possibility of working with
a very large system of characters.  Expressed as an expansion
in terms of these characters, our partition function (\ref{exampleZ})
would therefore have many terms whose properties would not be immediately
transparent.  This is unfortunate, especially since (\ref{exampleZ})
was constructed for its relative simplicity, and does not approach the
complexity to be found in more realistic situations.

It is nevertheless possible to bypass these difficulties by
recognizing that the only properties of the ``characters'' $\chi_i$
and $\chibar_i$ demanded in the derivations in Sects.~2 and 3 is
that they satisfy (\ref{Mtrans}) and (\ref{chiform}):
they must have $q$-expansions with non-negative coefficients,
they must transform covariantly under modular transformations,
and in particular they must transform as eigenfunctions
under $T:\tau\to \tau+1$.
Indeed, it is not necessary that they be true conformal-field-theoretic
characters at all, and in particular {\it linear combinations}\/
of such true characters can also serve as ``characters'' for the purposes
of our analyses, provided that they meet the above conditions.

Thus, by appropriately choosing linear combinations of the many characters
of the above $c=6$ conformal field theory, we will
be able to find a relatively small number of ``pseudo-characters''
which close into each other under modular transformations
and which thereby serve as a reduced set of ``characters'' with which to work.
Such a decomposition into pseudo-characters is not unique,
and in principle there exist many different sets of such pseudo-characters
in terms of which a given partition functions may be expressed.
The particular choice of pseudo-characters will not affect our final
results, of course, and merely amounts to reorganizing the calculation
in different ways.
There are some minor consequences of working with this reduced set of
pseudo-characters, however, and we shall point them out as they arise.

Thus, our first step is to find a reduced set of pseudo-characters which
meet the above conditions, and in terms of which
our partition function (\ref{exampleZ}) manifestly satisfies
the condition (\ref{choicetwo}).
Since our partition function (\ref{exampleZ}) is devoid of physical
tachyons, such a set of pseudo-characters must always exist, and indeed
it turns out that the following reduced set of only {\it nine}\/
chiral pseudo-characters suffices:
\beqn
    A~&\equiv&~ {\textstyle{1\over 8}}\,\,\eta^{-6}~
         \ttwo^2\,\tthree\tfour~(\tthree+\tfour)\nonumber\\
    B_1~&\equiv&~ {\textstyle{1\over 32}}\,\,\eta^{-6}~
          \ttwo^4~(\tthree+\tfour)\nonumber\\
    B_2~&\equiv&~ {\textstyle{1\over 12}}\,\,\eta^{-6}~
          \tthree\tfour~(\tthree^3-\tfour^3)\nonumber\\
    C~&\equiv&~ {\textstyle{1\over 32}}\,\,\eta^{-6}~
            \ttwo~(\tthree^4-\tfour^4)\nonumber\\
    D~&\equiv&~ {\textstyle{1\over 16}}\,\,\eta^{-6}~
          \ttwo^2\, \tthree\tfour~(\tthree-\tfour)\nonumber\\
    E_1~&\equiv&~ {\textstyle{1\over 64}}\,\,\eta^{-6}~
          \ttwo^4~(\tthree-\tfour)\nonumber\\
    E_2~&\equiv&~ {\textstyle{1\over 2}}\,\,\eta^{-6}~
          \tthree\tfour~(\tthree^3+\tfour^3)\nonumber\\
    F_1~&\equiv&~ {\textstyle{1\over 128}}\,\,\eta^{-6}~
          \ttwo~(\tthree^2-\tfour^2)^2\nonumber\\
    F_2~&\equiv&~ {\textstyle{1\over 8}}\,\,\eta^{-6}~
          \ttwo~(\tthree^2+\tfour^2)^2~.
\label{examplechars}
\eeqn
We have defined these pseudo-characters in such a way that they are each
normalized, with their first non-vanishing $q$-expansion coefficients
equal to one.  Note that the names of these characters
indicate their respective vacuum energies $H_i$, with
the letters $A$ through $F$ respectively signifying vacuum energies
$H_i=0,1/4,3/8,1/2,3/4,$ and $7/8$ (modulo 1).
Indeed, these characters have the following explicit $q$-expansions:
\beqn
A ~&=& ~ q^0 (
1 + 4 q + 14 {q^2} + 40 {q^3} + 100 {q^4} + 232 {q^5} + 480 {q^6} +...)
          \nonumber\\
B_1 ~&=& ~ q^{1/4} (
1 + 10 q + 59 {q^2} + 270 {q^3} + 1044 {q^4} + 3572 {q^5} + 11111 {q^6}
        +...)
          \nonumber\\
3B_2 ~&=& ~ q^{1/4} (
3 + 10 q + 41 {q^2} + 94 {q^3} + 260 {q^4} + 548 {q^5} + 1173 {q^6} +...)
          \nonumber\\
C ~&=& ~ q^{3/8} (
1 + 11 q + 67 {q^2} + 308 {q^3} + 1190 {q^4} + 4059 {q^5} +
     12574 {q^6} +...)
          \nonumber\\
D ~&=& ~ q^{1/2} (
1 + 4 q + 12 {q^2} + 32 {q^3} + 77 {q^4} + 172 {q^5} + 340 {q^6} +...)
          \nonumber\\
E_1 ~&=& ~ q^{3/4} (
1 + 10 q + 57 {q^2} + 250 {q^3} + 931 {q^4} + 3082 {q^5} + 9308 {q^6} +...)
          \nonumber\\
E_2 ~&=& ~ q^{-1/4} (
1 + 14 q + 37 {q^2} + 134 {q^3} + 305 {q^4} + 786 {q^5} + 1594 {q^6} +...)
          \nonumber\\
F_1 ~&=& ~ q^{7/8} (
1 + 7 q + 37 {q^2} + 154 {q^3} + 557 {q^4} + 1806 {q^5} + 5367 {q^6} +...)
          \nonumber\\
F_2 ~&=& ~ q^{-1/8} (
1 + 15 q + 113 {q^2} + 590 {q^3} + 2467 {q^4} + 8908 {q^5} +
    28877 {q^6} +...)~,\nonumber\\
\label{qexpansions}
\eeqn
from which we see that only the pseudo-characters $E_2$ and $F_2$
are tachyonic.  Indeed,
$E_2$ serves as the effective vacuum character in this reduced system,
with vacuum energy $H_i= -1/4= -c/24$.

Under the $S$ modular transformation, these pseudo-characters have
the following mixing matrix:
\beq
 S_{ij}~=~ {e^{i\pi/4}\over 4}
\pmatrix{ 2 & 0 & 0 & 0 & -4 & 0 & 0 & -16 & 1 \cr 0 & 0 &
 -{\textstyle{3\over 4}} & -2 & 0
   & 0 & \textstyle{1\over 8} & 4 & \textstyle{1\over 4} \cr 0 &
   -\textstyle{{16}\over 3} & 0 &
  \textstyle{{16}\over 3} & 0 & -\textstyle{{32}\over 3} & 0 &
\textstyle{{32}\over 3} & \textstyle{2\over 3} \cr 0 &
  -2 & \textstyle{3\over 4} & 0 & 0 & 4 & \textstyle{1\over 8} & 0 & 0 \cr -1 &
0 & 0 & 0 & 2 & 0
   & 0 & -8 & \textstyle{1\over 2} \cr 0 & 0 & -\textstyle{3\over 8} & 1 & 0 &
0 & \textstyle{1\over {16}}
   & -2 & -\textstyle{1\over 8} \cr 0 & 32 & 0 & 32 & 0 & 64 & 0 & 64 & 4 \cr
  -\textstyle{1\over 4} & \textstyle{1\over 2} & \textstyle{3\over {16}} & 0 &
-\textstyle{1\over 2} & -1 &
  \textstyle{1\over {32}} & 0 & 0 \cr 4 & 8 & 3 & 0 & 8 & -16 &
\textstyle{1\over 2} & 0 & 0 \cr
   }~.
\label{Smatrix}
\eeq
Thus, since (\ref{qexpansions}) indicates that each pseudo-character
is an eigenfunction of $T$, we see that this set of pseudo-characters
is closed under all modular transformations.

Written in terms of these nine pseudo-characters,
our modular-invariant partition function (\ref{exampleZ}) now
takes the simple form
\beqn
   Z~&=&~ ({\rm Im}\,\tau)^{-1/2} \,\biggl\lbrace
    |A|^2 ~-~ 3\,\left( B_1^\ast\,B_2 + B_2^\ast\,B_1\right)
      ~+~ 8 \,|C|^2  \nonumber\\
    &&~~~~+~  4\,|D|^2~+~
   \left( E_1^\ast \,E_2 + E_2^\ast\,E_1\right) ~-~ 4\,
     \left( F_1^\ast\,F_2 + F_2^\ast\,F_1\right) \biggr\rbrace~.
\label{examplerewritten}
\eeqn
Note that since only the pseudo-characters $E_2$ and $F_2$ are tachyonic
with $H_i<0$, the condition (\ref{choicetwo}) is indeed satisfied.
Indeed, from (\ref{examplerewritten}) we see that this partition
function $Z$ contains only unphysical tachyons, and that these
come in two distinct sets:  those with $m\geq 3/4$ and $n=-1/4$
(and vice versa) which arise from the $E$ terms, and those
with $m\geq 7/8$ and $n= -1/8$ (and vice versa) which arise from the
$F$ terms.

There are, however, two obvious features which indicate
that these pseudo-characters are not true characters in and of themselves.
The first appears, for example, in the $q$-expansion of $B_2$
in (\ref{qexpansions}):  not all of its coefficients are integers
(or equivalently, trivially rescaling $B_2\to 3B_2$, we find that
not all of its coefficients are divisible by its first non-zero coefficient).
For a true character, this divisibility property is essential, since the
first non-zero coefficient is the multiplicity of the {\it vacuum state}\/
in that sector of the conformal field theory, and all higher states
in that sector must share that multiplicity.  The breaking of this
property in our case, however, simply reflects the fact that $B_2$
represents the added contributions of many such characters
whose highest weights may indeed differ by integers.
Thus this divisibility condition, valid for each character individually,
need no longer hold for the sum.  This does not affect the validity
of the asymptotic expansions, however.

The second feature which indicates that our pseudo-characters are
not true characters is the appearance of certain vanishing
elements in the $S_{ij}$-matrix (\ref{Smatrix}).  As we have stated
in previous sections, one must have $S_{ij}\not =0$ for all $i$
where $j$ represents the vacuum sector;  such a condition is necessary
in order to obtain meaningful fusion rules in which any sector
fused with the identity (vacuum) sector reproduces that sector \cite{Verlinde}.
For our pseudo-characters, the effective vacuum sector corresponds
to $E_2$, yet we see that several of the corresponding matrix elements
in (\ref{Smatrix}) vanish.  This again
represents the fact that we have taken linear combinations when forming our
pseudo-characters, and causes no fundamental problem.
The effects of these vanishing matrix elements will be discussed below.

Given our partition function decomposed into products of pseudo-characters
as in (\ref{examplerewritten}),
it is now straightforward to examine the behavior of the
state degeneracies $a_{nn}^{(\ibar i)}$ in each of its six
sectors ($A$ through $F$).  Let us first focus on the
expected asymptotic behavior of each
of the nine chiral pseudo-characters individually, and determine the values
of the inverse Hagedorn temperature $C$ that they each
separately exhibit in (\ref{qexpansions}).
Since $E_2$ and $F_2$ are the only tachyonic
pseudo-characters, with $H_{E_2}= -1/4$ and $H_{F_2}=-1/8$,
we see that the dominant growth in the $q$-expansion coefficients
of each character $\chi_i$ depends on the elements
of the matrices $Q_{ij}^{(\alpha=1)}$ which couple
the character $\chi_i$ to the tachyonic sectors $E_2$ and $F_2$.
The strongest growth will come from the coupling to $E_2$, since
$E_2$ serves as the identity (vacuum) pseudo-sector with $H= -c/24= -1/4$;
such growth will be exponential with rate $C=4\pi \sqrt{1/4}=2\pi$.
Couplings to $F_2$, by contrast, produce exponential growth
with $C=4\pi \sqrt{1/8}= \sqrt{2}\pi$.  From (\ref{Qone}),
we see that $Q_{ij}^{(1)}\propto S_{ij}$,
and thus, denoting $S(\chi_i,\chi_j)\equiv S_{ij}$,
we find that this growth for any character $\chi_i$ is determined
by the values of the elements $S(\chi_i,E_2)$ and $S(\chi_i,F_2)$.
Specifically, the $q$-expansion coefficients
of a given $\chi_i$ will experience exponential growth with
$C=2\pi$ if $S(\chi_i, E_2)\not =0$,
and $C=\sqrt{2}\pi$ only if $S(\chi_i,E_2)=0$ and $S(\chi_i,F_2)\not =0$.
We thus immediately find that
\beq
    \lbrace \,B_1,\, C,\,E_1,\, F_1,\,F_2 \,\rbrace
         ~~~~~~~~~~ \Longleftrightarrow ~~~~~~~~~~
        C ~=~ 2\pi~,
\label{strongchars}
\eeq
whereas
\beq
    \lbrace \,A,\,B_2,\, D,\,E_2\,\rbrace
         ~~~~~~~~~~ \Longleftrightarrow ~~~~~~~~~~
        C ~=~ \sqrt{2}\,\pi~.
\label{weakchars}
\eeq

This division into ``strongly'' and ``weakly'' growing
pseudo-characters is actually already evident
in their $q$-expansions (\ref{qexpansions}).
Note that this division, however, is ultimately a consequence
of the fact that we are dealing with {\it pseudo}\/-characters, and not
true conformal field theory characters.
If these had been true characters, the $S_{ij}$-matrix elements
$S(\chi_i,E_2)$ would all have been non-vanishing, and all characters would
consequently experience the {\it same}\/ maximum asymptotic growth.
Physically, this implies that all sectors of a given conformal field theory
necessarily have the same Hagedorn temperature,
and it is only because we have here taken linear combinations
of these characters in forming our pseudo-characters and pseudo-sectors
that this dominant exponential has occasionally cancelled.
Thus, organizing our calculation in terms of pseudo-characters rather
than true characters has already yielded an early cancellation
of the sort that we are investigating.

Combining these chiral pseudo-characters together to form the
full left/right partition function (\ref{examplerewritten}),
we can now easily predict the rates of growth of the physical degeneracies
$a_{nn}^{(\ibar,i)}$ for the six different groupings of sectors [{\it i.e.},
for the $A$-terms contributing states with $n=0$ (modulo 1),
for the combined $B$-terms contributing states with $n=1/4$ (modulo 1),
for the $C$-terms with $n=3/8$ (modulo 1), and so forth].
Indeed, using (\ref{strongchars}) and (\ref{weakchars}),
we can separate these six groupings of
sectors according to the rates of exponential growth for their
separate degeneracies $a_{nn}^{(A)}$ through $a_{nn}^{(F)}$:
\beqn
     \lbrace  \,a_{nn}^{(C)},\, a_{nn}^{(F)}\,\rbrace
         ~~~~~~&\Longleftrightarrow&~~~~~~
       C_{\rm tot}~=~ 4\pi \nonumber\\
     \lbrace  \,a_{nn}^{(B)},\, a_{nn}^{(E)}\,\rbrace
           ~~~~~~&\Longleftrightarrow&~~~~~~
       C_{\rm tot}~=~ (2+\sqrt{2})\pi~\approx~ 3.41\,\pi \nonumber\\
     \lbrace  \,a_{nn}^{(A)},\, a_{nn}^{(D)}\,\rbrace
          ~~~~~~&\Longleftrightarrow&~~~~~~
       C_{\rm tot}~=~ 2\sqrt{2}\,\pi~\approx~ 2.83\,\pi~.
\label{shellgrowths}
\eeqn

These results are also easy to verify in terms of explicit $q$-
and $\qbar$-expansions.
In Figs.~2 and 3 we have plotted the net physical-state degeneracies $a_{nn}$
in this example as functions both of energy $n$ and spacetime mass $\sqrt{n}$,
where $Z=({\rm Im}\,\tau)^{-1/2}\sum_{mn} a_{mn}\qbar^m q^n$.
Those values of $a_{nn}$ with $n\in {\IZ}$ of course arise from the $A$
sector, those with $n\in {\IZ}+1/4$ from the $B$ sectors, {\it etc}.
As in Fig.~1, we have plotted
$\pm \log(|a_{nn}|)$ where the sign chosen is the sign of $a_{nn}$ itself;
note that base-10 logarithms are used for Fig.~2, while natural
logarithms are used for Fig.~3.  From these figures we see
that this partition function indeed exhibits the expected oscillations between
bosonic and fermionic surpluses as the energy $n$ is increased, with
wavelength $\Delta n=1$ (as is evident from Fig.~2).
Note, however, that the pattern of oscillation is now far more complex
than that for the toy example of Fig.~1;  this is due to the presence
of {\it six}\/ sectors $A$ through $F$ in this example,
as opposed to the mere two of the toy model.
Plotting these degeneracies versus $\sqrt{n}$ as in Fig.~3, we see that
the asymptotic exponential behavior for the separate sector degeneracies
$a_{nn}^{(A,...,F)}$ begins quite early, and indeed straight-line fits
on this logarithmic plot are accurate down to remarkably small values
of $n$.  From the different slopes of these lines, we can easily verify
(\ref{shellgrowths});  indeed, the fastest rates of growth are experienced
by the $C$- and $F$-sectors, and the slowest by the $A$- and $D$-sectors.
The signs of these individual sector contributions $a_{nn}^{(A,...,F)}$
are of course given by the signs in the partition
function (\ref{examplerewritten}).
Note that since the $A$- and $D$-sectors have both the same rate of
growth {\it and}\/ the same sign, they are fit remarkably well by
the {\it same}\/ asymptotic function.

Once again, we stress that each $a_{nn}^{(\ibar i)}$ must grow exponentially
with rate $C_{\rm tot}=4\pi$ when $\ibar$ and $i$ refer to true conformal
field theory characters.  Thus, given the results (\ref{shellgrowths}),
we see that our organization in terms of pseudo-characters
has already enabled certain cancellations to become evident.
In particular, the cancellations in the $A$, $B$, $D$, and $E$ sectors
can be viewed as a sort of ``aligned supersymmetry'', since they
represent the cancellations between different conformal field theory sectors
whose highest weights and vacuum energies {\it are}\/ aligned modulo 1.
Indeed, this is the alignment which
enabled us to take their linear combinations and thereby produce
pseudo-characters which were also eigenfunctions of the $T$
modular transformation.

Our goal, of course, is to verify analytically that
in fact {\it all}\/ vestiges of the dominant exponential
$C_{\rm tot}=4\pi$ behavior are cancelled in the full partition
function --- in this case by virtue of a ``misaligned supersymmetry''
between the misaligned $C$ and $F$ sectors.
While this certainly appears likely given the logarithmic plots
in Figs.~2 and 3, we must in fact verify that {\it all}\/
terms which have this exponential growth
are precisely cancelled when their appropriate
prefactors are taken into account.
This is not difficult, however.  From the
results (\ref{shellgrowths}), we see that such terms can only arise
in the $C$ and $F$ sectors.  Using the explicit forms of the asymptotic
expansions, we see that the relevant exponentially-growing
terms arising from the $C$-sector are given by
\beqn
     a_{nn}^{(C)} ~&\sim&~ N_{C^\ast C}\, \left| 2\pi
       \,Q_{C,E_2}^{(1)}\, f_{E_2}(n)\right|^2\nonumber\\
     &=&~ 8\, \left| 2\pi \,
        S(C,E_2)\,(4n)^{-3/4}\,I_{-3/2}(2\pi\sqrt{n})\right|^2~,
\label{annC}
\eeqn
while those from the $F$-sector are given by
\beqn
     a_{nn}^{(F)} ~&\sim&~ N_{F_1^\ast F_2}
       \left[2\pi \,Q_{F_2,E_2}^{(1)}\, f_{E_2}(n)\right]
       \left[2\pi \,Q_{F_1,E_2}^{(1)}\, f_{E_2}(n)\right]^\ast \nonumber\\
      && ~~~+~
     N_{F_2^\ast F_1}
       \left[2\pi \,Q_{F_1,E_2}^{(1)}\, f_{E_2}(n)\right]
       \left[2\pi \,Q_{F_2,E_2}^{(1)}\, f_{E_2}(n)\right]^\ast \nonumber\\
     &=&~ -8\,
        {\rm Re}\,\left\lbrace S(F_1,E_2) [S(F_2,E_2)]^\ast\right\rbrace \,
       \left| 2\pi \, (4n)^{-3/4}\,I_{-3/2}(2\pi\sqrt{n})\right|^2~.
\label{annF}
\eeqn
These expressions will be equal and opposite provided
\beq
        \left|S(C,E_2)\right|^2~=~
        {\rm Re}\,\left\lbrace S(F_1,E_2) [S(F_2,E_2)]^\ast\right\rbrace~,
\label{equalelements}
\eeq
and consulting the $S_{ij}$-matrix in (\ref{Smatrix})
we see that (\ref{equalelements}) is indeed satisfied.
Thus we have verified the cancellation of all terms with $C=4\pi$
exponential growth, and as a consequence
the sector-averaged number of states $\langle a_{nn}\rangle$
for this example must indeed grow more slowly than either of the separate
$C$- or $F$-sectors (or indeed than any other
true {\it conformal field theory}\/ sector of the theory).

The above cancellation is of course predicted by our general theorem,
since $C_{\rm tot}=4\pi$ for the present example,
and $C_{\rm eff}$ must be less than $C_{\rm tot}$.
However, our theorem can actually be used to predict much more.
We have just seen that the $C$- and $F$-sectors have leading
$C=4\pi$ terms which cancel.
However, from (\ref{shellgrowths}) we see that the next-largest growth
comes from the $B$- and $E$-sectors, with $C=(2+\sqrt{2})\pi$, and
these sectors {\it also}\/ appear to cancel in Fig.~2.
We might then wonder whether all exponential
growth also cancels for $C=(2+\sqrt{2})\pi$.
Such a cancellation would in principle be far more subtle than the
leading cancellation, for we have only demonstrated that
the $C$- and $F$-sector cancellation removes the  $C=4\pi$ terms;
in particular, residual {\it subleading}\/
$C=(2+\sqrt{2})\pi$ terms might nevertheless
survive from these sectors.   These residual terms would then have to combine
with the {\it leading}\/ $C=(2+\sqrt{2})\pi$ terms from
the $B$- and $E$-sectors in order to produce the next level of cancellation.

However, using our theorem, it is straightforward to demonstrate that this
secondary cancellation also takes place.
Note that the partition function $Z$ in this example
contains unphysical tachyons in the $E$-sectors;
these sectors have $H_i= -1/4,\,\Hbar_\ibar>0$ and vice versa.  From this
and the result (\ref{leadingterm}) we deduce
that the maximum possible growth that is allowed from the {\it leading}\/
term with $\alpha=\alphabar=1$ is $C=2\pi$.
We can now quickly survey the possible rates of growth that
can arise from subleading terms with general values of $\alpha$ and
$\alphabar$;  these values of $C$ are given in (\ref{possibleterm}),
and whether or not such rates of growth actually appear is determined
by the corresponding values of ${\cal P}^{(\alphabar,\alpha)}_{\jbar j}$
in (\ref{Pdef}).  It turns out that we do not even need to explicitly
calculate ${\cal P}^{(\alphabar,\alpha)}_{\jbar j}$.
In the present example, growth can occur only with
values $H_j,\Hbar_\jbar\in\lbrace -1/4, -1/8,0\rbrace$.
(The value $0$ corresponds to all $H,\Hbar>0$ situations and
also to all possible error terms, as discussed in Sect.~3;
note that for $\alpha\not= \alphabar$ we must of course permit
even the possibilities with $H_i=\Hbar_\ibar$.)
We easily find that there exists no combination of values of $\alpha$,
$\alphabar$, $H_i$, and $\Hbar_\ibar$ for which $C$
in (\ref{possibleterm}) exceeds $3\pi$.
Thus our theorem in this case actually precludes all exponential
growth with $C>3\pi$, and
as a consequence any remaining subleading $C=(2+\sqrt{2})\pi$ terms
from the $C$- and $F$-sectors must cancel exactly against the leading
$C=(2+\sqrt{2})\pi$ terms from the $B$- and $D$-sectors.

We have therefore been able to demonstrate that although $C_{\rm tot}=4\pi$
for this example, cancellations bring this down to
$C_{\rm eff}\leq 3\pi$, in accordance with our theorem.
This amounts to a remarkable total cancellation,
reducing the sector-averaged number of states $\langle a_{nn}\rangle$
by more than six orders of magnitude
for energies as low as $n\approx 20$.

\subsection{Second Example}

Our second example is more complicated than the first, but
in many ways more realistic.  In particular, this
partition function \cite{KRD}
resembles those found for actual non-supersymmetric
 {\it heterotic}\/ string theories compactified
to four spacetime dimensions, and contains precisely the
types of unphysical tachyons that such strings necessarily contain.
This function is special, however, in that
it has an exactly vanishing one-loop cosmological constant ---
 {\it i.e.}, even though this function is non-zero,
its integral over the fundamental domain
of the modular group vanishes identically \cite{KRD}.
We will discuss the relation between misaligned supersymmetry
and the cosmological constant in Sect.~5.

This function is as follows \cite{KRD}:
\beqn
    Z ~&\equiv&~{\textstyle{1\over 2}}\,
     ({\rm Im}\, \tau)^{-1}\,\eta^{-24} \,\overline{\eta}^{-12}\,
     \sum_{{i,j,k=2}\atop{i\not= j\not=k}}^4
       |\vartheta_i|^4 \, \biggl\lbrace
     \,{\vartheta_i}^4 {\vartheta_j}^4 {\vartheta_k}^4\,
      \left\lbrack
       \,2\,|\vartheta_j \vartheta_k|^8 -
      {\vartheta_j}^8\overline{\vartheta_k}^8 -
      \overline{\vartheta_j}^8  {\vartheta_k}^8 \right\rbrack \nonumber\\
      && ~~~~~
       ~~~~~~~ +\, {\vartheta_i}^{12} \,\left\lbrack
     \,4\, {\vartheta_i}^8 \overline{\vartheta_j}^4 \overline{\vartheta_k}^4
           + (-1)^i\,13\, |\vartheta_j \vartheta_k|^8 \right\rbrack
        \biggr\rbrace~.
\label{Qfunction}
\eeqn
Given the relations (\ref{thetaIsing}), we immediately see
that this function corresponds to left and right worldsheet theories
which are tensor products of two uncompactifed bosons and 44 or 20
Majorana fermions respectively.
These worldsheet theories thus have total central charges $c_{\rm left}=24$
and $c_{\rm right}=12$, and correspond to those of light-cone
ten-dimensional heterotic strings compactified to four dimensions.
We thus have $C_{\rm tot}=2(2+\sqrt{2})\pi$ for this example
(and indeed for {\it all}\/ heterotic superstrings, regardless of
spacetime dimension).
The modular weight is $k= -1$.
A decomposition into pseudo-characters satisfying all of the conditions
of Sect.~3 and Sect.~4 can be achieved by defining
the nine holomorphic pseudo-characters:
\beqn
       A_1 ~&\equiv&~ \eta^{-24}\,
      \thetatwo^4\thetathree^6\thetafour^6\,
            (\thetathree^6 -  \thetafour^6) \nonumber\\
       A_2 ~&\equiv&~ \eta^{-24}\,
      \thetatwo^{12}\thetathree^4\thetafour^4\,
            (\thetathree^2 -  \thetafour^2) \nonumber\\
       A_3 ~&\equiv&~ \eta^{-24}\, \lbrace
      \,2\,\thetatwo^{8}\thetathree^6\thetafour^6\,
            (\thetathree^2 +  \thetafour^2)\,
      +\,4\, (\thetathree^{22} +  \thetafour^{22}) \nonumber\\
       ~&&\phantom{\equiv~ \eta^{-24}\,}
      -\,13\,(\thetatwo\thetathree\thetafour)^4\,
            (\thetathree^{10} -  \thetafour^{10}) \rbrace\nonumber\\
       B ~&\equiv&~ \eta^{-24}\,
      \thetatwo^{6}\thetathree^4\thetafour^4\,
            (\thetathree^8 -  \thetafour^8) \nonumber\\
       C_1 ~&\equiv&~ \eta^{-24}\, \lbrace
      -2\,\thetatwo^{8}\thetathree^6\thetafour^6\,
            (\thetathree^2 -  \thetafour^2)\,
      +\,4\, (\thetathree^{22} -  \thetafour^{22}) \nonumber\\
       ~&&\phantom{\equiv~ \eta^{-24}\,}
      -\,13\,(\thetatwo\thetathree\thetafour)^4\,
            (\thetathree^{10} +  \thetafour^{10}) \,
      +\,2\,\thetatwo^4\thetathree^6\thetafour^6\,
            (\thetathree^{6} +  \thetafour^{6}) \rbrace\nonumber\\
       C_2 ~&\equiv&~ \eta^{-24}\,
      \thetatwo^{12}\thetathree^4\thetafour^4\,
            (\thetathree^2 +  \thetafour^2) \nonumber\\
       C_3 ~&\equiv&~ \eta^{-24}\,
      \thetatwo^{4}\thetathree^6\thetafour^6\,
            (\thetathree^6 +  \thetafour^6) \nonumber\\
       D_1 ~&\equiv&~ \eta^{-24}\, \lbrace
         2 \,\thetatwo^{6}\thetathree^8\thetafour^8
         \,+\, 4 \,\thetatwo^{22}
         \,+\, 13\,\thetatwo^{14}\thetathree^4\thetafour^4 \nonumber\\
       ~&&\phantom{\equiv~ \eta^{-24}\,}
       \,-\, \thetatwo^{6}\thetathree^4\thetafour^4\,
            (\thetathree^8 +  \thetafour^8) \rbrace\nonumber\\
       D_2 ~&\equiv&~ \eta^{-24}\, \lbrace
         2 \,\thetatwo^{6}\thetathree^8\thetafour^8
         \,+\, 4 \,\thetatwo^{22}
         \,+\, 13\,\thetatwo^{14}\thetathree^4\thetafour^4 \nonumber\\
       ~&&\phantom{\equiv~ \eta^{-24}\,}
       \,+\, \thetatwo^{6}\thetathree^4\thetafour^4\,
            (\thetathree^8 +  \thetafour^8)\rbrace
\label{whateverone}
\eeqn
and the nine anti-holomorphic pseudo-characters:
\beqn
   \overline{A_1}~&\equiv&~ \eta^{-12}\,
        \thetatwo^8\,(\thetathree^2-\thetafour^2)    \nonumber\\
   \overline{A_2}~&\equiv&~ \eta^{-12}\,
       (\thetathree\thetafour)^2\, (\thetathree^6 -  \thetafour^6) \nonumber\\
   \overline{A_3}~&\equiv&~ \eta^{-12}\,
      \thetatwo^4\thetathree^2\thetafour^2\,(\thetathree^2+\thetafour^2)
             \nonumber\\
   \overline{B}~&\equiv&~ \eta^{-12}\,
         \thetatwo^2 \,(\thetathree^8-\thetafour^8)\nonumber\\
   \overline{C_1}~&\equiv&~ \eta^{-12}\,
      \thetatwo^4\thetathree^2\thetafour^2\,(\thetathree^2-\thetafour^2)
             \nonumber\\
   \overline{C_2}~&\equiv&~ \eta^{-12}\,
        \lbrace \thetatwo^8\,(\thetathree^2+\thetafour^2)
    ~-~ 2\, \thetatwo^4\thetathree^2\thetafour^2\,(\thetathree^2-\thetafour^2)
          \rbrace \nonumber\\
   \overline{C_3}~&\equiv&~ \eta^{-12}\,
       (\thetathree\thetafour)^2\, (\thetathree^6 +  \thetafour^6) \nonumber\\
   \overline{D_1}~&\equiv&~ \eta^{-12}\,
         \thetatwo^2 \,(\thetathree^4-\thetafour^4)^2\nonumber\\
   \overline{D_2}~&\equiv&~ \eta^{-12}\,
         \thetatwo^2 \,(\thetathree^4+\thetafour^4)^2~.
\label{whatevertwo}
\eeqn
Unlike the pseudo-characters of the previous example,
we have not chosen these to be normalized.
As discussed at the end of Sect.~2, this will be ultimately reflected
as a rescaling of the elements of the
modular-transformation representation matrices,
and our above normalizations will render these
matrices particularly simple.

Like the pseudo-characters of the previous example,
these pseudo-characters have been organized according to their vacuum
energies $H$ (modulo 1), so that those lettered $A$ through $D$
have vacuum energies $H_i=0, 1/4, 1/2$, and $3/4$ (modulo 1) respectively.
Explicit $q$-expansions for these pseudo-characters are as follows:
\beqn
A_1 ~&=&~ 128\,(
         3 + 52 q + 292 {q^2} + 1440 {q^3} + ...)\nonumber\\
A_2 ~&=&~ 32768\,q\,(
         1 + 20 q + 216 {q^2} + ...)\nonumber\\
A_3 ~&=&~ 8\,q^{-1}\,(
         1 + 36 q + 78720 {q^2} + 6803824 {q^3} +
               230743038 {q^4} + ...)\nonumber\\
 B  ~&=&~ 2048\,q^{1/4}\,(
         1 + 42 q + 633 {q^2} + ...)\nonumber\\
C_1 ~&=&~ 512\,q^{1/2}\,(
         53 + 13464 q + 669874 {q^2} + ...)\nonumber\\
C_2 ~&=&~ 8192\,q^{1/2}\,(
        1 + 24 q + 298 {q^2} + ...)\nonumber\\
C_3 ~&=&~ 32\,q^{-1/2}\,(
         1 + 64 q + 510 {q^2} + 2688 {q^3} + ...)\nonumber\\
D_1 ~&=&~ 65536\,q^{3/4}\,(
         3 + 322 q + 12541 {q^2} + ...)\nonumber\\
D_2 ~&=&~ 256\,q^{-1/4}\,(
         1 + 894 q + 85251 {q^2} + 3243130 {q^3} + ...)\nonumber\\
\overline{A_1} ~&=&~ 2048\,q\,(
         1 + 20 q + 216 {q^2} + ...)\nonumber\\
\overline{A_2} ~&=&~ 8\,(
         3 + 52 q + 292 {q^2} + 1440 {q^3} + ...)\nonumber\\
\overline{A_3} ~&=&~ 32\,(
         1 + 12 q + 76 {q^2} + 352 {q^3} + ...)\nonumber\\
\overline{B  } ~&=&~ 128\,q^{1/4}\,(
         1 + 42 q + 633 {q^2} + 6042 {q^3} + ...)\nonumber\\
\overline{C_1} ~&=&~ 128\,q^{1/2}\,(
         1 + 8 q + 42 {q^2} + 176 {q^3} + ...)\nonumber\\
\overline{C_2} ~&=&~ 256\,q^{1/2}\,(
         1 + 40 q + 554 {q^2} + 4976 {q^3} + ...)\nonumber\\
\overline{C_3} ~&=&~ 2\,q^{-1/2}\,(
         1 + 64 q + 510 {q^2} + 2688 {q^3} + 11267 {q^4} + ...)\nonumber\\
\overline{D_1} ~&=&~ 1024\,q^{3/4}\,(
         1 + 22 q + 255 {q^2} + ...)\nonumber\\
\overline{D_2} ~&=&~ 16\,q^{-1/4}\,(
         1 + 62 q + 1411 {q^2} + 16314 {q^3} + ...)~.
\label{Qcharsexp}
\eeqn
Note that the pseudo-characters which are tachyonic are
$\lbrace A_3, C_3,D_2, \overline{C_3}, \overline{D_2}\rbrace$;
indeed, the pseudo-characters corresponding to the ``vacuum''
sectors for left- and right-moving
systems are $A_3$ and $\overline{C_3}$ respectively.

These two sets of pseudo-characters close separately
under the $S$ modular transformation, with mixing matrices
\beq
   S_{ij}~=~ {i\over 4} \pmatrix{ 0 & -2 & 0 & 2 & 0 & -2 & 0 & -1 & 1 \cr
      -2 & 0 & 0 & 2 & 0 & 0 & 2 & 1 & -1 \cr
      0 & 0 & 2 & 0 & 2 & 0 & -4 & 2 & 2 \cr
      2 & 2 & 0 & 0 & 0 & -2 & 2 & 0 & 0 \cr
      0 & 4 & 2 & 4 & 2 & 4 & -4 & -4 & 0 \cr
      -2 & 0 & 0 & -2 & 0 & 0 & 2 & -1 & 1 \cr
      0 & 2 & 0 & 2 & 0 & 2 & 0 & -1 & 1 \cr
      -2 & 2 & 2 & 0 & -2 & -2 & 2 & 0 & 0 \cr
      2 & -2 & 2 & 0 & -2 & 2 & 6 & 0 & 0 \cr  }
\label{Smatrixholo}
\eeq
and
\beq
     S_{\ibar\jbar}~=~ {i\over 4}
      \pmatrix{ 0 & -2 & 0 & 2 & 0 & 0 & 2 & -1 & -1 \cr
     -2 & 0 & 0 & 2 & -4 & -2 & 0 & 1 & 1 \cr
      0 & 0 & 2 & 0 & -2 & 0 & 0 & -1 & 1 \cr
      2 & 2 & 0 & 0 & -4 & -2 & 2 & 0 & 0 \cr
      0 & 0 & -2 & 0 & 2 & 0 & 0 & -1 & 1 \cr
      0 & -2 & 4 & -2 & -4 & 0 & 2 & 3 & -1 \cr
      2 & 0 & 0 & 2 & 4 & 2 & 0 & 1 & 1 \cr
      -2 & 2 & -4 & 0 & 0 & 2 & 2 & 0 & 0 \cr
      -2 & 2 & 4 & 0 & 8 & 2 & 2 & 0 & 0 \cr  }~.
\label{Smatrixantiholo}
\eeq
Thus, since each pseudo-character is also an eigenfunction of $T$,
each set of pseudo-characters is separately
closed under all modular transformations.
Note that the relative simplicity of the matrices (\ref{Smatrixholo})
and (\ref{Smatrixantiholo}) is due to the chosen normalizations
of the pseudo-characters.

In terms of these pseudo-characters, our modular-invariant
partition function (\ref{Qfunction})
now takes the relatively simple form
\beqn
    Z~&=&~ ({\rm Im}\,\tau)^{-1}\,\biggl\lbrace
        \half\,\left(
        \overline{A_1}^{\,\ast} A_1
        +\overline{A_2}^{\,\ast} A_2
        +\overline{A_3}^{\,\ast} A_3\right) ~+~
         \half\, \overline{B}^{\,\ast} B \nonumber\\
    &&~~~~ -\,\half\,
        \left(\overline{C_1}^{\,\ast} C_1
        +\overline{C_2}^{\,\ast} C_3
        +\overline{C_3}^{\,\ast} C_2\right) ~+~ {\textstyle {1\over 4}}\,
        \left(\overline{D_2}^{\,\ast} D_1
        -\overline{D_1}^{\,\ast} D_2 \right)\biggr\rbrace~.
\label{Qrewritten}
\eeqn
In this form it is easy to see that $Z$ contains unphysical tachyons
but lacks physical tachyons, and indeed the condition
(\ref{choicetwo}) is satisfied.

As in the previous example,
we can now easily determine the rates of exponential growth
for each separate pseudo-character, and then for each of the
four combined sectors ($A$ through $D$) in this partition function.
The results for the separate pseudo-characters are:
\beqn
      \lbrace A_3,\,C_1,\,D_1,\,D_2\rbrace ~&\Longleftrightarrow&~
          C=4\pi \nonumber\\
      \lbrace A_2,\,B,\,C_2,\,\overline{A_1},\,\overline{B},\,
   \overline{C_2},\,\overline{D_1},\,\overline{D_2}\rbrace
    ~&\Longleftrightarrow&~ C=2\sqrt{2}\,\pi \nonumber\\
      \lbrace A_1,\,C_3,\,\overline{A_2},\,\overline{A_3},\,
   \overline{C_1},\,\overline{C_3}\rbrace
      ~&\Longleftrightarrow&~ C=2\pi~,
\label{pseudoC}
\eeqn
and once again this division into separate groups
(which arises from our use of pseudo-characters rather than true characters)
is reflected in the actual $q$-expansions (\ref{Qcharsexp}).  From
(\ref{pseudoC}), then, we can immediately conclude that the combined
$A$-, $B$-, and $C$-sectors of the partition function experience
the following exponential rates of growth:
\beqn
         \lbrace a_{nn}^{(A)},\, a_{nn}^{(C)} \rbrace ~&\Longleftrightarrow&~
           C~=~ 6\pi\nonumber\\
         \lbrace a_{nn}^{(B)}\rbrace ~&\Longleftrightarrow&~
           C~=~ 4\sqrt{2}\,\pi~.
\label{ABCshells}
\eeqn
Indeed, it is the $\overline{A_3}^{\,\ast} A_3$ and
$\overline{C_1}^{\,\ast} C_1$ terms in the partition
function (\ref{Qrewritten}) which dominate in
producing the above asymptotic behavior.

Determining the rate of growth for the $D$-sector degeneracies $a_{nn}^{(D)}$
is a bit more subtle, however, since the two $D$-sector terms
$\overline{D_2}^{\,\ast}D_1$ and $\overline{D_1}^{\,\ast}D_2$
each separately have $C=2(2+\sqrt{2})\pi$ growth but appear in the partition
function (\ref{Qrewritten}) with opposite signs.
Thus, a cancellation within the $D$-sector terms may take place which
entirely removes this leading exponential behavior for $a_{nn}^{(D)}$.
The fastest way to see that this indeed occurs (and to simultaneously
determine the rate of the largest non-cancelling exponential growth)
is to define ${\cal D}_{1,2}\equiv D_2 \pm D_1$ and
$\overline{{\cal D}_{1,2}}\equiv \overline{D_2} \pm \overline{D_1}$
respectively.
The $D$-sector terms in the partition function then become
\beq
         {\textstyle {1\over 4}}\,
        \left(\overline{D_2}^{\,\ast} D_1
        -\overline{D_1}^{\,\ast} D_2 \right)~=~
         {\textstyle {1\over 8}}\,
        \left(\overline{{\cal D}_2}^{\,\ast} {\cal D}_1
        -\overline{{\cal D}_1}^{\,\ast} {\cal D}_2 \right)~,
\label{newDterms}
\eeq
but consulting the matrices
(\ref{Smatrixholo}) and (\ref{Smatrixantiholo}),
we now find
\beqn
   &&~ S({\cal D}_1,A_3)~\not=~0~,\nonumber\\
   &&~ S({\cal D}_2,A_3)~=~0~,~~~~~S({\cal D}_2,C_3)~\not=~0~,\nonumber\\
   &&~ S(\overline{{\cal D}_1},\overline{C_3})~\not=~0~,\nonumber\\
   &&~ S(\overline{{\cal D}_2},\overline{C_3})~=~0~,~~~~~
       S(\overline{{\cal D}_1},\overline{D_2})~=~0~.
\label{newSelements}
\eeqn
This means that these new ${\cal D}$ and $\overline{\cal D}$
combinations have the following exponential rates of growth:
\beqn
        {\cal D}_1 ~~&\Longleftrightarrow&~~ C= 4\pi\nonumber\\
        \lbrace {\cal D}_2,\,\overline{{\cal D}_1}\rbrace
    ~~&\Longleftrightarrow&~~ C= 2\sqrt{2}\,\pi\nonumber\\
    \overline{{\cal D}_2} ~~&\Longleftrightarrow&~~ C\leq \sqrt{2}\,\pi~.
\label{newCs}
\eeqn
Note that the growth for $\overline{{\cal D}_2}$ is extraordinarily
suppressed relative to the others, with all leading $\alphabar=1$
terms in its expansion vanishing.
We thus easily\footnote{
    The reader may therefore wonder why we did not define ${\cal D}_{1,2}$
    and $\overline{{\cal D}_{1,2}}$ to be our pseudo-characters
    originally.  There are two reasons.  The first is that if we had
    expressed our partition function solely in terms of the ${\cal D}$
    pseudo-characters, the condition (\ref{choicetwo}) would not have
    been satisfied, since {\it all}\/ of the ${\cal D}$ pseudo-characters
    are tachyonic with $H<0$ (as opposed to only {\it two}\/ of the
    $D$ pseudo-characters).  Thus, the structure of the physical and unphysical
    tachyons in this example would have been less apparent, and one would have
    needed to verify explicitly via $q$- and $\qbar$-expansions of the
    partition function that all physical tachyons are indeed cancelled.
    The second reason concerns the ${\cal D}$-combinations themselves,
    for their $q$-expansion coefficients are not all of the same sign for
    all $n$ (indeed, both ${\cal D}_2$ and $\overline{{\cal D}_2}$
    have $q$-expansion coefficients which are negative at small $n$
    but positive for large $n$).  This is related to the fact that
    the leading asymptotic terms vanish for these ${\cal D}$ combinations,
    and that the various subleading terms have different signs.
    It thus requires greater values of $n$ for one of these subleading terms
    to become dominant, and thereby fix a sign for the coefficients
    as $n\to\infty$.}
find that the dominant growth comes from the
$\overline{{\cal D}_1}^{\,\ast} {\cal D}_2$ term in (\ref{newDterms}),
with
\beq
    \lbrace a_{nn}^{(D)}\rbrace ~~\Longleftrightarrow~~ C=4\sqrt{2}\,\pi~.
\label{Dshell}
\eeq

These rates of growth for the individual sector degeneracies
$a_{nn}^{(A,B,C,D)}$ which are given in (\ref{ABCshells}) and (\ref{Dshell})
can also be verified by explicitly expanding
$Z$ as a function $q$ and $\qbar$,
$Z=({\rm Im}\,\tau)^{-1}\sum a_{mn}\qbar^m q^n$.
In Fig.~4 we have plotted the resulting physical degeneracies $a_{nn}$
as a function of $n$, following the conventions
of our earlier figures.  We once again observe the appearance of
a ``misaligned supersymmetry'', with alternating signs for the
net numbers of physical states.  The largest rate of growth comes
from the $A$- and $C$- sectors, in agreement with (\ref{ABCshells}),
and the smaller rate of growth is experienced by the $B$- and $D$-
sectors, in agreement with (\ref{ABCshells}) and (\ref{Dshell}).
The presence of only four sectors in this example renders the pattern
of oscillation significantly simpler than that for the first example
we considered.

Let us now determine the extent to which the ``misaligned supersymmetry''
in this example implies an exact cancellation of functional forms for
the sector-averaged number of states $\langle a_{nn}\rangle$.
As we have seen, this partition function corresponds to
a worldsheet conformal field theory with
$c_{\rm left}=24$ and $c_{\rm right}=12$, and therefore
each individual conformal field theory sector $(\ibar i)$
of the theory separately experiences leading exponential growth with
$C_{\rm tot}=2(2+\sqrt{2})\pi \approx 6.83 \pi$.
We thus easily see that the $C_{\rm eff}$ for this theory is less
than $C_{\rm tot}$, for our organization in terms of the above
pseudo-characters
has made the cancellation of this leading $C=2(2+\sqrt{2})\pi$
term manifest within each of the four sector-groupings, $A$ through $D$.
Hence, the strictest prediction of our theorem is in fact already
trivially verified for this example, with
$C_{\rm tot}=2(2+\sqrt{2})\pi$ and $C_{\rm eff}\leq 6\pi$.

However, just as in the previous example, we might suspect that even
further cancellations necessarily follow as a result of our theorem,
and this is indeed the case.  The next largest rates of growth are, of course,
$C=6\pi$ from the $A$- and $C$-sector groupings, and
$C=4\sqrt{2}\,\pi\approx 5.66\pi$ from the $B$- and $D$-sector groupings;
cancellations of these respective rates of growth would require first that
the {\it leading}\/ terms within the $A$- and $C$-sectors cancel directly, and
then that the remaining {\it subleading}\/ terms with $C=4\sqrt{2}\,\pi$ from
these sectors cancel against the {\it leading}\/ terms from the $B$- and
$D$-sectors.  However, our theorem can also be used to demonstrate that both of
these cancellations also occur,  for the theorem guarantees that the largest
contribution from the $\alpha=\alphabar=1$ leading terms
is that due to the unphysical tachyon in the $\overline{A_3}^{\,\ast}A_3$
term, yielding only $C_{\rm eff}=4\pi$, while the largest subdominant term
with $\alpha>1$ and/or $\alphabar>1$ is easily found to be that with
$(\alpha,\alphabar)=(1,2)$ and $(H_j,\Hbar_\jbar)= (-1,-1/2)$, yielding only
$C=(4+\sqrt{2})\pi\approx 5.41\pi$.

Thus we conclude that the leading terms from all four of our sectors cancel
completely in this case, with the contributions from the $A$-, $B$-, $C$-,
and $D$-sectors all cancelling with each other as a result of the misaligned
supersymmetry.  Indeed, despite the separate rates of growth given in
(\ref{ABCshells}) and (\ref{Dshell}), we find that
$C_{\rm eff}\leq (4+\sqrt{2})\pi$.
This is an extraordinary cancellation for $\langle a_{nn}\rangle$,
amounting to nearly eight orders of magnitude for energies
as small as $n\approx 10$.

\subsection{$C_{\rm eff}$ and the Subleading Terms}

In the previous two examples, we witnessed some remarkable cancellations,
with all of the leading $\alpha=\alphabar=1$ terms cancelling in the
summation over sectors leading to $\langle a_{nn}\rangle$.
Indeed, in each case we were able to explicitly demonstrate that
the value of $C_{\rm eff}$ was {\it at most}\/ that of the
largest {\it subleading}\/ term with either $\alpha >1$ or $\alphabar >1$.
Given these results, then, it is naturally tempting to take the next step,
and determine whether any {\it further}\/ cancellations occur between
the purely subdominant contributions with either $\alpha$
or $\alphabar$ greater than $1$.  Indeed, we shall see in Sect.~5 that
this second example is somewhat special by virtue of its vanishing one-loop
cosmological constant \cite{KRD}, and we would
therefore anticipate that many more
cancellations should occur for this case in particular.
(This function is special for other reasons as well;
a detailed discussion of this function and its properties can be found
in Ref.~\cite{KRD}.)
Moreover, we conjectured at the end of Sect.~3 that {\it all}\/ subleading
terms should cancel in general,
resulting in an ultimate value $C_{\rm eff}=0$.
This would of course require an {\it infinite}\/ number of additional
cancellations, for there are an infinite number of subleading
terms contributing smaller and smaller values of $C$.

Regrettably, however, it is not possible to proceed any further
for these examples, and check explicitly whether such additional subleading
cancellations actually occur using the asymptotic-expansion
formalism presented in Sects.~2 and 3.  The reason, as briefly mentioned at
the end of Sect.~3, concerns the suitability of these asymptotic expansions
for situations (such as our calculation of $\langle a_{nn}\rangle$) in
which the energy $n$ is treated as a continuous variable.
In particular, recall the form of the final result (\ref{finalexpansion}):
the coefficients ${\cal P}^{(\alphabar,\alpha)}_{\jbar j}$
with either $\alpha>1$ or $\alphabar>1$
are precisely the coefficients which
determine whether such subleading terms survive the sector-averaging
process. It turns out, however, that the definition of these coefficients
given in (\ref{Pdef}) is ultimately unsuitable for
the $\alpha,\alphabar>1$ cases.

In order to see why this is so,
let us first recall the original derivation of the asymptotic expansions
in Sect.~2.  The asymptotic expansions of the coefficients
of any individual character $\chi_i(q)$ are given in
(\ref{finalone}), and in particular the coefficients $Q_{ij}^{(\alpha)}$
are given in (\ref{Qdef}).  Later in Sect.~2, we evaluated these
coefficients explicitly, with the results listed
in (\ref{Qone}), (\ref{Qtwo}), (\ref{Qthree}), and (\ref{Qfourfivesix}).
It was important for the consistency of these results and for the
validity of the asymptotic expansions in general that these
coefficients $Q^{(\alpha)}_{ij}$ be real quantities;
since each value of $\alpha$ ultimately corresponds to
a different rate of exponential growth for the chiral degeneracies $a_n^{(i)}$,
there is no way that complex coefficients $Q_{ij}^{(\alpha)}$
can combine to yield a value of $a_n^{(i)}$ which is real
for each value of $n$.
However, while the dominant coefficient $Q_{ij}^{(1)}$ is indeed real for
all values of $n$, the results given for all of the $Q_{ij}^{(\alpha)}$ with
$\alpha >1$ implicitly assumed that $n\in {\IZ}$.  For example,
tracing again the steps leading to $Q^{(3)}_{ij}$ and {\it not}\/
assuming that $n\in {\IZ}$, we actually obtain
\beq
     Q_{ij}^{(3)}~=~ e^{-\pi i n}\,2\,{\rm Re}\,
     \left\lbrace e^{\pi i k/2} \,(ST^3S)_{ij}\,
    \exp\left[ {{2\pi i}\over 3}\, (n+ H_i+H_j)\right]
     e^{-\pi i n} \right\rbrace~
\label{Qthreenew}
\eeq
instead of (\ref{Qthree}).
While this result of course reduces to (\ref{Qthree})
for $n\in {\IZ}$, for other values of $n$ this quantity is
actually complex, with phase $e^{-i \pi n}$.
This phase is in fact the same for {\it all}\/ of the $\alpha >1$
coefficients, with the modifications to the other values of
$Q^{(\alpha)}_{ij}$ taking the same general form as (\ref{Qthreenew}).

This observation is very important, because ultimately these
coefficients $Q^{(\alpha)}_{ij}$ became the building blocks
of the coefficients $P^{(\alphabar,\alpha)}_{\jbar j}$
defined in (\ref{Pdef}), and these are precisely the quantities
whose values determine whether or not the subleading terms cancel.
There is thus no guarantee that the coefficients
$P^{(\alphabar,\alpha)}_{\jbar j}$ with $\alpha>1$ and/or $\alphabar>1$
will be real, and in fact
the first non-cancelling subleading term in the second example
we examined turns out to be complex.
We thus see that the very definitions of
$P^{(\alphabar,\alpha)}_{\jbar j}$ for $\alpha>1$ and/or $\alphabar>1$
are unsuitable as coefficients in the asymptotic expansions for
the sector-averaged quantities $\langle a_{nn}\rangle$,
and that an intrinsically different sort of asymptotic expansion is necessary.
Note that this is the same conclusion we reached at the end of Sect.~3.
Indeed, what is needed is an asymptotic expansion
which is calculated directly for sector-averaged quantities such
as $\langle a_{nn}\rangle$ in which the energy $n$ is to be regarded as
a continuous or ``sector-averaged'' variable.  Such an expansion would
also hopefully have many of these cancellations built in at an early stage,
and thereby enable us to efficiently determine the precise value of
$C_{\rm eff}$ and the behavior of $\langle a_{nn}\rangle $ as $n\to\infty$.
We shall outline the steps by which such an
expansion might be obtained in Sect.~6.

Note, however, that
the lack of such a suitable asymptotic expansion does not
affect the primary cancellations of the {\it leading}\/
terms with $(\alpha,\alphabar)=(1,1)$, and in particular the validity
of our theorem relies upon cancellations between only these terms.
Thus, our result that $C_{\rm eff}<C_{\rm tot}$, and its implications
concerning ``misaligned supersymmetry'', remain unaltered.

\vfill\eject
\setcounter{footnote}{0}
\section{Finiteness and the Cosmological Constant}

As discussed in the Introduction, modular
invariance and the absence of physical tachyons are the conditions
which guarantee finite loop amplitudes in string theory.
Since these are also the conditions which yield the ``misaligned
supersymmetry'', it is natural to interpret the resulting boson/fermion
oscillation as the mechanism by which the net numbers of states
in string theory distribute themselves level-by-level so as to produce
finite amplitudes.  In this section we will provide some evidence
for this by focusing on the simplest loop amplitude in string
theory, namely the one-loop vacuum polarization amplitude
or cosmological constant, defined as
\beq
    \Lambda ~\equiv ~ \int_{\cal F} {d^2\tau\over{({\rm Im}\,\tau)}^2}
        ~Z(\tau,\overline{\tau}) ~=~
    \int_{\cal F} {d^2\tau\over{({\rm Im}\,\tau)}^2}
        ~({\rm Im}\,\tau)^k \,\sum_{m,n} a_{mn}\,\qbar^m\,q^n~.
\label{lambdadef}
\eeq
Here ${\cal F}$ is the fundamental domain of the modular group,
\beq
   {\cal F}~\equiv~\left\lbrace \tau~|~
      {\rm Im}\,\tau>0,~
     -\half \leq {\rm Re}\,\tau \leq \half,~
      |\tau| \geq 1\right\rbrace~,
\label{fundamentaldomain}
\eeq
and one is instructed to integrate over $\tau_1\equiv {\rm Re}\,\tau$
in (\ref{lambdadef}) before integrating over $\tau_2\equiv {\rm Im}\,\tau$
in the $\tau_2>1$ region.

The first thing we notice about (\ref{lambdadef}) is that
contributions to $\Lambda$ come from both the physical states
with $m=n$, and the unphysical states with $m\not=n$.  This
arises because the fundamental domain ${\cal F}$ consists of
two distinct regions in the complex $\tau$-plane:  an infinite
rectangular-shaped region ${\cal F}_1$ with ${\rm Im}\,\tau\geq 1$, and
a curved region ${\cal F}_2$ with ${\rm Im}\,\tau\leq 1$.
Integrating over the rectangular region ${\cal F}_1$, we see that
only the physical states with $m=n$ contribute, for the contributions
from terms in (\ref{lambdadef}) with $m\not= n$ are cancelled
in the integration over $\tau_1$.  However, since the
curved second region ${\cal F}_2$ does not extend the over the full
range $-\half\leq \tau_1 \leq \half$, the unphysical contributions from this
part of the integration are not completely cancelled,
and unphysical (or ``off-shell'') states with $m\not =n$ thereby
contribute to loop amplitudes.

This would in principle cause a problem for us, since our result concerning
misaligned supersymmetry applies to only the {\it physical}\/ states
with $m=n$.\footnote{
      It is a trivial exercise, however, to extend the result in Sect.~3
      to apply to unphysical states as well;  indeed, these unphysical
      states also experience analogous ``misaligned supersymmetries''
      with analogous boson/fermion oscillations.}
Furthermore, even after the $\tau_1$-integration is performed in
(\ref{lambdadef}), one is left with the integration over $\tau_2$, and
it is not readily clear how our result concerning the coefficients $a_{nn}$
will then translate into a result concerning the finiteness of $\Lambda$.
In particular, note that (\ref{lambdadef}) is manifestly finite if the
spectrum is free of physical tachyons, since the presumed modular invariance
of the theory has already been used to truncate the region of
$\tau$-integration to the fundamental domain ${\cal F}$ and thereby avoid the
dangerous ultraviolet $\tau\to 0$ region.  Thus the finiteness of
(\ref{lambdadef}) only implicitly rests on the behavior of the state
degeneracies, and it is difficult to see how to directly relate the two.

All of these problems can be circumvented,
however, due to a remarkable result of Kutasov and Seiberg \cite{kutsei}
which expresses the one-loop cosmological constant directly in terms
of only the {\it physical}\/-state degeneracies, and which
does so {\it without}\/ subsequent $\tau_2$-integrations.
Their result is as follows.
Let us first define the quantity
\beq
     g(\tau_2)~\equiv~  \int_{-1/2}^{+1/2} d\tau_1 \, Z(\tau_1,\tau_2)~.
\label{gdef}
\eeq
Substituting the general $q$-expansion for $Z$ given
in (\ref{lambdadef}) and performing the $\tau_1$-integration, we easily
find that $g(\tau_2)$ receives contributions from only the physical states:
\beq
          g(\tau_2)~=~ {\tau_2}^k\, \sum_{n} \, a_{nn}\,
              \exp(-4\pi n\tau_2)~.
\label{gann}
\eeq
Indeed, we can interpret $g(\tau_2)$ as being a regulated measure of the
total number of physical states in the theory, with $\tau_2$ serving
as the cutoff which regulates this divergent quantity.  Therefore
$\lim_{\tau_2\to 0} g(\tau_2)$ might be interpreted as
giving the total number of states {\it without}\/ any cutoff.
One would naively expect this quantity to diverge, since
$k$ is negative in most situations of interest and since $\sum_n a_{nn}$ is
formally a divergent quantity.  However, Kutasov and Seiberg
show \cite{kutsei} that this limit is
actually {\it finite}\/ in any modular-invariant theory which
is free of physical tachyons, and moreover
\beq
            \lim_{\tau_2\to 0}\, g(\tau_2) ~=~ {3\over \pi}\,\Lambda~.
\label{KSresult}
\eeq
This result is quite general, and applies to all partition functions
$Z$ which are free of physical tachyons ({\it i.e.}, which
have $a_{nn}=0$ for all $n<0$), and whose {\it unphysical}\/ tachyons
are not {\it too}\/ tachyonic.  Explicitly, this latter condition
states that $Z$ must have non-zero values of $a_{mn}$ only for
$m\geq m_0$ and $n\geq n_0$ where $m_0,n_0>-1$.  Results similar
to (\ref{KSresult}) can nevertheless
be obtained for cases in which this last condition is violated \cite{kutsei}.

The result (\ref{KSresult}) is quite powerful, since it enables us
to formally calculate the complete one-loop
cosmological constant (\ref{lambdadef})
given knowledge of only the {\it physical}\/ state degeneracies.
Indeed, this implies that the assumed modular invariance of $Z$ is sufficiently
strong a constraint that these physical degeneracies themselves determine
the contributions to $\Lambda$ from the {\it unphysical}\/ states
as well.\footnote{
       Note that this does not fix the actual
       number and distribution of unphysical
       states, but instead determines only their total integrated
       contribution to $\Lambda$.}
However, (\ref{KSresult}) can also be interpreted as a severe constraint
on the distributions of the {\it physical}\/ states in any
tachyon-free modular-invariant theory, for somehow the net degeneracies
$\lbrace a_{nn}\rbrace$ in (\ref{gann}) must arrange themselves in such a way
that $\lim_{\tau_2\to 0} g(\tau_2)$ is finite.
Indeed, such an arrangement is evidently precisely what is required to
yield finite amplitudes.  Thus, via this result (\ref{KSresult}),
we are furnished with a mathematical condition
on a modular-invariant set of
degeneracies $\lbrace a_{nn}\rbrace$ which is necessary and sufficient
to yield a finite one-loop cosmological constant:
\beq
         \lim_{\tau_2\to 0}\,g(\tau_2) ~<~ \infty~,
\label{KScondition}
\eeq
or explicitly,
\beq
     \lim_{\tau_2\to 0}\,{\tau_2}^k \,\sum_{n} a_{nn}\,\exp(-4\pi \tau_2 n)
    ~<~\infty~.
\label{KSconditiontwo}
\eeq
For physical situations in which the modular weight $k$ is
negative ({\it i.e.}, for $D>2$), this in fact requires
\beq
     \lim_{\tau_2\to 0} \,\sum_{n} \,a_{nn}\,\exp(-4\pi \tau_2 n)
    ~=~0~.
\label{KSconditionthree}
\eeq

 {\it A priori}, there are any number of conceivable distributions
$\lbrace a_{nn}\rbrace$ which might satisfy the finiteness condition
(\ref{KSconditionthree}), and this condition alone is therefore
not sufficiently restrictive to {\it predict}\/ the resulting behavior
for the actual distribution of physical states.
For example, simple distributions which trivially
satisfy (\ref{KSconditionthree}) with $n\in{\IZ}$ include
$a_{nn}= (n^2-n-\quarter)/(n^2-\quarter)^2$ for $n\geq 1$,
or $a_{nn}=r^n-(1-r)^{-1}\delta_{n,0}$ for all $n\geq 0$ and $r<1$.
The condition (\ref{KSconditionthree}) can
nevertheless be used to {\it rule out}\/ certain behavior for
the net degeneracies $a_{nn}$.
For example, it is straightforward to
show that if $a_{nn}\sim n^{-B} e^{C\sqrt{n}}$ with $C> 0$
as $n\to\infty$,
then
\beq
        g(\tau_2) ~\sim~
     (\tau_2)^{k+2B-3/2}\,\exp\left({C^2\over 16\pi\tau_2}\right)~
    ~~~~~{\rm as}~\tau_2\to 0~.
\label{badresult}
\eeq
Thus the finiteness condition
(\ref{KSconditionthree}) cannot be satisfied for any
$B$ and $C$, and as a consequence all direct exponential growth for
the net degeneracies $\lbrace a_{nn}\rbrace$ is prohibited.

Without knowledge of misaligned supersymmetry and the
resulting boson/fermion oscillation,
this last result would seem quite remarkable, since we know that
each individual sector contributes a set of
degeneracies $\lbrace a_{nn}^{(\ibar i)}\rbrace$
which {\it does}\/ grow exponentially with $C=C_{\rm tot}$.  Indeed,
$C_{\rm tot}$ is the inverse Hagedorn temperature, and it is precisely
this exponential growth which is responsible for the famous
Hagedorn phenomenon which is thought to signal a phase transition
in string theory.

However, misaligned supersymmetry and the resulting boson/fermion
oscillation now provide a natural alternative solution
which reconciles the finiteness condition (\ref{KSconditionthree}) with
growing behavior for which $|a_{nn}|\to\infty$ as $n\to\infty$.
Indeed, we can easily see that such oscillations
permit {\it many}\/ growing solutions to (\ref{KSconditionthree});
for example, simple distributions such as
$a_{nn}=(-1)^n n^2$, $a_{nn}=(-1)^n n^4$,
$a_{nn}=(-1)^n (n- n^5)$, and $a_{nn}=(-1)^n (2n^3 +n^5)$
all non-trivially satisfy (\ref{KSconditionthree}),
where the factor of $(-1)^n$ is meant to illustrate the
alternating-sign behavior for $a_{nn}$ which
is implicit in the boson/fermion oscillations.
In fact, we can easily show
that (\ref{KSconditionthree}) is satisfied for {\it any}\/ $a_{nn}= (-1)^n
f(n)$
provided the function $f(n)$ is even in $n$, with a Taylor-expansion
of the form $f(n)=\sum_{k=1}^\infty c_k n^{2k}$ with no constant term.
The argument goes as follows.  Since
\beq
       \sum_{n=0}^\infty \,(-1)^n e^{-4\pi \tau_2 n}
     ~=~ {1\over 1+e^{-4\pi\tau_2}}
         ~=~ {1\over 2} ~+~ {1\over 2} \left(
      {\sinh 4\pi\tau_2 \over 1+\cosh 4\pi\tau_2} \right)~,
\label{hyperidentity}
\eeq
we have
\beq
   g(\tau_2)~=~\sum_{n=0}^\infty \, (-1)^n f(n) e^{-4\pi \tau_2 n} ~=~
   \sum_{k=1}^\infty
     c_k \left({-1\over 4\pi} {d\over d\tau_2}\right)^{2k} \left\lbrace
     {1\over 2}~+~ {1\over 2} \left(
      {\sinh 4\pi\tau_2 \over 1+\cosh 4\pi\tau_2} \right) \right\rbrace~.
\label{fplushyperidentity}
\eeq
But now we see that the derivatives remove the contributions from the
additive factor of $\half$ within the braces, and of course the
remaining term within the braces is odd in $\tau_2$.
Thus we find that $g(\tau_2)$ itself is an odd function of $\tau_2$,
and since $g(\tau_2)$ is continuous at $\tau_2 = 0$, we have
$\lim_{\tau_2\to 0} g(\tau_2)=0$.

This heuristic argument demonstrates that the non-trivial
critical ingredient in the success of all of these functions
is their pattern of oscillation which is represented schematically
by $(-1)^n$.  Indeed, it is precisely this oscillation which permits
solutions for which $|a_{nn}|\to\infty$ as $n\to\infty$ to satisfy
(\ref{KSconditionthree}), and which therefore renders
the Hagedorn-type growth of the separate numbers of bosonic
and fermionic states in string theory consistent with the finiteness
condition (\ref{KSconditionthree}).

Of course, our complete result predicts much more than this, yielding
not only the overall boson/fermion oscillation,
but also detailed information concerning both the actual
asymptotic forms of the degeneracies $a_{nn}$ as $n\to \infty$,
and their cancellations upon sector-averaging.
Indeed, our result demands that all traces of the leading exponential
behavior with $C=C_{\rm tot}$ must cancel, and we have seen
that our theorem very often implies that
many subleading terms with smaller values of $C$ must cancel as well.
The examples presented in Sect.~4 showed that these
cancellations can in fact be quite subtle,
with the subleading terms from one sector cancelling against
the leading terms from another sector, and we found that the
combined effect of these successive cancellations was often quite dramatic.

However, the result (\ref{KSconditionthree}) now presents us with a
slightly different tool in interpreting these cancellations.
The vanishing of the limit (\ref{KSconditionthree})
depends crucially on two seemingly-separate properties
of the degeneracies $\lbrace a_{nn}\rbrace$:  their {\it high}\/-energy
behavior for large $n$ must be such that this limit {\it converges},
and then the value at which this convergence takes
place must exactly balance the contributions to the limit which
come from the {\it low}-energy states.
Let us focus for the moment on the high-energy states, and consider
the question of convergence.
We have seen that oscillatory asymptotic behavior for
the $\lbrace a_{nn}\rbrace$ generically yields convergence,
with the rapid fluctuations in some sense cancelling each other
and yielding no net contribution to (\ref{KSconditionthree}).
This is precisely the content of our main result, which asserts
that all traces of the $C=C_{\rm tot}$ behavior of each individual sector
are removed in the sector-averaged quantity $\langle a_{nn}\rangle$.
In this sense, then, we might approximate the $a_{nn}$
for asymptotically high values of $n$ in (\ref{KSconditionthree}) by
$\langle a_{nn}\rangle$, and work only with the net functional
form that this quantity represents.

However, assuming that this replacement is justified at high $n$,
we now reach an interesting conclusion.
The asymptotic behavior of $\langle a_{nn}\rangle$ is governed, of course,
not by $C_{\rm tot}$, but by $C_{\rm eff}$.
Thus, if $C_{\rm eff}\not= 0$, then even the value of $\langle a_{nn}\rangle$
experiences exponential growth, for there must remain subleading terms in
$\langle a_{nn}\rangle$ with $C=C_{\rm eff}$
whose contributions remain uncancelled.
But even this residual exponential growth would be dangerous for the
convergence of (\ref{KSconditionthree}), for we have seen
in (\ref{badresult}) that {\it any}\/ net exponential growth causes
$g(\tau_2)$ to diverge as $\tau_2\to 0$.
Thus we would conjecture that we must in fact have
\beq
           C_{\rm eff}~ {\buildrel {?} \over =}~ 0
\label{conjecture}
\eeq
for any modular invariant theory which is free of physical tachyons.
Indeed, this would require that {\it all}\/ exponential
growth cancel in the sector-averaging process, whether this
growth is leading or subleading.  This would of course necessitate
an {\it infinite}\/ number of cancellations, for
there are indeed an infinite number of such subleading terms
whose cancellations would necessarily become more and more intricate
and inter-related.

Given the conjecture $C_{\rm eff} {\buildrel {?}\over =} 0$,
let us now carry this argument one step further, and proceed to
examine the situation with {\it polynomial}\/ growth,
\beq
     \langle a_{nn}\rangle ~\sim~ n^\epsilon~,~~~~~~~ \epsilon \geq 0~.
\label{conjsteptwo}
\eeq
For theories without physical tachyons, the cosmological constant
$\Lambda$ is finite, and thus the finiteness
condition (\ref{KSconditiontwo}) implies
\beq
         \lim_{\tau_2\to 0}\,\sum_{n} \,a_{nn}\,\exp(-4\pi\tau_2 n)~\sim~
             {\tau_2}^{-k}~.
\label{conjstepone}
\eeq
Once again assuming that we can replace $a_{nn}$ by the
sector-average $\langle a_{nn}\rangle$,
we thereby obtain the constraint
\beq
    \lim_{\tau_2\to 0}\,\sum_{n} \,n^\epsilon\,\exp(-4\pi\tau_2 n)~\sim~
         {\tau_2}^{-k}~,
\label{conjstepthree}
\eeq
and since
\beq
    \lim_{\tau_2\to 0}\,
   \left({-1\over 4\pi}{d\over d\tau_2}\right)^\epsilon\,
      \sum_n \, \exp(-4\pi\tau_2 n)~\sim~
   \left({-1\over 4\pi}{d\over d\tau_2}\right)^\epsilon\,
      \left({1\over  4\pi\tau_2}\right)~\sim~(\tau_2)^{-(1+\epsilon)}~,
\label{conjstepfour}
\eeq
we find $\epsilon= k-1$.
Similar arguments can also be used to show that for $k\leq 0$,
we must have $\langle a_{nn}\rangle\to 0$ as $n\to\infty$
(indeed, for $k=0$ the fastest allowed growth for $\langle a_{nn}\rangle$
is $\sim r^n$ with $r<1$).
Thus, as $n\to\infty$, we have
\beqn
    \langle a_{nn}\rangle ~{\buildrel {?}\over \sim}~ \cases{
            0 &       for $k \leq  0$\cr
            n^{k-1} & for $k\geq 1$~.\cr}
\label{conjecturetwo}
\eeqn

These conjectures (\ref{conjecture}) and (\ref{conjecturetwo}) are certainly
appealing on aesthetic grounds.  Moreover, these are in fact the
 {\it same} conjectures that we made at the end of Sect.~3 on
the basis of a comparison between our two-variable
theorem and its one-variable counterpart.
This remarkable agreement indicates that there exists
a certain logical internal consistency to these conjectures,
and it would be interesting to see if a proof could be constructed
using what would necessarily be a fundamentally different type of
asymptotic expansion for $\langle a_{nn}\rangle$.
We shall briefly outline the inital steps by which such an
expansion might be obtained in Sect.~6.

\section{Concluding Remarks}

In this paper we have been able to prove a general theorem
concerning the distributions of physical bosonic and fermionic states
in any theory which is modular-invariant and which
contains no physical tachyons.
This result is therefore especially applicable to string theories lacking
spacetime supersymmetry, and we demonstrated that despite the absence
of such supersymmetries, a ``misaligned supersymmetry'' must nevertheless
survive in which any surplus of bosonic states at any mass level
implies a surplus of fermionic states at a higher mass
level, which in turn implies a surplus of bosonic states at a still higher
level, {\it etc}.
We demonstrated that this oscillation may be responsible
for the finiteness of string amplitudes by considering the case of
the one-loop cosmological constant, and
also showed that our general theorem
is the natural mathematical generalization of some powerful one-variable
theorems in modular-function theory to the physically relevant
cases of partition functions $Z(q,\qbar)$ of two variables.
Our analysis also introduced a new quantity, the so-called
``sector-averaged'' number of states $\langle a_{nn}\rangle$,
which may well be relevant for the asymptotic behavior of string
amplitudes at high energy, and we made a conjecture concerning
its behavior as $n\to\infty$.

There are nevertheless a number of possible extensions to our
results which we shall now briefly indicate;  these will hopefully
become the subjects of future research.
First, we have only considered the general question of finiteness
as it pertains to the case of the one-loop cosmological constant,
yet a demonstration that our misaligned supersymmetry implies
finiteness to all orders for all $n$-point functions
(or even finiteness for {\it all}\/ one-loop amplitudes)
would be a far more difficult task.  In particular, such functions
depend on knowledge of the interactions, and not just
the numbers of physical states.
Second, such investigations would almost certainly also
involve the examination of the behavior of the
degeneracies of the {\it unphysical}\/ (or so-called ``off-shell'')
states in string theory;  indeed, our focus in this paper has been on
the physical states, and it was only due to the result
of Kutasov and Seiberg discussed in Sect.~5 that this was sufficient
for discussing the finiteness of the one-loop cosmological constant.
We expect that the unphysical states also experience analogous
asymptotic cancellations and misaligned supersymmetries, however,
for the degeneracies and distributions of the unphysical states are
closely tied by modular invariance to those of the physical states.

A third issue which we have mentioned at various points in this paper
concerns the suitability
of the asymptotic expansions presented in Sect.~2 as building
blocks for the asymptotic expansions of
the sector-averaged number of states $\langle a_{nn}\rangle$;
not only are there problems with the behavior of certain
subleading coefficients, but the forms of these expansions
themselves are somewhat cumbersome for our purposes,
with successive cancellations occurring in highly non-trivial
ways which do not lend themselves to general analysis.
The structure of the error terms in these expansions also prohibited
a more powerful conclusion concerning the ultimate value of $C_{\rm eff}$,
and in particular we have conjectured on other grounds that in
fact $C_{\rm eff}=0$.
One might therefore hope instead for a reformulation of these asymptotic
expansions, for a new derivation which would proceed directly
from the definition of $\langle a_{nn}\rangle$ in a manner
analogous to that of Refs.~\cite{HR} and \cite{KV}.  We can
in fact see fairly quickly how such a derivation might be formulated.
Starting directly from the contour integrals in (\ref{contour}),
we would immediately combine the holomorphic and anti-holomorphic
sectors of the theory,
\beq
        a_{nn}^{(\ibar i)}~=~
      \abar_n^{(\ibar)} a_n^{(i)}~=~ {-1\over 4\pi^2}\,
     \oint d\qbar \oint dq ~ { \chibar_\ibar (\qbar)\,\chi_i(q) \over
        (\qbar q)^{n+1}}~.
\label{contourproduct}
\eeq
Note we are here using the shifted variables $n$ from
the beginning, as discussed in Sect.~3,
and therefore $n$ is not necessarily an integer.
Also note that $q$ and $\qbar$ are to be regarded as {\it independent}\/
variables in this analysis, with each independently
taking values along its respective contour.
We would then multiply (\ref{contourproduct}) by $N_{\ibar i}$ and
sum over $(\ibar,i)$ in order to build an expansion for
$\langle a_{nn}\rangle$ directly, yielding
\beq
        \langle a_{nn}\rangle~=~
      {-1\over 4\pi^2}\,
     \oint d\qbar \oint dq ~ \sum_{\ibar i} N_{\ibar i}\,
   { \chibar_\ibar (\qbar)\,\chi_i(q) \over
        (\qbar q)^{n+1}}~=~
     {-1\over 4\pi^2} \oint d\qbar \oint dq ~ {\sum_{\ibar i} N_{\ibar i}
    \,\chibar_\ibar (\qbar)\,\chi_i(q) \over
        (\qbar q)^{n+1}}~.
\label{contourproducttwo}
\eeq
Note that the second equality above follows from the presumed independence
of $n$ and $i$, which is the essence of the sector-averaging.
However, we now recognize that the numerator of the final expression
in (\ref{contourproducttwo}) is nothing but the partition
function $Z(q,\qbar)$ without its factor $({\rm Im}\,\tau)^k$.
We thus simply have
\beq
        \langle a_{nn}\rangle~=~
     {-1\over 4\pi^2} \oint d\qbar \oint dq ~
     {Z(q,\qbar) \over (\qbar q)^{n+1}  [(\tau-\overline{\tau})/(2i)]^k  }~
\label{anncontour}
\eeq
where we have generalized ${\rm Im}\,\tau\rightarrow
(\tau-\overline{\tau})/(2i)$ (as is necessary
for the invariance of $Z$ under simultaneous identical
modular transformations of $q$ and $\qbar$).
Thus, we see already that the behavior of $\langle a_{nn}\rangle$ now
directly depends on the behavior of the partition function $Z(q,\qbar)$
as $q$ and $\qbar$ approach the singular points on their respective
unit circles, and all cancellations between the expansions of the
individual characters comprising $Z$ have already been incorporated.
One could therefore hope to begin afresh from (\ref{anncontour}),
and derive an asymptotic expansion directly for
$\langle a_{nn}\rangle$ in which a minimum of unnecessary
cancellations appear and the error terms are minimized.
Such a derivation would
presumably parallel the steps and analyses given in Ref.~\cite{KV}
for the simpler case of the one-variable function $\chi_i(q)$,
although the appearance of two independent variables $q$ and $\qbar$ will
of course introduce new subtleties.

Finally, let us mention some potential applications of our results.
Perhaps the most important concerns the general question of
supersymmetry-breaking
in string theory;  our results concerning a residual misaligned supersymmetry
are of course quite general, and thus it should be possible to use
them to constrain the possible supersymmetry-breaking scenarios in string
theory.  For example, it would be interesting to understand on a deeper
level the relation between our misaligned-supersymmetry theorem and the various
soft-supersymmetry breaking theorems in string theory.  This would no doubt
entail developing a dynamical understanding of the types of symmetry-breaking
terms which, while breaking supersymmetry, nevertheless preserve the
cancellation of the functional forms which describe the separate
distributions of bosonic and fermionic states.
Note, in this regard, that our results clearly preclude any
supersymmetry-breaking scenario in which, for example,
the energies of fermionic states are merely shifted
(even infinitesimally) relative to
those of their bosonic counterparts.  Rather,
misaligned supersymmetry requires
that any such energy shifts must be
simultaneously accompanied by the introduction
or removal of a certain number $\Phi(n+\Delta n)-\Phi(n)$ of extra states,
where $\Phi(n)$ is the asymptotic function describing the density
of bosonic or fermionic
states and $\Delta n$ is magnitude of the induced energy shift.

Another application of our results concerns string theory at
the {\it hadronic}\/ scale, and the implications of our results as they relate
to the effective ``QCD strings'' which model hadron dynamics.  There have in
fact been some recent developments in this area.  Interpreting the various
bosonic states in such hadronic string theories as individual mesons, and
assuming that one can similarly model the corresponding fermionic string
states as baryons \cite{FR}, it has recently been
demonstrated \cite{cudelldien} that the actual numbers and distributions
of the experimentally observed meson and baryon states are not in disagreement
with the oscillations resulting from misaligned supersymmetry.
Indeed, since it is never possible to experimentally survey
the {\it infinite}\/ range of energies necessary in order to test
modular invariance directly, such oscillations generally serve
as the only profound yet indirect experimental ``signature'' of modular
invariance which is ``local'' in energy (acting level-by-level)
and thereby experimentally accessible.  It has even been possible, using
both the Kutasov-Seiberg result and the misaligned supersymmetry, to make
concrete predictions \cite{cudelldien,FR} concerning the appearance
and structure of new hadronic states.

The forms of these asymptotic expansions themselves have also played an
important role in recent analyses of the hadronic spectrum \cite{diencudell}.
It has long been known that the Hagedorn-type rise of the number
of meson states as $n\to\infty$ is indicative of an
effective string-like picture underlying the color flux tube in mesons,
and by fitting the simple Hagedorn exponential form
$a_n\sim n^{-b} e^{C\sqrt{n}}$ to the meson spectrum one
obtains \cite{Hagedorn} the well-known result $T_H\equiv C^{-1}\approx
160$ MeV [where $\alpha'$, the Regge slope, is taken to be
0.85 (GeV)$^{-2}$].  From this one can directly calculate the central
charge of the effective ``QCD string'' underlying meson spectroscopy,
obtaining $c\approx 7$.  However, this result is in strong disagreement
with the majority of QCD string proposals which have $c\approx 2$,
and indeed alternative analyses of the central charge
of the QCD string which compare the results of the effective
static-quark potential with data also predict $c\approx 2$.
It has recently been demonstrated \cite{diencudell}, however,
that the conflict between these two results arises from the incorrect
application of the simple Hagedorn form $n^{-b} e^{C\sqrt{n}}$
to data which is {\it not}\/ sufficiently asymptotic in energy.
Indeed, using the complete asymptotic forms discussed in this paper
(most notably, the replacement of the above Hagedorn exponentials by
Bessel functions) profoundly alters the results, and yields new
estimates of the Hagedorn temperature which are now in agreement
with the result $c\approx 2$.
Thus these asymptotic expansions, derived from the principles of
modular invariance and conformal invariance,
also turn out to have wide-ranging applications beyond
their primarily theoretical interest.
Taken together, then, these recent results demonstrate that the observed
hadronic spectrum is consistent with an underlying string theory in
which modular invariance plays a significant role.

There have also been other recent efforts to compare the rigorous
predictions of string theory with the predictions of QCD.
Those which are most closely related to the ideas of this paper
include attempts to construct a toy model of QCD which
analytically exhibits an infinite number of Regge trajectories,
an exponential rise in the number of meson states, and
a Hagedorn deconfining transition.
It has recently been shown \cite{QCDtwo} that
two-dimensional QCD coupled to adjoint matter has precisely
these properties, and this has sparked efforts \cite{QCDtwo} to
determine whether the spectrum of states predicted in such models
is consistent with the Kutasov-Seiberg constraint (\ref{KSconditionthree}).
Since this is in general a difficult constraint to verify,
it might be interesting (and simpler) to determine whether the spectrum
of this model exhibits a misaligned supersymmetry.
This would indeed be powerful evidence for the applicability of string-like
ideas to the realm of QCD.

Finally, of course, we remark that our result concerning misaligned
supersymmetry and the cancellation of the separate bosonic and fermionic
functional forms is, due to its generality, relevant for all
physical situations in which $\langle a_{nn}\rangle$, rather than
$a_{nn}$, plays a role.  This thereby includes all areas, such
as string thermodynamics or the high-energy behavior of string
scattering amplitudes,
in which the physics is determined by the high-energy asymptotic behavior of
the physical-state degeneracies.
Indeed, the full consequences of the reduction of the effective
degeneracy exponential growth rate from $C_{\rm tot}$ to $C_{\rm eff}$
are likely to be significant and far-reaching.
An investigation of some of these issues is currently underway.

\bigskip
\medskip
\leftline{\large\bf Acknowledgments}
\medskip

I am pleased to thank A.~Anderson, D.~Kutasov, C.S.~Lam, R.~Myers,
D.~Spector, H.~Tye, and E.~Witten for fruitful conversations.
I also wish to acknowledge the hospitality of the Aspen Center
for Physics, where portions of this paper were written.
This work was supported in part by NSERC (Canada) and les fonds FCAR
(Qu\'ebec).

\bigskip
\medskip
\vfill\eject

\bibliographystyle{unsrt}

\vfill\eject


%

 ~
\vfill
\noindent {\bf Fig.\ 1}:
The net number of physical states $a_{nn}$ for the toy model
(\ref{toyZ}), plotted versus the energy $n$ [equivalently the
spacetime (mass)$^2$].  Negative values of $\pm \log_{10} (|a_{nn}|)$ are
plotted for $a_{nn}<0$.  Also sketched is the sector-averaged number of
states $\langle a_{nn}\rangle$, assuming a cancellation scenario with
$C_{\rm eff}=C_{\rm tot}/4$.
\eject
\bigskip

 ~
\vfill
\noindent {\bf Fig.\ 2}:
The net number of physical states $a_{nn}$ for the partition function
(\ref{examplerewritten}), plotted versus worldsheet energy $n$ as in
Fig.~1.
The complex pattern of oscillation is due to the presence of six
sectors in this theory;  the ``wavelength'' of this oscillation
is nevertheless $\Delta n=1$.
\eject

 ~
\vfill
\bigskip
\noindent {\bf Fig.\ 3}:
The net number of physical states $a_{nn}$ for the partition function
(\ref{examplerewritten}), now plotted versus spacetime mass $\sqrt{n}$.
Note that the asymptotic exponential growth for each sector
begins quite early, with the sign of each $a_{nn}$ determined from the
partition function (\ref{examplerewritten}).
\eject

 ~
\vfill
\bigskip
\noindent {\bf Fig.\ 4}:
The net number of physical states $a_{nn}$ for the partition function
(\ref{Qrewritten}), plotted versus energy $n$ as in Fig.~1.
This partition function has a one-loop cosmological constant which
vanishes identically (see Ref.~\cite{KRD}), and --- like the partition
functions of heterotic string theories built from periodic/anti-periodic
worldsheet fermions --- contains exactly four sectors.
\eject

\end{document}